\documentclass[aps,preprint,floatfix,nofootinbib,showpacs]{revtex4-1}
\pdfoutput=1
\usepackage{graphicx,color,rotating}
\usepackage{hyperref}
\usepackage{epsfig}

\newcommand{\wt}{\widetilde}

\newcommand{\lsim}{\raisebox{-0.13cm}{~\shortstack{$<$ \\[-0.07cm] $\sim$}}~}
\newcommand{\gsim}{\raisebox{-0.13cm}{~\shortstack{$>$ \\[-0.07cm] $\sim$}}~}

\begin{document}
\renewcommand{\thefootnote}{\arabic{footnote}}

{\small
\begin{flushright}
CNU-HEP-15-01
\end{flushright} }

%\title{Higgcision in the Minimal Supersymmetric Standard Model}
\title{
Higgs data constraints \\ on the minimal supersymmetric standard model}
\author{
Kingman Cheung$^{1,2,3}$, Jae Sik Lee$^{3,4}$, and
Po-Yan Tseng$^1$}
\affiliation{
$^1$ Department of Physics, National Tsing Hua University,
Hsinchu 300, Taiwan \\
$^2$ Division of Quantum Phases and Devices, School of Physics,
Konkuk University, Seoul 143-701, Republic of Korea \\
$^3$ Physics Division, National Center for Theoretical Sciences,
Hsinchu, Taiwan \\
$^4$ Department of Physics, Chonnam National University, \\
300 Yongbong-dong, Buk-gu, Gwangju, 500-757, Republic of Korea
}
%12.60.Jv   Supersymmetric models (see also 04.65.+e Supergravity)
%14.80.Da   Supersymmetric Higgs bosons
\date{\today}

\begin{abstract}
  We perform global fits to 
  the most recent data (after summer 2014)
  on Higgs boson signal strengths
  in the framework of the minimal supersymmetric standard model
  (MSSM). 
We further impose the existing limits on the masses
of charginos, staus, stops and sbottoms together with
the current Higgs mass constraint $|M_{H_1} - 125.5\,{\rm GeV}| < 6$ GeV.
The heavy supersymmetric (SUSY) particles such as squarks
  enter into the loop factors of the $Hgg$
  and $H\gamma\gamma$ vertices while other SUSY
  particles such as sleptons and charginos also enter into that of the
  $H\gamma\gamma$ vertex. We also take into account the possibility of
  other light particles such as
  other Higgs bosons and neutralinos, such that
  the 125.5 GeV Higgs boson can decay into. We use the
  data from the ATLAS, CMS, and the Tevatron, with
  existing limits on SUSY particles, to constrain on the relevant
  SUSY parameters. We obtain allowed regions in the
  SUSY parameter space of squark, slepton and chargino masses, and the $\mu$
  parameter.
We find that 
$|\Delta S^\gamma/S^\gamma_{\rm SM}|\lsim 0.1$ at $68\%$ confidence level when 
$M_{\tilde{\chi}^{\pm}_1}>300$ GeV and $M_{\tilde{\tau}_1}>300$ GeV,
irrespective of the squarks masses. Furthermore,
$|\Delta S^\gamma/S^\gamma_{\rm SM}|\lsim 0.03 $
when $M_{\tilde{\chi}^\pm_1,{\tilde{\tau}_1}} > 500$ GeV
and $M_{{\tilde t}_1,{\tilde b}_1} \gsim 600$ GeV.
\end{abstract}

\maketitle

\section{Introduction}

The celebrated particle observed by the ATLAS \cite{atlas} and the
CMS \cite{cms}  Collaborations at the Large Hadron Collider (LHC) in July 2012
is mostly consistent with the standard model (SM) Higgs boson
than any other extensions of the SM \cite{Cheung:2013kla,Cheung:2014noa}, 
at least
in terms of some statistical measures. The SM Higgs boson was proposed
in 1960s \cite{higgs}, but only received the confirmation recently
through its decays into $\gamma\gamma$ and $ZZ^* \to 4 \ell$ modes.

Although the data on Higgs signal strengths are best described by the SM,
the other extensions are still viable options to explain the data.
Numerous activities occurred in the constraining the SM boson
\cite{Cheung:2013kla,r1,r2,r3,r4,r5,r6,r7,r8,r9,r10,r11,r12,r13,r14,r15,r16,r17,r18}, higher dimension operators of the Higgs boson
\cite{anom1,anom2,anom3,anom4,anom5,anom6}, the two-Higgs doublet models
\cite{2hdm0,2hdm1,2hdm2,2hdm3,2hdm4,2hdm5,2hdm6,2hdm7,2hdm8,2hdm9,Cheung:2013rva,
2hdm10,2hdm11,2hdm12},
and in the supersymmetric framework
\cite{susy1,susy2,susy3,susy4,susy5,susy6,susy7,susy8,susy9,susy10}.
A very recent update to all the data as of summer 2014
was performed in Ref.~\cite{Cheung:2014noa}. We shall describe
the most significant change to the data set in Sec. III.
In this work, we perform the fits in the framework of the
minimal supersymmetric standard model (MSSM) to all the most updated data
on Higgs signal strengths as of summer 2014.

In our previous analysis of the two-Higgs-doublet model (2HDM)
 \cite{Cheung:2013rva}, we do not
specify which neutral Higgs boson is the observed Higgs boson, so that
the whole scenario can be described by a small set of parameters.
The bottom and leptonic Yukawa couplings are determined through the
top Yukawa coupling, and the $HWW$ coupling is determined via $\tan\beta$
and top Yukawa, so that a minimal set of parameters includes only
$\tan\beta$ and the top Yukawa coupling.  We can easily include the
effects of the charged Higgs boson
by the loop factor in the $H\gamma\gamma$ vertex,
and include possibly very light Higgs bosons
by the factor $\Delta \Gamma_{\rm tot}$.
Here we follow the same strategy for the global fits in the framework of
MSSM, the Higgs sector of which is the same as the Type II of the 2HDM,
in order to go along with a minimal set of parameters,
unless we specifically investigate the spectrum of
supersymmetric particles, e.g., the chargino mass.

In this work, we perform global fits in the MSSM under various
initial conditions to the most
updated data on Higgs boson signal strengths. A few specific features
are summarized here.
\begin{enumerate}
\item We use a minimal set of parameters without specifying the spectrum
of the SUSY particles. For example, all up-, down- and lepton-type
Yukawa couplings and the gauge-Higgs coupling are given in terms of
the top Yukawa coupling, $\tan\beta$, and $\kappa_d$, where $\kappa_d$ is
the radiative correction in the bottom Yukawa coupling defined later.

\item Effects of heavy SUSY particles appear in the loop factors $\Delta S^g$
and $\Delta S^\gamma$ of the $Hgg$ and $H\gamma\gamma$ vertices, respectively.

\item Effects of additional light Higgs bosons or light neutralinos that
the 125.5 GeV Higgs boson
can decay into are included by the deviation
$\Delta \Gamma_{\rm tot}$ in the Higgs boson width.

\item CP-violating effects can occur in Yukawa couplings, which are
quantified by the CP-odd part of the top-Yukawa coupling. Effects of
other CP sources can appear in the loop factor of $Hgg$ and $H\gamma\gamma$
vertices. We label them as $\Delta P^g$ and $\Delta P^\gamma$, respectively.
In Ref.~\cite{Cheung:2014oaa}, we have computed all the
Higgs-mediated CP-violating contributions to the 
electric dipole moments (EDMs) and compared to
existing constraints from the EDM measurements of Thallium,
neutron, Mercury, and Thorium monoxide.
Nevertheless, we are content with CP-conserving fits in this work.

\item We impose the existing limits of chargino and stau masses when
we investigate specifically their effects on the vertex of $H\gamma\gamma$.
The current limit on chargino and stau masses are \cite{pdg}
\[
  M_{\tilde{\chi}^\pm} > 103.5 \;{\rm GeV},\qquad
  M_{\tilde{\tau}_1} > 81.9 \;{\rm GeV} \;.
\]
Similarly, the current limits for stop and sbottom masses quoted in PDG
are \cite{pdg}
\[
M_{{\tilde t}_1} > 95.7\;{\rm GeV}\,, \qquad
M_{{\tilde b}_1} > 89 \;{\rm GeV}\,,
\]
which will be applied in calculating the effects in $H\gamma\gamma$
and $Hgg$ vertices.
Note that the current LHC limits on the stop and sbottom masses
are $M_{{\tilde t}_1} > 650$ GeV and $M_{{\tilde b}_1} > 600$ GeV
at $95$\% confidence level in a simplified model
with $M_{\tilde{\chi}_1^0} = 0$ GeV~\cite{pdg}.
However, there often exist underlying assumptions of 
search strategies and the mass of the lightest neutralino.
Therefore, we conservatively take the 
above mass limits on the stops and sbottoms in most of
the analysis.
%except for 
%the MSSM-3 case in which all SUSY particles are involved
%and we shall show the 
%effect of imposing $M_{\tilde{t}_1} \agt 600$ GeV and the current
%Higgs mass constraint.
%

\item Since we shall try to find the implication of 
the current Higgs signal strength data 
on the SUSY spectrum, which in practice affects the lightest Higgs boson
mass, we therefore also calculate the corresponding Higgs boson mass
and impose the current Higgs mass constraint of $M_{H_1} \sim 125.5 \pm 6$
GeV, taking at a roughly $3$-$\sigma$ level. 

\end{enumerate}

The organization of the work is as follows. In the next section,
we describe the convention and formulas for all the couplings used
in this work.
In Sec. III, we describe various CP-conserving fits and present the results.
In Sec. IV, we specifically investigate the SUSY parameter
space of charginos, staus, stops, and sbottoms.
We put the synopsis and conclusions in Sec. V.

\section{Formalism}

For the Higgs couplings to SM particles we assume that the observed
Higgs boson is a generic CP-mixed state without carrying any
definite CP-parity. We follow the conventions and notation of
{\tt CPsuperH}~\cite{Lee:2003nta}. 

\subsection{Yukawa couplings}
The Higgs sector of the MSSM
is essentially the same as the Type II of the 2HDM. More details of the
2HDM can be found in Ref.~\cite{Cheung:2013rva}.
In the MSSM, the first Higgs doublet couples to the down-type quarks
and charged leptons while the second Higgs doublet couples to the up-type
quarks only. After both doublets take on vacuum-expectation values (VEV)
we can rotate the neutral components $\phi^0_1,\,\phi^0_2$ and $a$
into mass eigenstates $H_{1,2,3}$ through a mixing matrix $O$ as follows:
\[
 (\phi^0_1,\, \phi^0_2,\, a )_\alpha^T = O_{\alpha i} (H_1,\,H_2,\, H_3)_i^T \;,
\]
with the mass ordering $M_{H_1} \le M_{H_2} \le M_{H_3}$.
We do not specify which Higgs boson is the observed one, in fact,
it can be any of the $H_{1,2,3}$.
We have shown in Ref.~\cite{Cheung:2013rva} that the bottom and lepton Yukawa
couplings can be expressed in terms of the top Yukawa coupling in general
2HDM. We can therefore afford a minimal set of input parameters.

The effective Lagrangian governing the interactions of the neutral
Higgs bosons with quarks and charged leptons is
\begin{equation}
\label{eq1}
{\cal L}_{H\bar{f}f}\ =\ - \sum_{f=u,d,l}\,\frac{g m_f}{2 M_W}\,
\sum_{i=1}^3\, H_i\, \bar{f}\,\Big( g^S_{H_i\bar{f}f}\, +\,
ig^P_{H_i\bar{f}f}\gamma_5 \Big)\, f\ .
\end{equation}
At the tree level, $(g^S,g^P) = (O_{\phi_1 i}/c_\beta, -O_{ai}
\tan\beta)$ and $(g^S, g^P) = (O_{\phi_2 i}/s_\beta, -O_{ai}
\cot\beta)$ for $f=(\ell,d)$ and $f=u$, respectively, and
$\tan\beta\equiv v_2/v_1$ is the ratio of the VEVs of the two doublets.
Threshold corrections to the down-type Yukawa couplings
change the relation between the Yukawa coupling $h_d$ and mass $m_d$ as
\footnote{
  In general settings, $\kappa_{d}$ and $\kappa_{s}$ are usually the same,
but $\kappa_{b}$ could be very different because of the third generation
squarks. However, our main concern in this work is the third-generation
Yukawa couplings. Thus, we shall focus on $\kappa_b$ although we are using
the conventional notation $\kappa_d$.
}
\begin{equation}
\label{eq2}
h_d =\frac{\sqrt{2} m_d}{v \cos\beta}\,\frac{1}
{1+\kappa_d\tan\beta}\,.
\end{equation}
Thus, the Yukawa couplings of neutral Higgs-boson mass eigenstates $H_i$ to
the down-type quarks are modified as
\begin{eqnarray}
  \label{gSHbb}
g^S_{H_i\bar{d}d} & =& {\rm Re}\, \bigg(\,
\frac{1}{1\, +\, \kappa_d\,\tan\beta}\,\bigg)\,
\frac{O_{\phi_1 i}}{\cos\beta}
\ +\ {\rm Re}\, \bigg(\, \frac{\kappa_d}{1\, +\,
\kappa_d\, \tan\beta}\,\bigg)\
\frac{O_{\phi_2 i}}{\cos\beta}\nonumber\\
&& +\: {\rm Im}\, \bigg[\,
\frac{ \kappa_d\, (\tan^2\beta\, +\, 1)}{1\, +\,
\kappa_d\, \tan\beta}\,\bigg]\
O_{ai}\, , \nonumber\\[0.35cm]
  \label{gPHbb}
g^P_{H_i\bar{d}d} & =& -\, {\rm Re}\, \bigg(\,
\frac{ \tan\beta\, -\, \kappa_d}{1\, +\, \kappa_d \tan\beta}\,\bigg)\, O_{ai}
\ +\ {\rm Im}\, \bigg(\, \frac{\kappa_d\,\tan\beta}{1\, +\,
\kappa_d\, \tan\beta}\,\bigg)\
\frac{O_{\phi_1 i}}{\cos\beta}\nonumber\\
&&-\: {\rm Im}\, \bigg(\,
\frac{\kappa_d}{1\, +\, \kappa_d\, \tan\beta}\,\bigg)\
\frac{O_{\phi_2 i}}{\cos\beta}\ ,
\end{eqnarray}

In the MSSM, 
neglecting the electroweak corrections and taking the most dominant
contributions,
$\kappa_b$ can be split into \cite{Lee:2004} 
\[
\kappa_b 
= \epsilon_g + \epsilon_H ,
\]
where $\epsilon_g$ and $\epsilon_H$ are the contributions from the
sbottom-gluino exchange diagram and from stop-Higgsino diagram,
respectively. Their explicit expressions are
\[
\epsilon_g = \frac{2\alpha_s}{3\pi}M^*_3\mu^* 
I(m^2_{\tilde{b}_1},m^2_{\tilde{b}_2},\vert M_3\vert^2) , \qquad
\epsilon_H = \frac{\vert h_t\vert^2}{16\pi^2}A^*_t\mu^* 
I(m^2_{\tilde{t}_1},m^2_{\tilde{t}_2},\vert \mu \vert^2)\ ,
\]
where $M_3$ is the gluino mass, $h_t$ and $A_t$ are the 
top-quark Yukawa and trilinear coupling, respectively.

\subsection{Couplings to gauge bosons}
\begin{itemize}
\item
Interactions of the Higgs bosons with the gauge bosons $Z$ and $W^\pm$
are described by
\begin{equation}
{\cal L}_{HVV}  =  g\,M_W \, \left(W^+_\mu W^{- \mu}\ + \
\frac{1}{2c_W^2}\,Z_\mu Z^\mu\right) \, \sum_i \,g_{_{H_iVV}}\, H_i
\end{equation}
where
\begin{equation}
g_{_{H_iVV}} = c_\beta\, O_{\phi_1 i}\: +\: s_\beta\, O_{\phi_2 i}\,.
\end{equation}

\item
Couplings to two photons:
the amplitude for the decay process
$H_i \rightarrow \gamma\gamma$ can be written as
\begin{equation} \label{hipp}
{\cal M}_{\gamma\gamma H_i}=-\frac{\alpha M_{H_i}^2}{4\pi\,v}
\bigg\{S^\gamma(M_{H_i})\,
\left(\epsilon^*_{1\perp}\cdot\epsilon^*_{2\perp}\right)
 -P^\gamma(M_{H_i})\frac{2}{M_{H_i}^2}
\langle\epsilon^*_1\epsilon^*_2 k_1k_2\rangle
\bigg\}\,,
\end{equation}
where $k_{1,2}$ are the momenta of the two photons and
$\epsilon_{1,2}$ the wave vectors of the corresponding photons,
$\epsilon^\mu_{1\perp} = \epsilon^\mu_1 - 2k^\mu_1 (k_2 \cdot
\epsilon_1) / M^2_{H_i}$, $\epsilon^\mu_{2\perp} = \epsilon^\mu_2 -
2k^\mu_2 (k_1 \cdot \epsilon_2) / M^2_{H_i}$ and $\langle \epsilon_1
\epsilon_2 k_1 k_2 \rangle \equiv \epsilon_{\mu\nu\rho\sigma}\,
\epsilon_1^\mu \epsilon_2^\nu k_1^\rho k_2^\sigma$.
The decay rate of $H_i\to \gamma\gamma$ is
proportional to $|S^\gamma|^2 + |P^\gamma|^2$.
The form factors are given by
\begin{eqnarray}
S^\gamma(M_{H_i})&=&2\sum_{f=b,t,\tau} N_C\,
Q_f^2\, g^{S}_{H_i\bar{f}f}\,F_{sf}(\tau_{f})
- g_{_{H_iVV}}F_1(\tau_{W})
+ \Delta S^\gamma_i \,, \nonumber \\
P^\gamma(M_{H_i})&=&2\sum_{f=b,t,\tau}
N_C\,Q_f^2\,g^{P}_{H_i\bar{f}f}\,F_{pf}(\tau_{f})
+ \Delta P^\gamma_i \,,
\end{eqnarray}
where $\tau_{x}=M_{H_i}^2/4m_x^2$, $N_C=3$ for quarks and $N_C=1$ for
taus, respectively.
In MSSM, the factors $\Delta S^\gamma_i$ and $\Delta P^\gamma_i$
receive contributions from charginos, sfermion, and charged Higgs boson:
\begin{eqnarray}
\Delta S^\gamma_i &=& \sqrt{2} g\,\sum_{f=\tilde{\chi}^\pm_1,\tilde{\chi}^\pm_2} \,
g^{S}_{H_i\bar{f}f}\,\frac{v}{m_f} F_{sf}(\tau_{if})
\nonumber \\ &&
- \sum_{\tilde{f}_j=\tilde{t}_1,\tilde{t}_2,\tilde{b}_1,\tilde{b}_2,
           \tilde{\tau}_1,\tilde{\tau}_2}
N_C\, Q_f^2g_{H_i\tilde{f}^*_j\tilde{f}_j}
\frac{v^2}{2m_{\tilde{f}_j}^2} F_0(\tau_{i\tilde{f}_j})
%\nonumber \\ &&
- g_{_{H_iH^+H^-}}\frac{v^2}{2 M_{H^\pm}^2} F_0(\tau_{iH^\pm})
\,, \nonumber \\[3mm]
\Delta P^\gamma_i &=&\sqrt{2}g\,\sum_{f=\tilde{\chi}^\pm_1,\tilde{\chi}^\pm_2}
g^{P}_{H_i\bar{f}f} \,\frac{v}{m_f}
F_{pf}(\tau_{if})   \label{eq4}
 \,,
\end{eqnarray}
where the couplings to charginos, sfermions, and charged Higgs  are
defined in the interactions:
\begin{eqnarray}
{\cal L}_{H\wt{\chi}^+\wt{\chi}^-}
&=&-\frac{g}{\sqrt{2}}\sum_{i,j,k} H_k
\overline{\wt{\chi}_i^-}
\left(g_{H_k\tilde{\chi}^+_i\tilde{\chi}^-_j}^{S}+i\gamma_5
g_{H_k\tilde{\chi}^+_i\tilde{\chi}^-_j}^{P}\right)
\wt{\chi}_j^-\,, \nonumber \\
{\cal L}_{H\tilde{f}\tilde{f}}&=&v\sum_{f=u,d}\,g_{H_i\tilde{f}^*_j\tilde{f}_k}
(H_i\,\tilde{f}^*_j\,\tilde{f}_k)\,,
\nonumber \\
{\cal L}_{3H} & = & \: v\,\sum_{i=1}^3 g_{_{H_iH^+H^-}}\,H_iH^+H^-\,.
\end{eqnarray}
We shall describe the couplings of the Higgs boson to the charginos,
sfermions, and charged Higgs boson a little later.

\item Couplings to two gluons:
similar to $H\to\gamma\gamma$,
the amplitude for the decay process
$H_i \rightarrow gg$ can be written as
\begin{equation} \label{higg}
{\cal M}_{gg H_i}=-\frac{\alpha_s\,M_{H_i}^2\,\delta^{ab}}{4\pi\,v}
\bigg\{S^g(M_{H_i})
\left(\epsilon^*_{1\perp}\cdot\epsilon^*_{2\perp}\right)
 -P^g(M_{H_i})\frac{2}{M_{H_i}^2}
\langle\epsilon^*_1\epsilon^*_2 k_1k_2\rangle
\bigg\}\,,
\end{equation}
where $a$ and $b$ ($a,b=1$ to 8) are indices of the eight $SU(3)$
generators in the adjoint representation.
The decay rate of $H_i\to gg $ is
proportional to $|S^g|^2 + |P^g|^2$.
The fermionic contributions and additional loop contributions from
squarks in the MSSM to
the scalar and pseudoscalar form factors are given by
\begin{eqnarray}
S^g(M_{H_i})&=&\sum_{f=b,t}
g^{S}_{H_i\bar{f}f}\,F_{sf}(\tau_{f}) +
\Delta S^g_i\,,
\nonumber \\
P^g(M_{H_i})&=&\sum_{f=b,t}
g^{P}_{H_i\bar{f}f}\,F_{pf}(\tau_{f}) +
\Delta P^g_i \,,
\end{eqnarray}
with
\begin{eqnarray}
\Delta S^g_i &=& -\sum_{\tilde{f}_j=\tilde{t}_1,\tilde{t}_2,\tilde{b}_1,\tilde{b}_2}
g_{H_i\tilde{f}^*_j\tilde{f}_j}
\frac{v^2}{4m_{\tilde{f}_j}^2} F_0(\tau_{i\tilde{f}_j}) \,, \nonumber \\
\Delta P^g_i &=& 0\,,  \label{eq3}
\end{eqnarray}
where the $\Delta P^g=0$ because
there are no colored SUSY fermions in the MSSM that can
contribute to $\Delta P^g$ at one loop level.
\end{itemize}

\subsection{Interactions of neutral Higgs bosons with charginos, sfermions,
and charged Higgs}

The interactions between the Higgs bosons and charginos are described
by the following Lagrangian:
\begin{eqnarray}
{\cal L}_{H\wt{\chi}^+\wt{\chi}^-}
&=&-\frac{g}{\sqrt{2}}\sum_{i,j,k} H_k
\overline{\wt{\chi}_i^-}
\left(g_{H_k\tilde{\chi}^+_i\tilde{\chi}^-_j}^{S}+i\gamma_5
g_{H_k\tilde{\chi}^+_i\tilde{\chi}^-_j}^{P}\right)
\wt{\chi}_j^-\,,
%=-g\sum_{\alpha,i,j,k}
%X^\alpha_{k;ij}\,\overline{\wt{\chi}_j^+}P_\alpha \wt{\chi}_i^+ \, H_k\,:
\nonumber \\
g_{H_k\tilde{\chi}^+_i\tilde{\chi}^-_j}^{S}&=&\frac{1}{2}\left\{
[(C_R)_{i1}(C_L)^*_{j2}G^{\phi_1}_k+(C_R)_{i2}(C_L)^*_{j1}G^{\phi_2}_k]
+[i\leftrightarrow j]^* \right\}\,,
\nonumber \\
g_{H_k\tilde{\chi}^+_i\tilde{\chi}^-_j}^{P} &=&\frac{i}{2}\left\{
[(C_R)_{i1}(C_L)^*_{j2}G^{\phi_1}_k+(C_R)_{i2}(C_L)^*_{j1}G^{\phi_2}_k]
-[i\leftrightarrow j]^* \right\}\,,    \label{eq13}
\end{eqnarray}
where $G^{\phi_1}_k=(O_{\phi_1 k}-is_\beta O_{ak})$,
$G^{\phi_2}_k=(O_{\phi_2 k}-ic_\beta O_{ak})$,
$i,j=1,2$, and $k=1-3$.
The chargino mass matrix in the $(\tilde{W}^-,\tilde{H}^-)$ basis
\begin{eqnarray}
{\cal M}_C = \left(\begin{array}{cc}
     M_2              & \sqrt{2} M_W\, c_{\beta} \\[2mm]
\sqrt{2} M_W\, s_{\beta} & \mu
             \end{array}\right)\, ,       \label{eq14}
\end{eqnarray}
is diagonalized by two different unitary matrices
$ C_R{\cal M}_C C_L^\dagger ={\sf diag}\{M_{\tilde{\chi}^\pm_1},\,
M_{\tilde{\chi}^\pm_2}\}$, where
$M_{\tilde{\chi}^\pm_1} \leq M_{\tilde{\chi}^\pm_2}$.
The chargino mixing matrices $(C_L)_{i\alpha}$ and $(C_R)_{i\alpha}$
relate the electroweak eigenstates to the mass eigenstates, via
\begin{eqnarray}
\tilde{\chi}^-_{\alpha L} &=&
(C_L)^*_{i \alpha } \tilde{\chi}_{iL}^-\,,\qquad
\tilde{\chi}^-_{\alpha L}\ =\ (\tilde{W}^-, \tilde{H}^-)_L^T\,,\nonumber\\
\tilde{\chi}^-_{\alpha R} &=& (C_R)^*_{i \alpha} \tilde{\chi}_{iR}^-\,,\qquad
\tilde{\chi}^-_{\alpha R}\ =\ (\tilde{W}^-, \tilde{H}^-)_R^T\,.
\end{eqnarray}

The Higgs-sfermion-sfermion interaction can be written in terms of the sfermion mass 
eigenstates as
\begin{equation}
{\cal L}_{H \tilde{f} \tilde{f}}
= v \sum_{f=u,d} g_{H_i \tilde{f}^*_j \tilde{f}_k}
(H_i \tilde{f}^*_j \tilde{f}_k)\,,  \label{eq16}
\end{equation}
where
\[
v g_{H_i \tilde{f}^*_j \tilde{f}_k} = (\Gamma^{\alpha \tilde{f}^* \tilde{f}})_{\beta \gamma} 
O_{\alpha i} U^{\tilde{f}^*}_{\beta j}
U^{\tilde{f}}_{\gamma k}\,,
\]
with $\alpha = (\phi_1,\phi_2,a) = (1,2,3)$, \, $\beta, \gamma = L, R $, \, 
$ i = (H_1, H_2, H_3) = (1,2,3)$ and $j,k = 1,2$. The expressions for the couplings 
$\Gamma^{\alpha \tilde{f}^* \tilde{f}}$ are shown in \cite{Lee:2003nta}.
The stop and sbottom mass matrices may conveniently be written in the 
$(\tilde{q}_L,\tilde{q}_R)$ basis as
\begin{eqnarray}
\tilde{\cal M}^2_q = \left(\begin{array}{cc}
     M^2_{\tilde{Q}_3} + m^2_q + c_{2\beta}M^2_Z(T^q_z - Q_q s^2_W) & h^*_q v_q (A^*_q - \mu R_q)/
\sqrt{2} \\[2mm]
h_q v_q (A_q - \mu^* R_q)/\sqrt{2} & 
M^2_{\tilde{R}_3} + m^2_q + c_{2\beta} M^2_Z Q_q s^2_W
             \end{array}\right)\, ,     \label{eq17}
\end{eqnarray}
with $q=t,b, \, R=U,D, \, T^t_z=-T^b_z=1/2, \, Q_t=2/3, \, Q_b=-1/3, \, v_b=v_1, \, v_t=v_2,
 \, R_b=\tan\beta =v_2/v_1, \, R_t=\cot\beta$, and $h_q$ is the Yukawa coupling of 
the quark $q$. On the other hand, the stau mass matrix is written in the 
$(\tilde{\tau}_L,\tilde{\tau}_R)$ basis as
\begin{eqnarray}
\tilde{\cal M}^2_{\tau} = \left(\begin{array}{cc}
     M^2_{\tilde{L}_3} + m^2_{\tau} + c_{2\beta}M^2_Z(s^2_W - 1/2) & h^*_{\tau} v_1 
(A^*_{\tau} - \mu \tan\beta)/\sqrt{2} \\[2mm]
h_{\tau} v_1 (A_{\tau} - \mu^* \tan\beta)/\sqrt{2} & 
M^2_{\tilde{E}_3} + m^2_{\tau} + c_{2\beta} M^2_Z s^2_W
             \end{array}\right)\, .     \label{eq18}
\end{eqnarray}
The $2 \times 2$ sfermion mass matrix $\tilde{M}^2_f$ for $f=t,b$ and
$\tau$ is diagonalized by a unitary matrix $U^{\tilde{f}}$ :
$U^{\tilde{f} \dagger} \tilde{M}^2_f U^{\tilde{f}}= {\bf
  diag}(m^2_{\tilde{f}_1},m^2_{\tilde{f}_2})$ with $m^2_{\tilde{f}_1}
\leq m^2_{\tilde{f}_2}$. The mixing matrix $U^{\tilde{f}}$ relates the
electroweak eigenstates $\tilde{f}_{L,R}$ to the mass eigenstates
$\tilde{f}_{1,2}$, via
\[
(\tilde{f}_L,\tilde{f}_R)^T_{\alpha} = U^{\tilde{f}}_{\alpha i} (\tilde{f}_1,\tilde{f}_2)^T_i \, .
\]
Interactions between the Higgs bosons and the charged Higgs boson
 can be found in Ref.~\cite{Cheung:2013rva}.

\section{Data, Fits, and Results}
\subsection{Data}
Our previous works \cite{Cheung:2013kla,Cheung:2013rva,Cheung:2014oaa}
were performed
with data of the Summer 2013.
Very recently we have also updated the
model-independent fits using the data of the
Summer 2014 \cite{Cheung:2014noa}.
The whole set of Higgs strength data on $H\to \gamma\gamma$, $ZZ^* \to 4 \ell$,
$WW^* \to \ell\nu \ell\nu$, $\tau\tau$, and $b\bar b$ are listed
in Ref.~\cite{Cheung:2014noa}. The most significant changes
since summer 2013
are the $H\to \gamma\gamma$ data from both ATLAS and CMS.
The ATLAS Collaboration updated their best-measured value from
$\mu_{ggH+ttH} = 1.6 \pm 0.4$
to $\mu_{\rm inclusive}=1.17 \pm 0.27$ \cite{atlas_zz_2014},
while the CMS $H\to\gamma\gamma$ data entertained a very dramatic change
from $\mu_{\rm untagged}=0.78\,^{+0.28}_{-0.26}$
to $\mu_{ggH}= 1.12 \,^{+0.37}_{-0.32}$ \cite{cms_aa_2014}.
Other notable differences can be found in Ref.~\cite{Cheung:2014noa}.
The $\chi^2_{\rm SM}/$d.o.f. for the SM is now at $16.76/29$,
which corresponds to a $p$-value of $0.966$.

\subsection{CP-Conserving (CPC) Fits}
We consider the CP-conserving MSSM and use the most updated Higgs
boson signal strengths to constrain a minimal set of parameters under
various conditions.  
Regarding the $i$-th Higgs boson $H_i$ as the
candidate for the 125 GeV Higgs boson,
the varying parameters are:
\begin{itemize}
\item 
the up-type  Yukawa coupling $C_u^S \equiv g^S_{H_i\bar{u}u}=
    O_{\phi_2 i}/s_\beta$, see Eq.~(\ref{eq1}),
\item 
  the ratio of the VEVs of the two Higgs doublets $\tan\beta\equiv v_2/v_1$,
\item the parameter $\kappa_d$ (assumed real) quantifying the modification
 between the down-type quark mass and Yukawa coupling due to radiative
  corrections, as shown in Eq.~(\ref{eq2}),
\item $\Delta S^\gamma\equiv \Delta S^\gamma_i$ as in Eq.~(\ref{eq4})
\item $\Delta S^g\equiv \Delta S^g_i$ as in Eq.~(\ref{eq3}), and
\item the deviation in the total decay width of the observed
Higgs boson: $\Delta \Gamma_{\rm tot}$.
\end{itemize}
The down-type and lepton-type Yukawa and the gauge-Higgs couplings
are derived as
\begin{eqnarray}
C_d^S&\equiv & g^S_{H_i\bar{d}d}=  \bigg(\,
\frac{O_{\phi_1 i}+\kappa_d O_{\phi_2 i}}{1\, +\, \kappa_d\,\tan\beta}\,\bigg)\,
\frac{1}{\cos\beta} \,, \nonumber\\[2mm]
C_\ell^S&\equiv & g^S_{H_i\bar{\ell}\ell} = 
\frac{O_{\phi_1 i}}{\cos\beta}\,, \nonumber\\[2mm]
C_v&\equiv & g_{H_iVV} 
= c_\beta\, O_{\phi_1 i}\: +\: s_\beta\, O_{\phi_2 i}   \label{eq5}
\end{eqnarray}
with
\begin{equation}
O_{\phi_1 i}=\pm\sqrt{1-s_\beta^2 (C_u^S)^2}\,, \ \ \
O_{\phi_2 i}=C_u^S s_\beta\,.
\end{equation}
In place of $\tan\beta$ we can use $C_v$ as a varying parameter, and
then $\tan\beta \;(t_\beta) $ would be determined by
%\begin{equation}
%\label{eq:sbsq}
%s_\beta^2=\frac{1-C_v^2}{1+(C_u^S)^2-2C_vC_u^S}
%=\frac{(1-C_v^2)}{(1-C_v^2)+(C_u^S-C_v)^2}\,.
%\end{equation}
\begin{equation}
\label{eq:tbsq}
t_\beta^2=\frac{(1-C_v^2)}{(C_u^S-C_v)^2}
=\frac{(1-C_v^2)}{\left[(C_u^S-1)+(1-C_v)\right]^2}
\,. \\[3mm]
\end{equation}
We note that $t_\beta= \infty$ when $(C_u^S-1)=-(1-C_v)<0$
\footnote{Note $C_v\leq 1$ and positive definite in our convention.}
while $t_\beta=1$ when $(C_u^S-1)=\pm\sqrt{1-C_v^2}-(1-C_v)$.
Therefore $t_\beta$ changes from $\infty$ to $1$ when $(C_u^S-1)$ deviates from
$-(1-C_v)$ by the amount of $\pm\sqrt{1-C_v^2}$. This implies that the value 
of $t_\beta$ becomes more and more sensitive to the deviation of $C_u^S$ from $1$ 
as $C_v$ approaches to its SM value $1$.

\medskip 

We are going to perform the following three categories of CPC fits 
varying the stated parameters
while keeping the others at their SM values.
\begin{itemize}
\item{\bf CPC.II}
\begin{itemize}
\item{\bf CPC.II.2}: $C_u^S$, $\tan\beta$
    ($\kappa_d=\Delta\Gamma_{\rm tot}=\Delta S^\gamma = \Delta S^g = 0$ )
\item{\bf CPC.II.3}: $C_u^S$, $\tan\beta$, $\kappa_d$
    ($\Delta\Gamma_{\rm tot}=\Delta S^\gamma = \Delta S^g = 0$ )
\item{\bf CPC.II.4}: $C_u^S$, $\tan\beta$, $\kappa_d$, $\Delta \Gamma_{\rm tot}$
    ($\Delta S^\gamma = \Delta S^g = 0$ )
\end{itemize}
\item{\bf CPC.III}
\begin{itemize}
\item{\bf CPC.III.3}: $C_u^S$, $\tan\beta$, $\Delta S^\gamma$
    ($\kappa_d=\Delta\Gamma_{\rm tot}=\Delta S^g = 0$ )
\item{\bf CPC.III.4}: $C_u^S$, $\tan\beta$, $\Delta S^\gamma$, $\kappa_d$
    ($\Delta\Gamma_{\rm tot}= \Delta S^g = 0$ )
\item{\bf CPC.III.5}: $C_u^S$, $\tan\beta$, $\Delta S^\gamma$, $\kappa_d$, $\Delta
\Gamma_{\rm tot}$
    ($\Delta S^g = 0$ )
\end{itemize}
\item{\bf CPC.IV}
\begin{itemize}
\item{\bf CPC.IV.4}: $C_u^S$, $\tan\beta$, $\Delta S^\gamma$, $\Delta S^g$
    ($\kappa_d=\Delta\Gamma_{\rm tot}= 0$ )
\item{\bf CPC.IV.5}: $C_u^S$, $\tan\beta$, $\Delta S^\gamma$, $\Delta S^g$, $\kappa_d$
    ($\Delta\Gamma_{\rm tot}=  0$ )
\item{\bf CPC.IV.6}: $C_u^S$, $\tan\beta$, $\Delta S^\gamma$, $\Delta S^g$, 
$\kappa_d$, $\Delta \Gamma_{\rm tot}$
\end{itemize}
\end{itemize}
Basically, the {\bf CPC.II}, {\bf CPC.III}, and {\bf CPC.IV}  fits vary 
($C_u^S$,$\tan\beta$),
($C_u^S$,$\tan\beta$,$\Delta S^\gamma$), and
($C_u^S$,$\tan\beta$,$\Delta S^\gamma$,$\Delta S^g$), respectively.
Each category of {\bf CPC} fits includes three fits:
the second fit adds $\kappa_d$ to the set of varying parameters and
$\Delta \Gamma_{\rm tot}$ is further varied in the third one.
The Arabic number at the end of each label denotes the 
total number of varying parameters.

The $\Delta S^\gamma$ is the deviation in the $H\gamma\gamma$
vertex factor other than the effects of changing the Yukawa and
gauge-Higgs couplings, and it receives contributions from any exotic
particles running in the triangular loop. For example, the charginos,
charged Higgs bosons, sleptons, and squarks in the MSSM.  Here we are
content with a varying $\Delta S^\gamma$ without specifying the
particle spectrum of the MSSM. Later in the next section we shall
specifically investigate the effects of charginos, staus,
stops, and sbottoms.

In the MSSM,
$\Delta S^g$ receives contributions only from colored SUSY particles--squarks
running in the $Hgg$ vertex. The current limits on squark masses are 
in general above TeV such
that $\Delta S^g$ is expected to be small. 
Nevertheless, we do not restrict the size of $\Delta S^g$ in this fit
in order to see the full effect of $\Delta S^g$.

The parameter $\kappa_d$
arises from the loop corrections to the down-type Yukawa couplings. It changes
the relation between the mass and the Yukawa coupling of the down-type quarks.
We limit the range of $|\kappa_d | < 0.1$ as it is much smaller than $0.1$ in
most of the MSSM parameter space.

Although the charginos are constrained to be heavier than
103.5 GeV and sleptons to be heavier than 81.9 GeV \cite{pdg}, there
are still possibilities that the decays of 
the 125.5 GeV Higgs boson into
neutralinos and another neutral Higgs
boson are kinematically allowed. These channels have not been
explicitly searched for, but we can take them into account by the
deviation $\Delta \Gamma_{\rm tot}$ in the total decay width of the
observed Higgs boson.

The best-fit points for the fits are summarized in Table~\ref{tab:best}.
We see that the $p$ values of the
{\bf CPC.II.2}, {\bf CPC.III.3}, and {\bf CPC.IV.4} fits are the 
highest in each category.
Also, the $p$ value of the {\bf CPC.III.3} fit is  slightly higher than that of
the {\bf CPC.IV.4} fit, followed by the {\bf CPC.II.2} fit.

\begin{sidewaystable}[thb!]
  \caption{\small \label{tab:best}
The best-fit values for various {\bf CPC} fits.
The SM chi-square per degree of freedom is $\chi^2_{\rm SM}$/d.o.f.$=16.76/29$,
and $p$-value$=0.966$.
  }
\begin{ruledtabular}
\begin{tabular}{ l|ccc|rcrrrrrrr}
Fits  & $\chi^2$ & $\chi^2$/dof & $p$-value &
\multicolumn{9}{c}{Best-fit values} \\
& & & & $C_u^S$ & $\tan\beta$ & 
$\Delta S^\gamma$ & $\Delta S^g$ & $\kappa_d$ & $\Delta{\Gamma}_{tot}$ &
   $C_v$ & $C_{d}^{S}$ & $C_{\ell}^{S}$  \\
\hline\hline
%CPC.II.2
{\bf CPC.II.2}
& $16.74$ & $0.620$ & $0.937$ & $1.011$ &
$0.111$ & $-$ & $-$ & $-$ & $-$ & $1.000$ & $1.000$ & $1.000$\\
%CPC.II.3
{\bf CPC.II.3}
& $16.74$ & $0.644$ & $0.917$ & $1.011$ &
$0.194$ & $-$ & $-$ & $0.099$ & $-$ & $1.000$ & $1.000$ & $1.000$\\
{\bf CPC.II.4}
& $16.72$ & $0.669$ & $0.892$ & $1.023$ &
$0.312$ & $-$ & $-$ & $-0.079$ & $0.103$ & $1.000$ & $0.997$ & $0.998$\\
\hline
%CPC.III.3
{\bf CPC.III.3}
& $15.50$ & $0.596$ & $0.947$ & $-0.930$ &
$0.194$ & $2.326$ & $-$ & $-$ & $-$  & $0.932$ & $1.003$ & $1.003$\\
%CPC.III.4
{\bf CPC.III.4}
& $15.48$ & $0.619$ & $0.929$ & $-0.948$ &
$0.180$ & $2.402$ & $-$ & $-0.097$ & $-$ &  $0.940$ & $1.036$ & $1.002$\\
%CPC.III.5
{\bf CPC.III.5}
& $15.43$ & $0.643$ & $0.907$ & $1.061$ &
$0.100$ & $-0.938$ & $-$ & $0.100$ & $0.557$ & $1.000$ & $1.000$ & $1.000$\\
\hline
%CPC.IV.4
{\bf CPC.IV.4}
& $14.85$ & $0.594$ & $0.945$ & $-1.219$ &
$0.154$ & $2.893$ & $1.547$ & $-$ & $-$ & $0.943$ & $0.994$ & $0.994$\\
& $14.85$ & $0.594$ & $0.945$ & $-1.219$ &
$0.154$ & $2.893$ & $0.204$ & $-$ & $-$ & $0.943$ & $0.994$ & $0.994$\\
%CPC.IV.5
{\bf CPC.IV.5}
& $14.83$ & $0.618$ & $0.926$ & $-1.224$ &
$0.164$ & $2.902$ & $1.540$ & $0.088$ & $-$ & $0.935$ & $0.962$ & $0.993$\\
& $14.83$ & $0.618$ & $0.926$ & $-1.225$ &
$0.164$ & $2.902$ & $0.217$ & $0.088$ & $-$ & $0.935$ & $0.962$ & $0.993$\\
%CPC.IV.6
{\bf CPC.IV.6}
& $14.83$ & $0.645$ & $0.901$ & $-1.213$ &
$0.173$ & $2.868$ & $1.528$ & $0.082$ & $-0.071$ & $0.929$ & $0.962$ & $0.993$\\
& $14.83$ & $0.645$ & $0.901$ & $-1.213$ &
$0.173$ & $2.870$ & $0.213$ & $0.079$ & $-0.075$ & $0.929$ & $0.963$ & $0.993$\\
& $14.83$ & $0.645$ & $0.901$ & $1.022$ &
$2.600$ & $-1.228$ & $-0.180$ & $0.005$ & $-0.839$ & $0.782$ & $-0.811$ & $-0.837$\\
& $14.83$ & $0.645$ & $0.901$ & $1.022$ &
$2.600$ & $-1.228$ & $-1.288$ & $0.005$ & $-0.840$ & $0.782$ & $-0.811$ & $-0.837$\\
\end{tabular}
\end{ruledtabular}
\end{sidewaystable}

\begin{sidewaystable}[thb!]
  \caption{\small \label{tab:best3}
The other local minima for various CPC fits.
  }
\begin{ruledtabular}
\begin{tabular}{ l|ccc|rcrrrrrrr}
Fits  & $\chi^2$ & $\chi^2$/dof & $p$-value &
\multicolumn{9}{c}{Best-fit values} \\
& & & & $C_u^S$ & $\tan\beta$ & 
$\Delta S^\gamma$ & $\Delta S^g$ & $\kappa_d$ & $\Delta{\Gamma}_{tot}$ &
     $C_v$ & $C_{d}^{S}$ & $C_{\ell}^{S}$  \\
\hline\hline
{\bf CPC.III.3}
& $15.68$ & $0.603$ & $0.944$ & $1.000$ &
$34.58$ & $-0.853$ & $-$ & $-$ & $-$ & $1.000$ & $1.039$ & $1.039$\\
{\bf CPC.III.4}
& $15.59$ & $0.624$ & $0.926$ & $0.999$ &
$9.332$ & $-1.026$ & $-$ & $-0.006$ & $-$ & $0.976$ & $-1.170$ & $-1.051$\\
\hline
{\bf CPC.IV.4}
& $15.23$ & $0.609$ & $0.936$ & $1.000$ &
$5.681$ & $-1.127$ & $-0.057$ & $-$ & $-$ & $0.940$ & $-1.002$ & $-1.002$\\
& $15.23$ & $0.609$ & $0.936$ & $1.000$ &
$5.695$ &  $-1.126$ & $-1.395$ & $-$ & $-$ & $0.940$ & $-1.002$ & $-1.002$\\
%\hline
{\bf CPC.IV.5}
& $15.22$ & $0.634$ & $0.914$ & $1.000$ &
$5.423$ & $-1.128$ & $-0.062$ & $0.002$ & $-$ & $0.934$ & $-0.980$ & $-0.999$\\
& $15.22$ & $0.634$ & $0.914$ & $1.000$ &
$5.429$ & $-1.127$ & $-1.387$ & $0.002$ & $-$ & $0.934$ & $-0.980$ & $0.999$\\

\end{tabular}
\end{ruledtabular}
\end{sidewaystable}

\subsection{Results}
\begin{figure}[t!]
\centering
\includegraphics[width=2.0in]{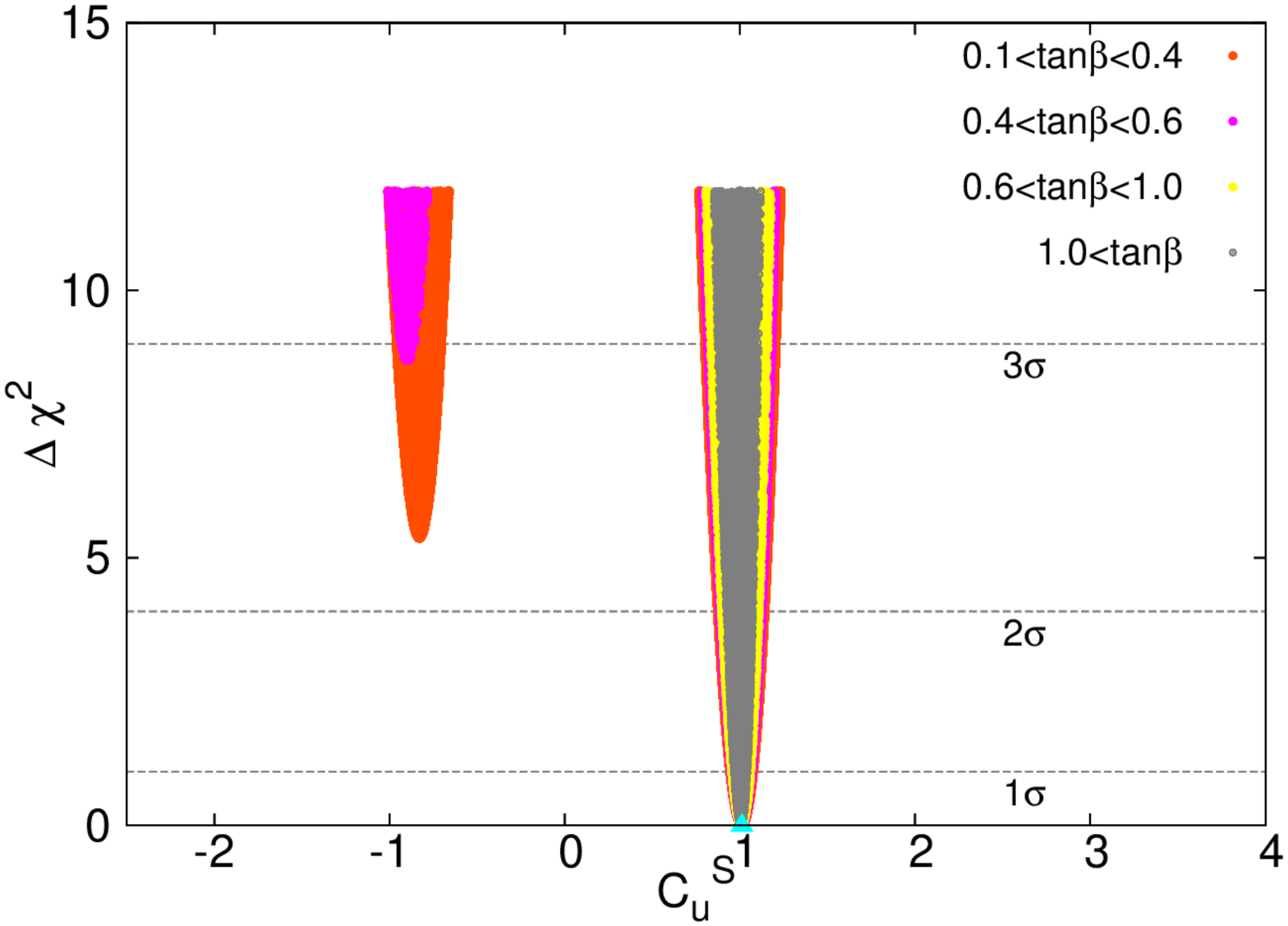}
\includegraphics[width=2.0in]{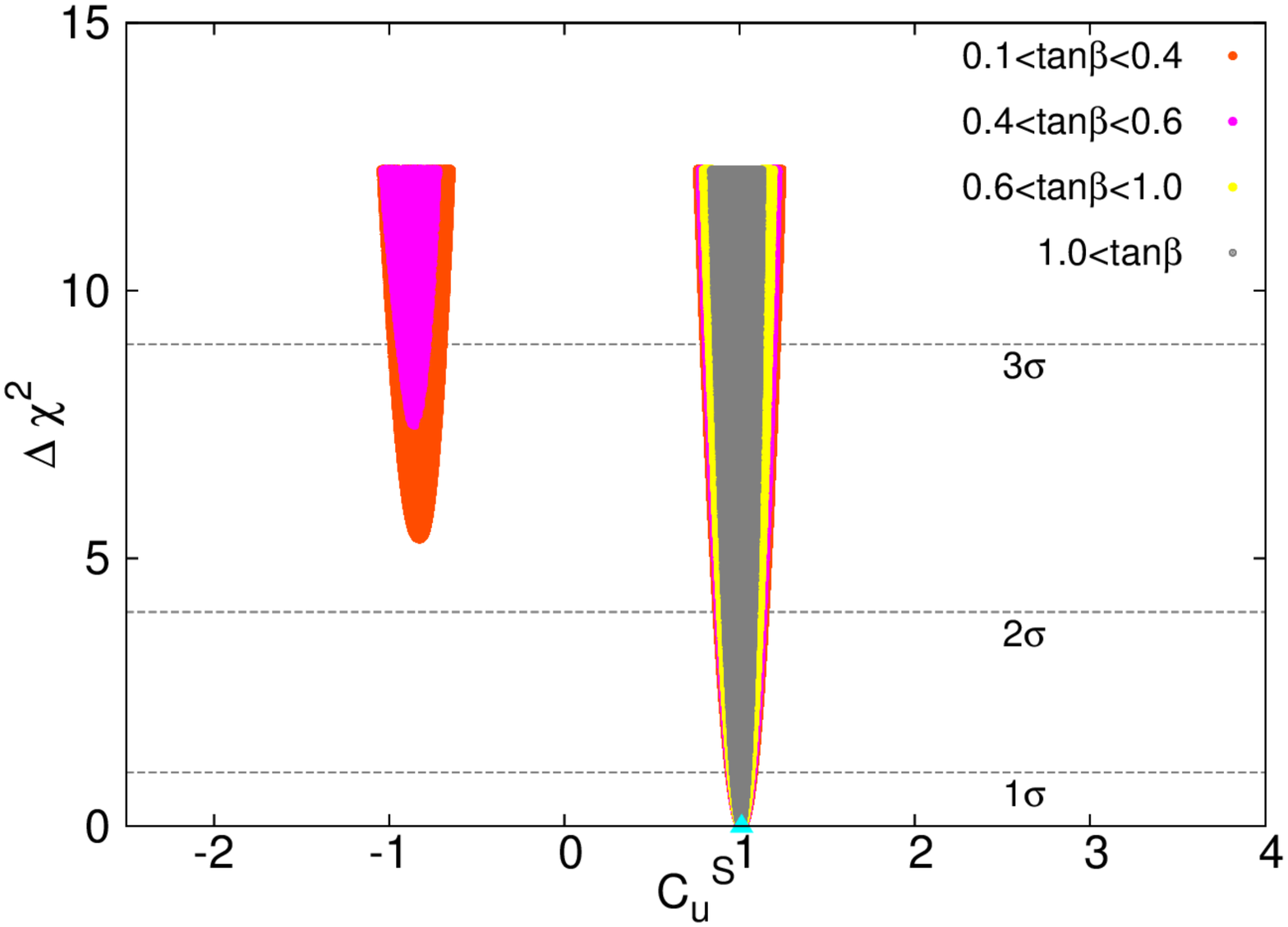}
\includegraphics[width=2.0in]{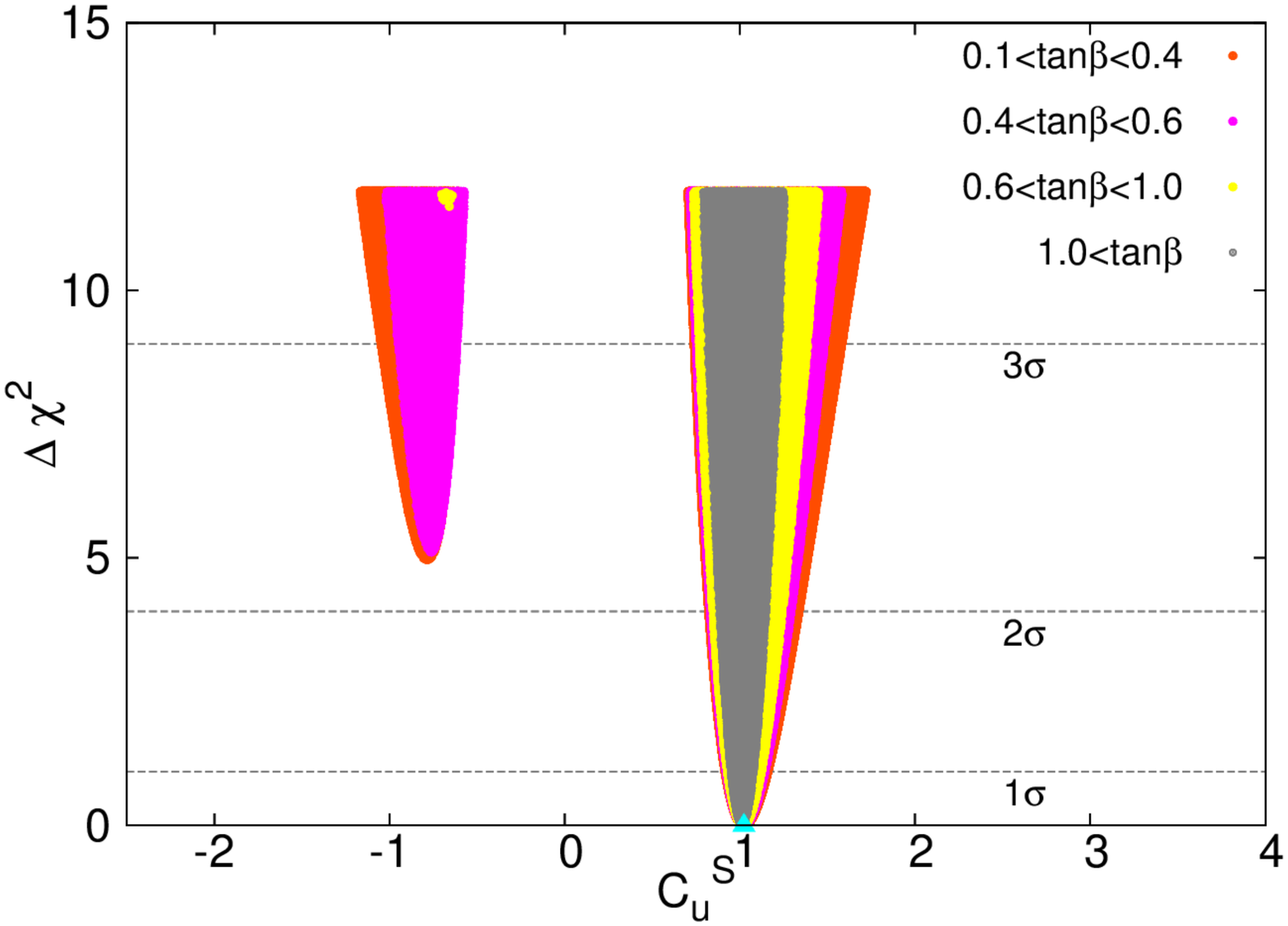}
\includegraphics[width=2.0in]{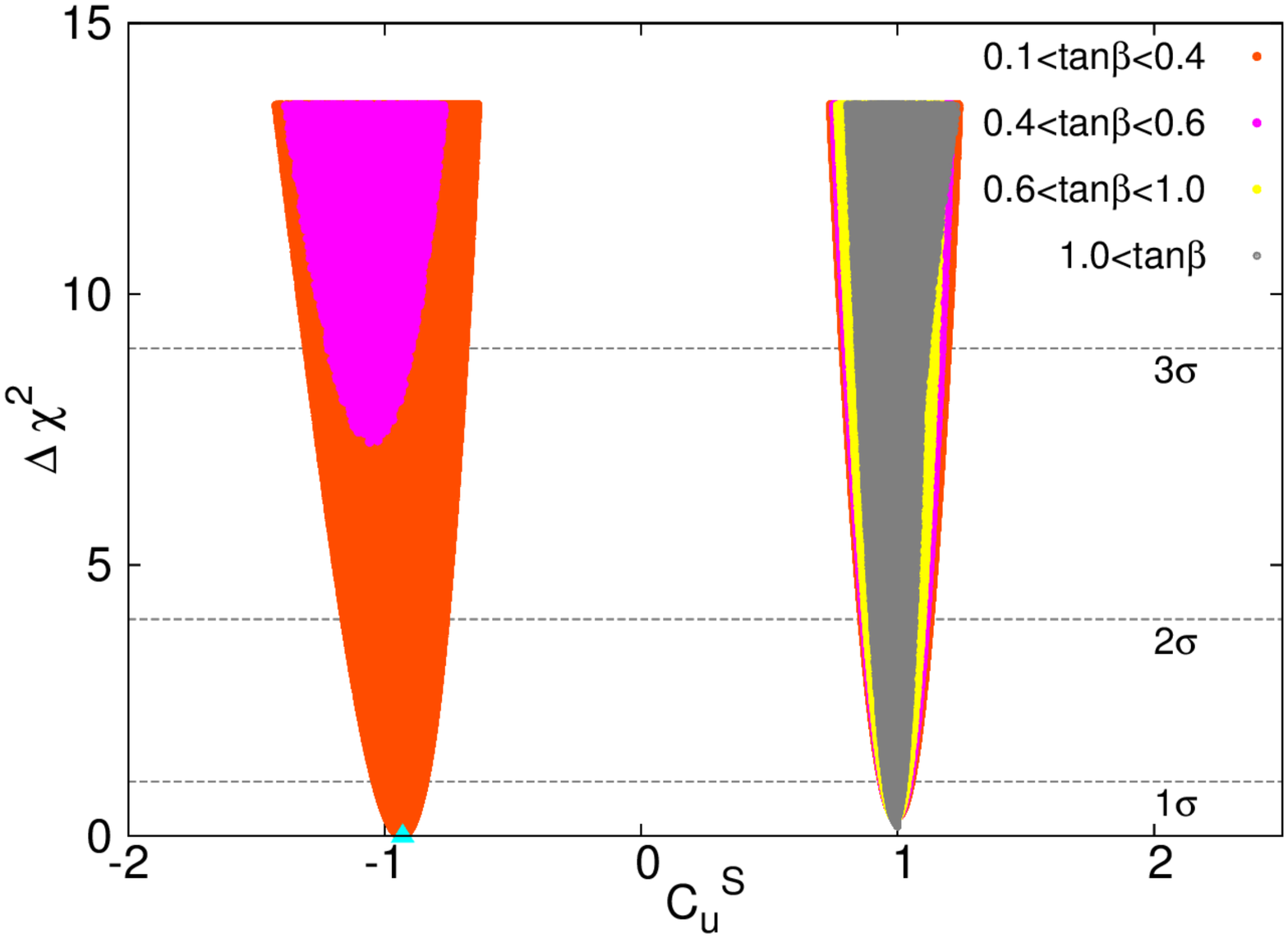}
\includegraphics[width=2.0in]{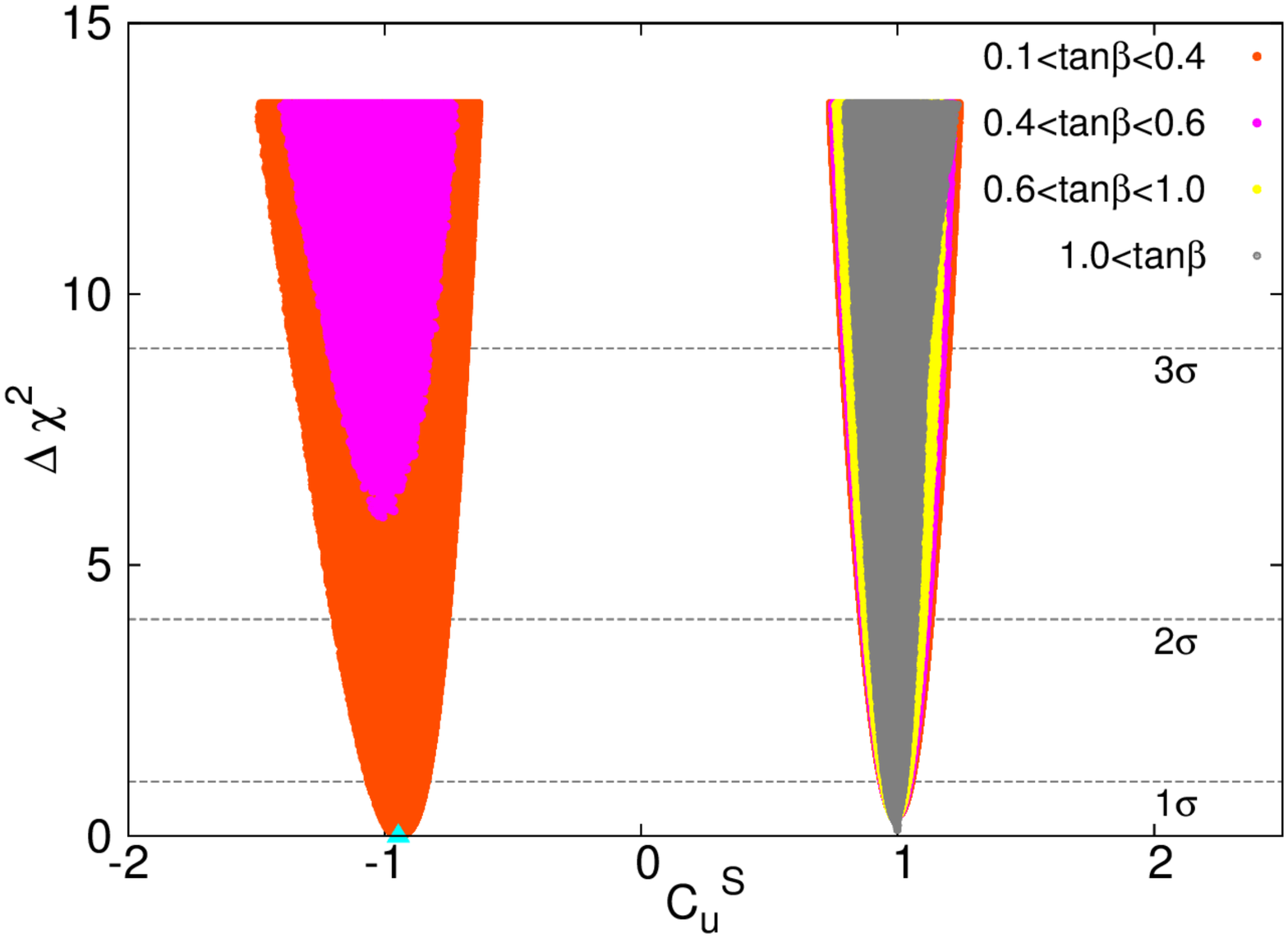}
\includegraphics[width=2.0in]{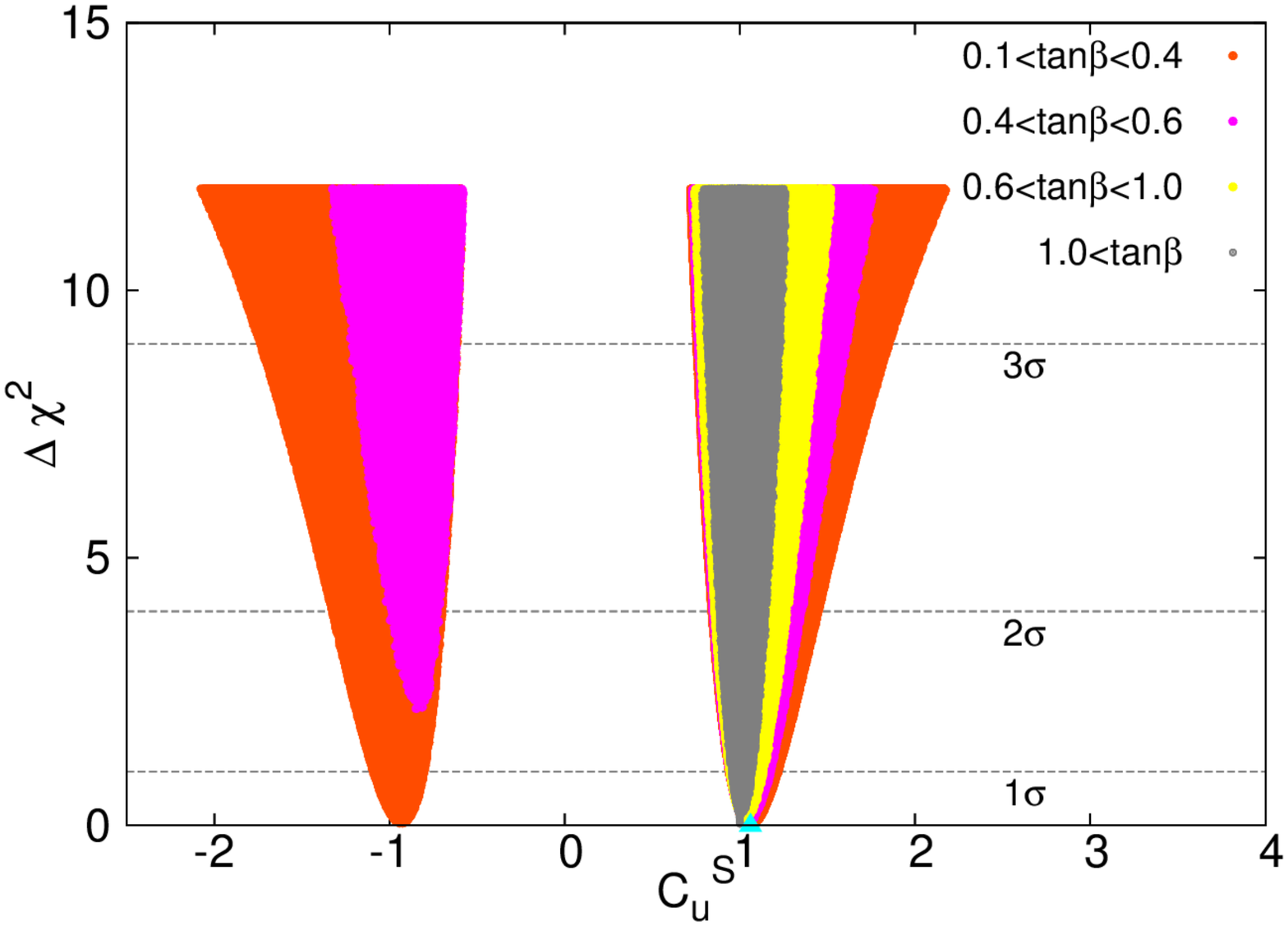}
\includegraphics[width=2.0in]{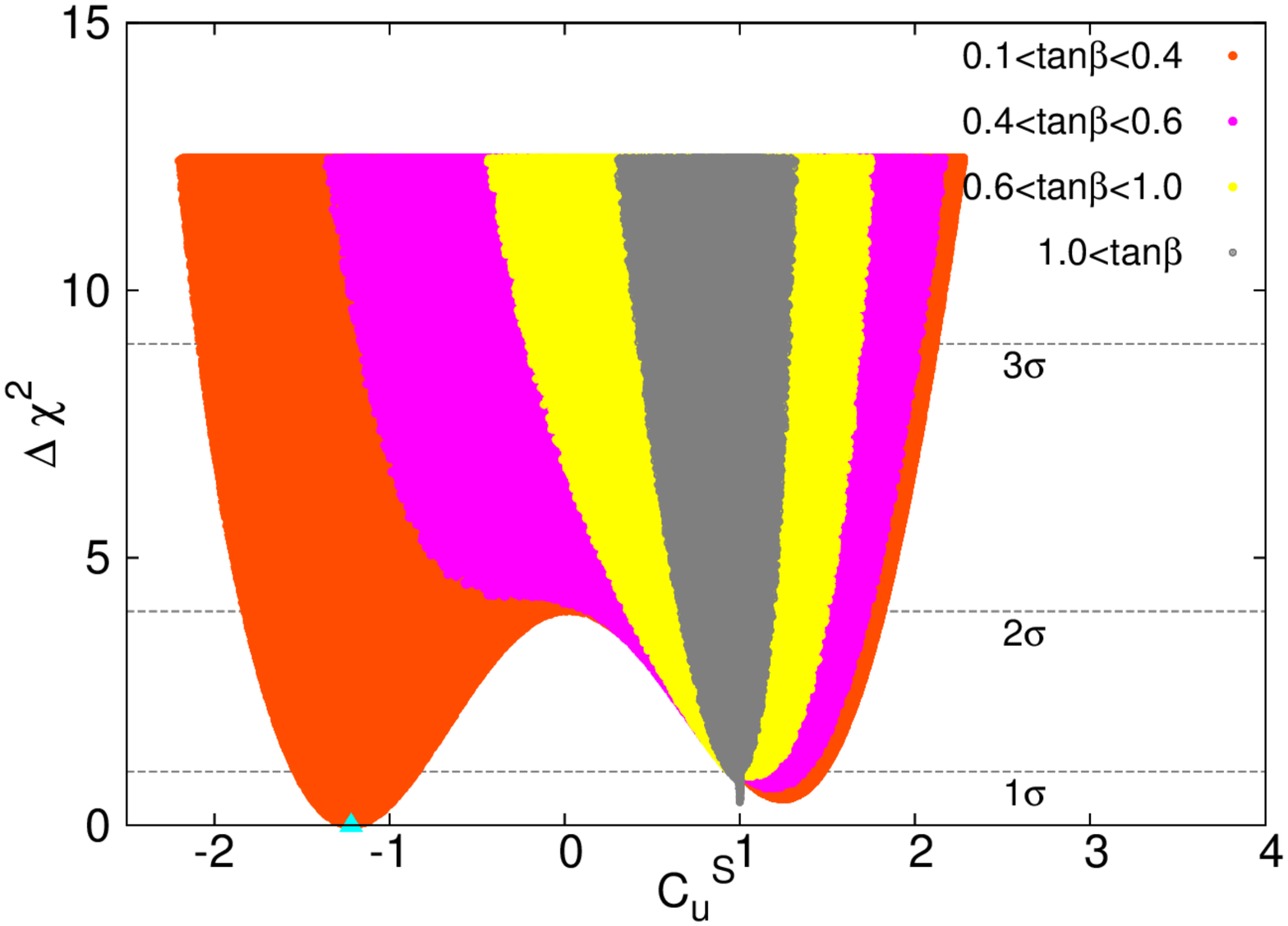}
\includegraphics[width=2.0in]{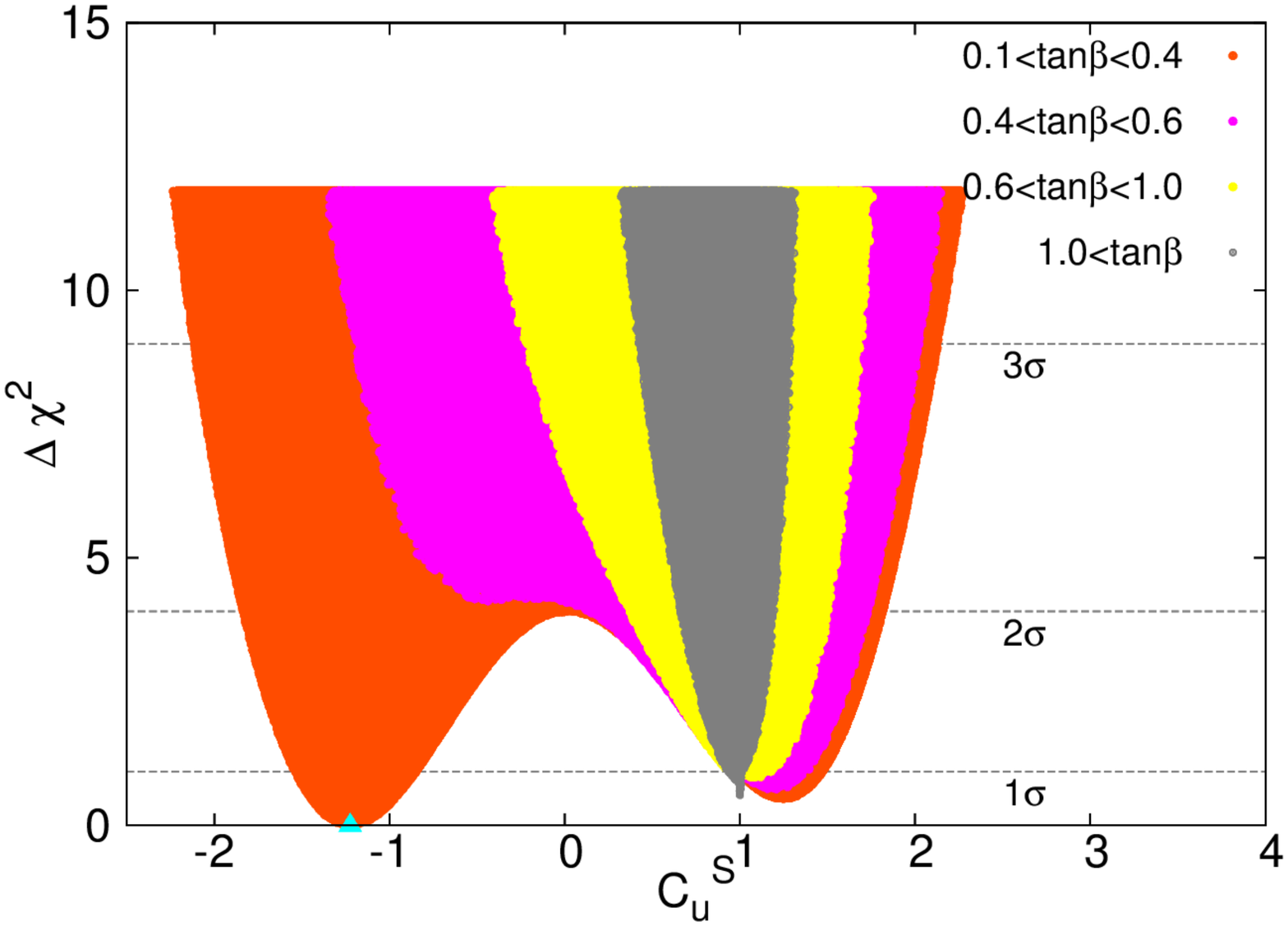}
\includegraphics[width=2.0in]{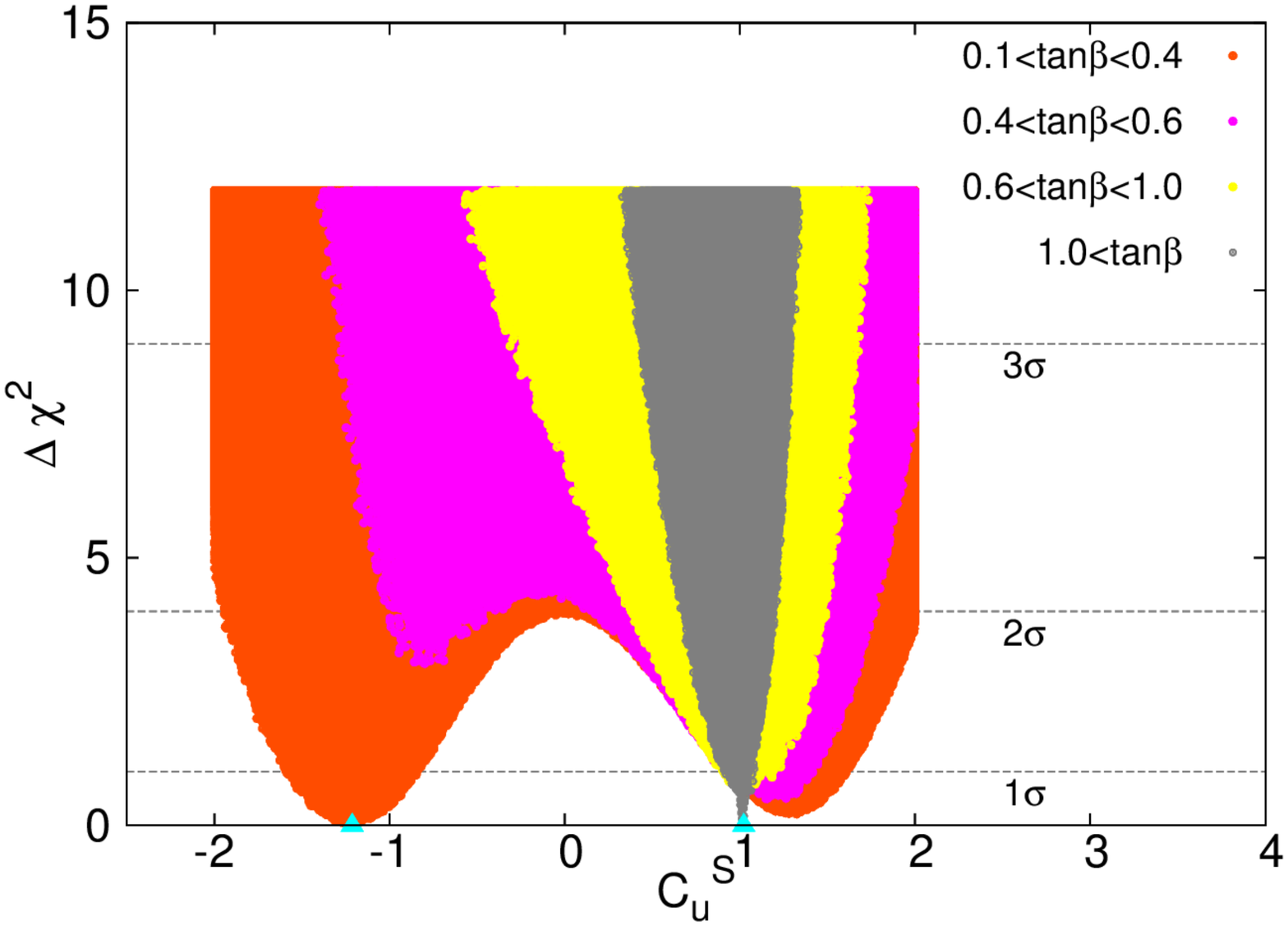}
\caption{\small \label{fig:chi2}
Plots of $\Delta\chi^2$ vs $C_u^S$ for three categories of CPC fits:
{\bf CPC.II} (upper row), {\bf CPC.III} (middle row), and
{\bf CPC.IV} (lower row).
The left frames show the cases of
{\bf CPC.II.2} (varying $C_u^S$, $\tan\beta$), 
{\bf CPC.III.3} (varying $C_u^S$, $\tan\beta$, $\Delta S^\gamma$), and 
{\bf CPC.IV.4} (varying $C_u^S$, $\tan\beta$, $\Delta S^\gamma$, $\Delta S^g$).
In the middle frames, 
the cases {\bf CPC.II.3, CPC.III.4}, {\bf CPC.IV.5} are shown by adding
$\kappa_d$ to the corresponding set of varying parameters. 
The right frames are for the cases of
{\bf CPC.II.4, CPC.III.5}, and {\bf CPC.IV.6} in which
$\Delta \Gamma_{\rm tot}$ is further varied.
In each frame, each different color corresponds to a different 
range of $\tan\beta$:
$0.1<\tan\beta<0.4$ (red),
$0.4<\tan\beta<0.6$ (magenta),
$0.6<\tan\beta<1$ (yellow), and
$1<\tan\beta$ (gray).
}
\end{figure}

Before we present descriptions of the confidence regions and
the correlations among the fitting parameters
$C_u^S$, $\tan\beta$, $\Delta S^\gamma$, $\Delta S^g$, $\kappa_d$, and
$\Delta\Gamma_{\rm tot}$, we look into the behavior of 
$\Delta\chi^2$ versus $C_u^S$
in each category of fits. 
In the {\bf CPC.II} fits, the minimum $\chi^2$ values are 
$16.74$ ({\bf CPC.II.2}, {\bf CPC.II.3}) and $16.72$ ({\bf CPC.II.4})
(see Table~\ref{tab:best}), and 
$\Delta\chi^2$ versus $C_u^S$ are shown in the upper row of Fig.~\ref{fig:chi2}.
The minima are located at 
$C_u^S=1.011$ ({\bf CPC.II.2}, {\bf CPC.II.3}) and
$C_u^S=1.023$ ({\bf CPC.II.4})
and the second local minima are 
developed around $C_u^S=-1$ but with $\Delta\chi^2\gsim 5$.  
It is clear that $C_u^S \approx 1$ is preferred much more than the negative
values.
The $\Delta\chi^2$
dependence on $C_u^S$ hardly changes by varying  $\kappa_d$
as shown in the upper-middle frame. 
With $\Delta\Gamma_{\rm tot}$ varying further,
we observe the dependence of $\Delta\chi^2$ on $C_u^S$
becomes broader by extending to the regions of $|C_u^S|>1$
as shown in the upper-right frame. 
We also observe that the second local
minimum around $C_u^S=-1$ disappears when $\tan\beta \gsim 0.6$.

In the {\bf CPC.III} fits, the minimum $\chi^2$ values are 
$15.50$ ({\bf CPC.III.3}), $15.48$ ({\bf CPC.III.4}), and $15.43$ ({\bf CPC.III.5}):
see Table~\ref{tab:best}, and
$\Delta\chi^2$ versus $C_u^S$ are shown in the middle row of Fig.~\ref{fig:chi2}.
The minima are located at 
$C_u^S=-0.930$ ({\bf CPC.III.3}), 
$C_u^S=-0.948$ ({\bf CPC.III.4}), and 
$C_u^S=1.061$ ({\bf CPC.III.5}),
and the second local minima are 
developed around $C_u^S=1$ ({\bf CPC.III.3} and {\bf CPC.III.4})
and $C_u^S=-1$ ({\bf CPC.III.5}), respectively.
In contrast to the {\bf CPC.II} fits, the 
$\Delta\chi^2$ difference between the true and local minima
is tiny, $\left.\Delta\chi^2\right|_{\rm local}-
\left.\Delta\chi^2\right|_{\rm true}\lsim 0.2$: see Table~\ref{tab:best3}.
The $\Delta\chi^2$ dependence on $C_u^S$ hardly changes by 
varying  $\kappa_d$ additionally (shown in the middle-middle frame),
but when $\Delta\Gamma_{\rm tot}$ is varied further,
the dependence of $\Delta\chi^2$ on $C_u^S$
becomes broader, the same as the {\bf CPC.II} fits (see the middle-right 
frame).
We observe the true/local
minima around $C_u^S=-1$ disappear when $\tan\beta \gsim 0.6$.

In the {\bf CPC.IV} fits, the minimum $\chi^2$ values are
$14.85$ ({\bf CPC.IV.4}), $14.83$ ({\bf CPC.IV.5} and {\bf CPC.IV.6}):
see Table~\ref{tab:best}, and
$\Delta\chi^2$ versus $C_u^S$ are shown in the lower row of Fig.~\ref{fig:chi2}.
The minima are located at 
$C_u^S=-1.219$ ({\bf CPC.IV.4}),
$C_u^S=-1.225$ ({\bf CPC.IV.5}), and
$C_u^S=-1.213\,,1.022$ ({\bf CPC.IV.6}).
The second local minima are
developed for {\bf CPC.IV.4} and {\bf CPC.IV.5} at $C_u^S=1$:
 see Table~\ref{tab:best3}.
Similar to the {\bf CPC.III} fits the
$\Delta\chi^2$ difference between the true and local minima
is tiny for {\bf CPC.IV.4} and {\bf CPC.IV.5}, 
$\left.\Delta\chi^2\right|_{\rm local}-
\left.\Delta\chi^2\right|_{\rm true}\sim 0.4$: see Table~\ref{tab:best3}.
On the other hand,
in contrast to the {\bf CPC.III} fits
any values of $C_u^S$ between $-2$ and $2$
are allowed at $2$-$\sigma$ level and higher.
The behavior of $\Delta\chi^2$ by additionally
varying  $\kappa_d$ and $\Delta\Gamma_{\rm tot}$
 is the same as in the previous cases.
%see the lower-middle 
%and lower-right frames.
We again observe the true
minima around $C_u^S=-1$ disappear when $\tan\beta \gsim 0.6$.

\begin{figure}[t!]
\centering
\includegraphics[width=2.0in]{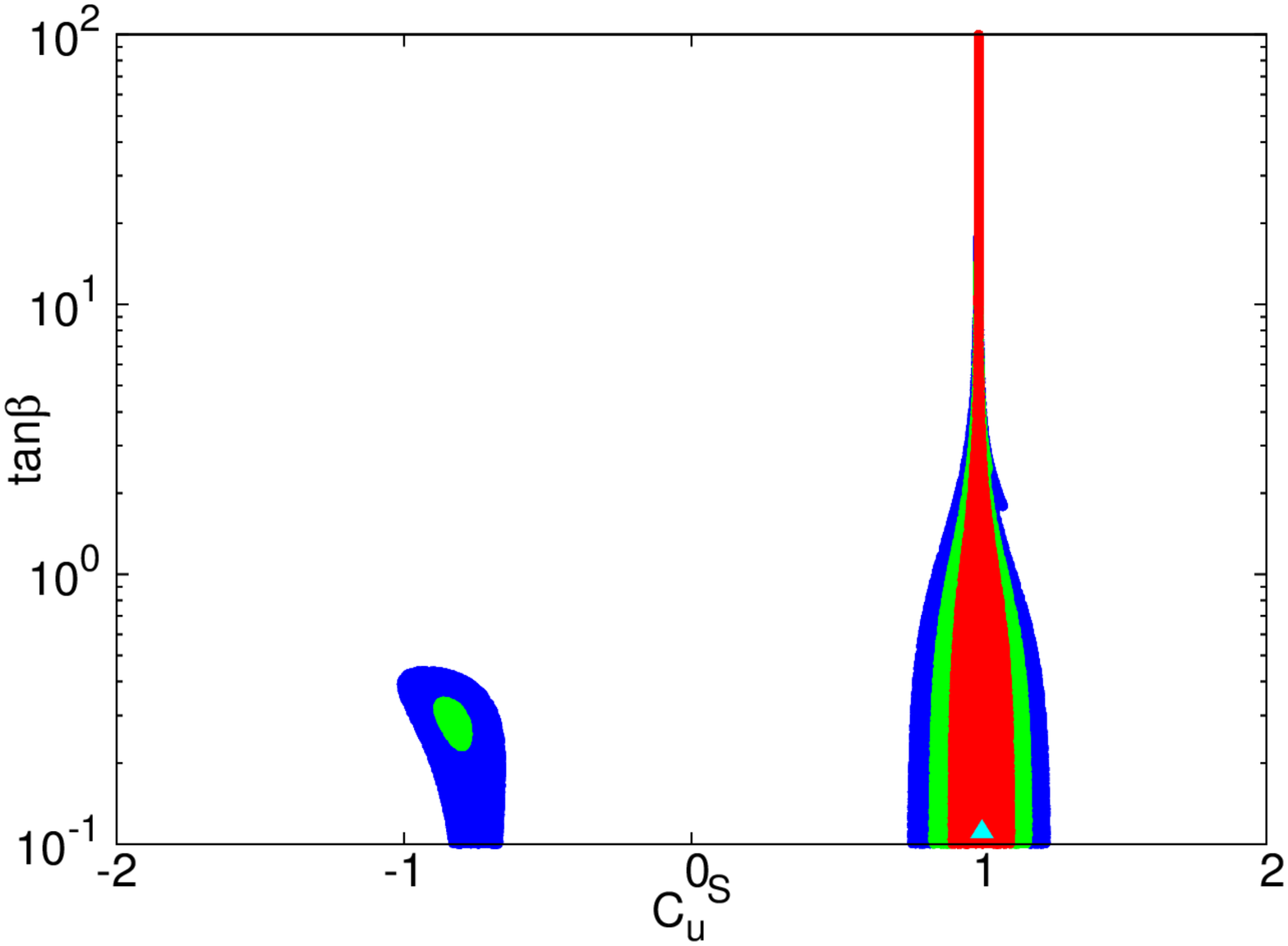}
\includegraphics[width=2.0in]{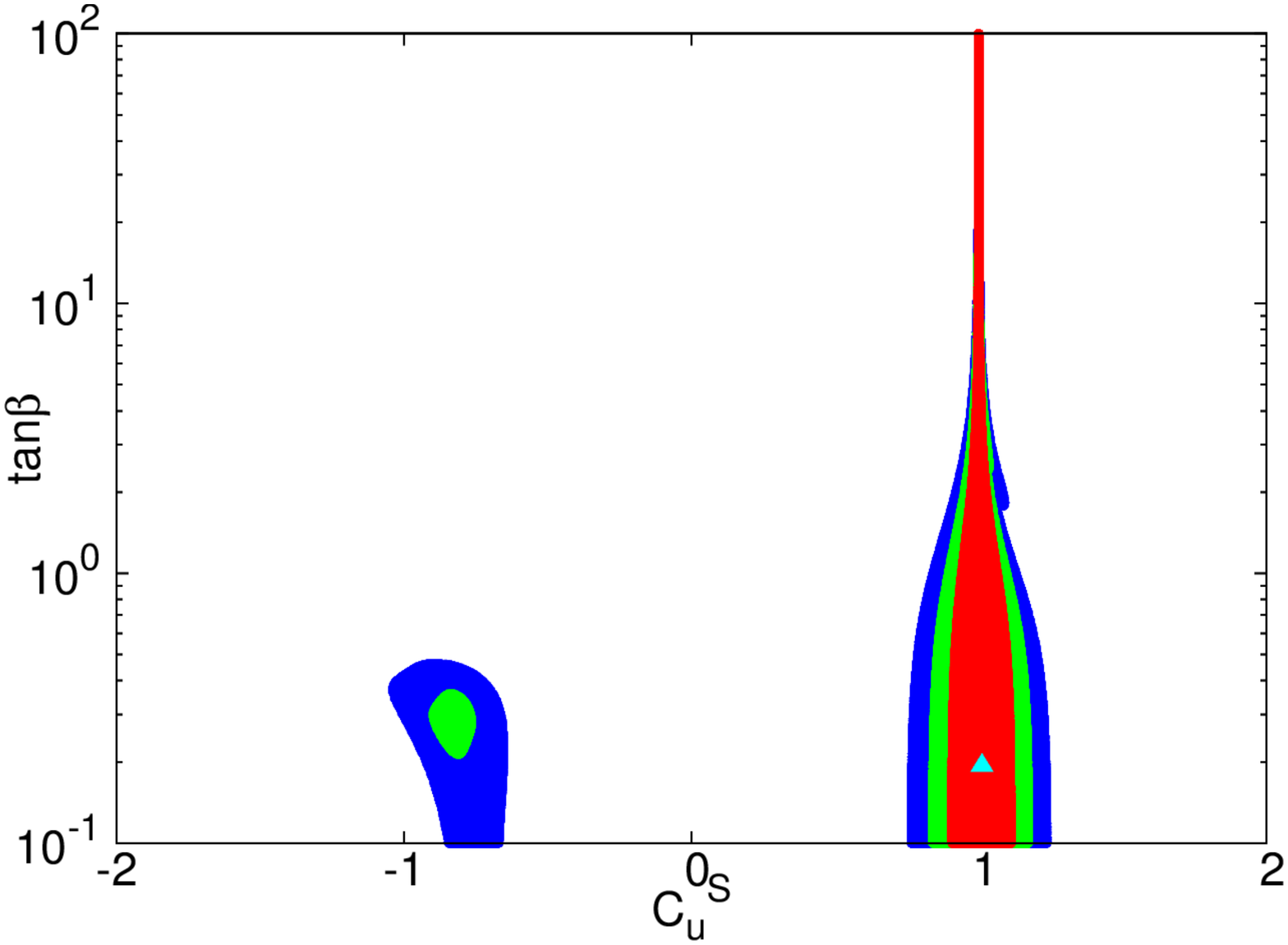}
\includegraphics[width=2.0in]{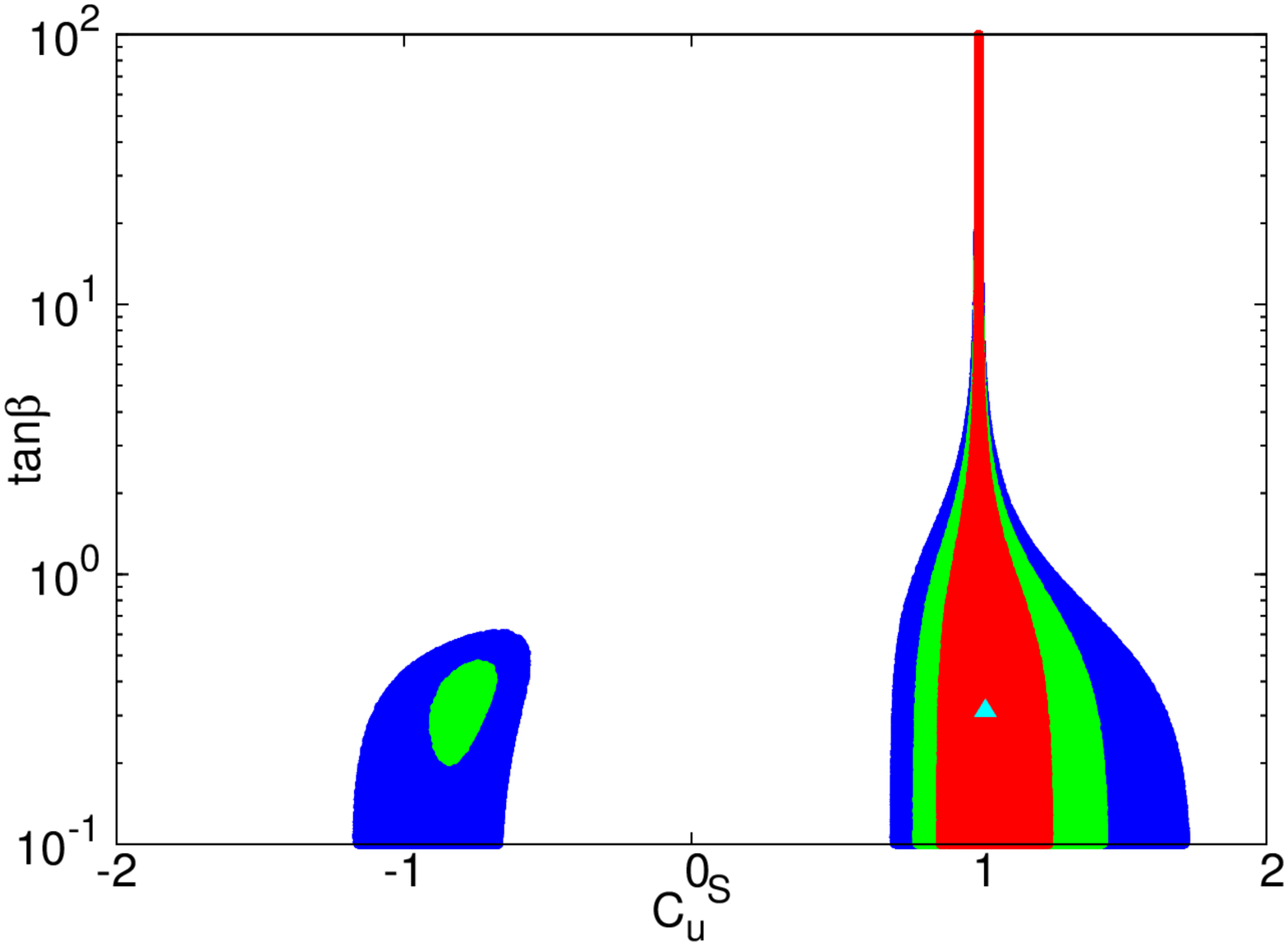}
\includegraphics[width=2.0in]{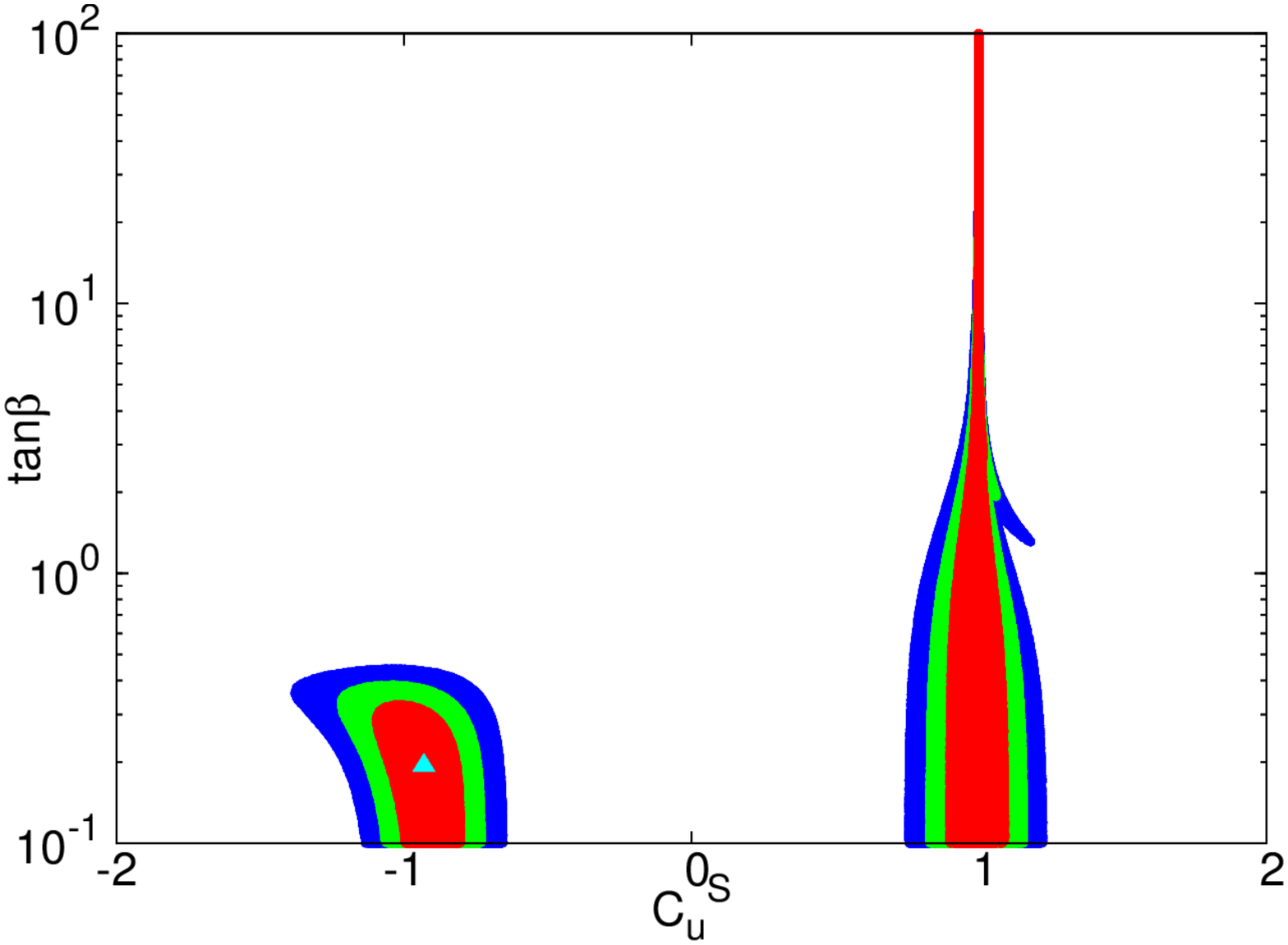}
\includegraphics[width=2.0in]{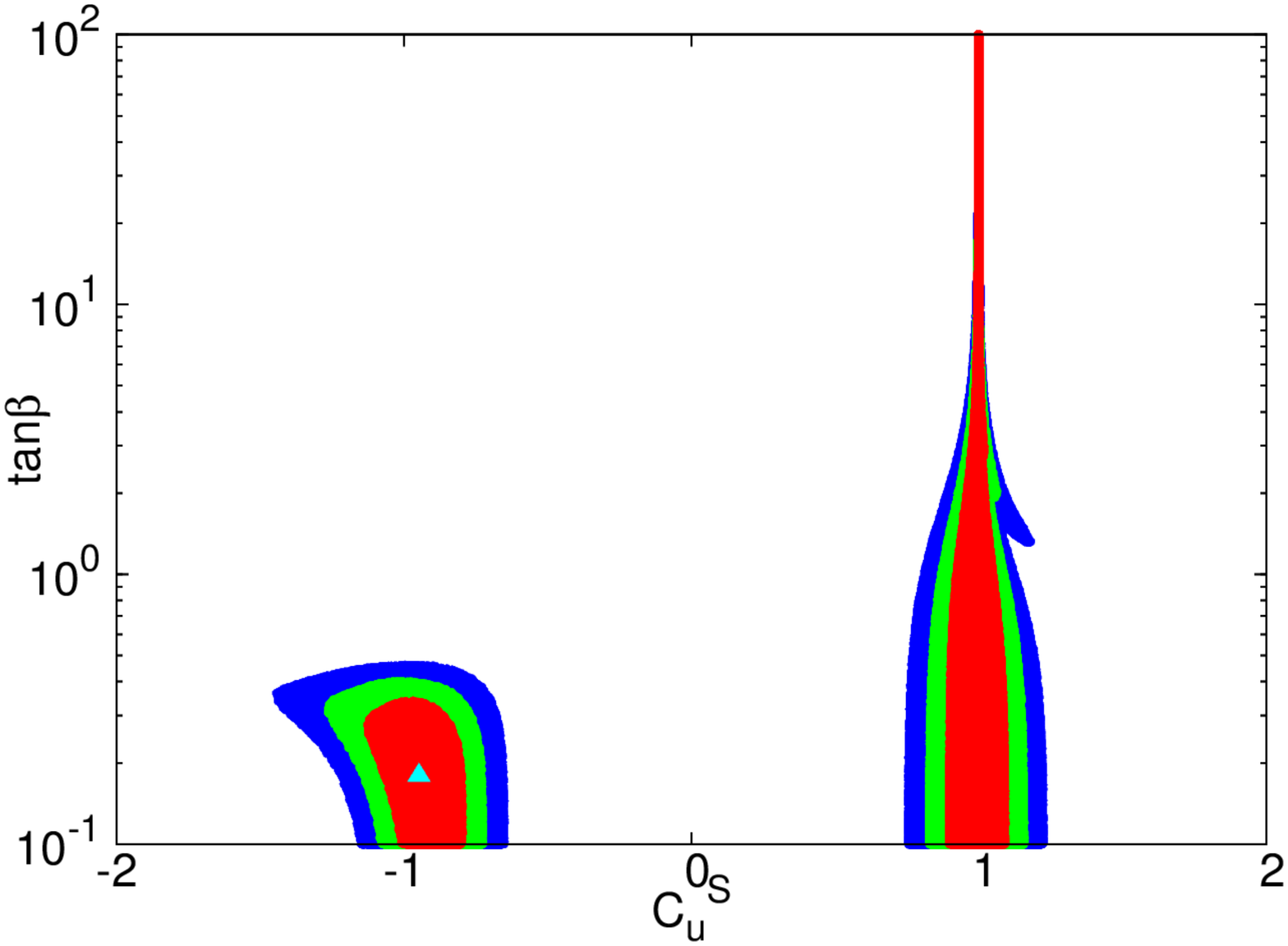}
\includegraphics[width=2.0in]{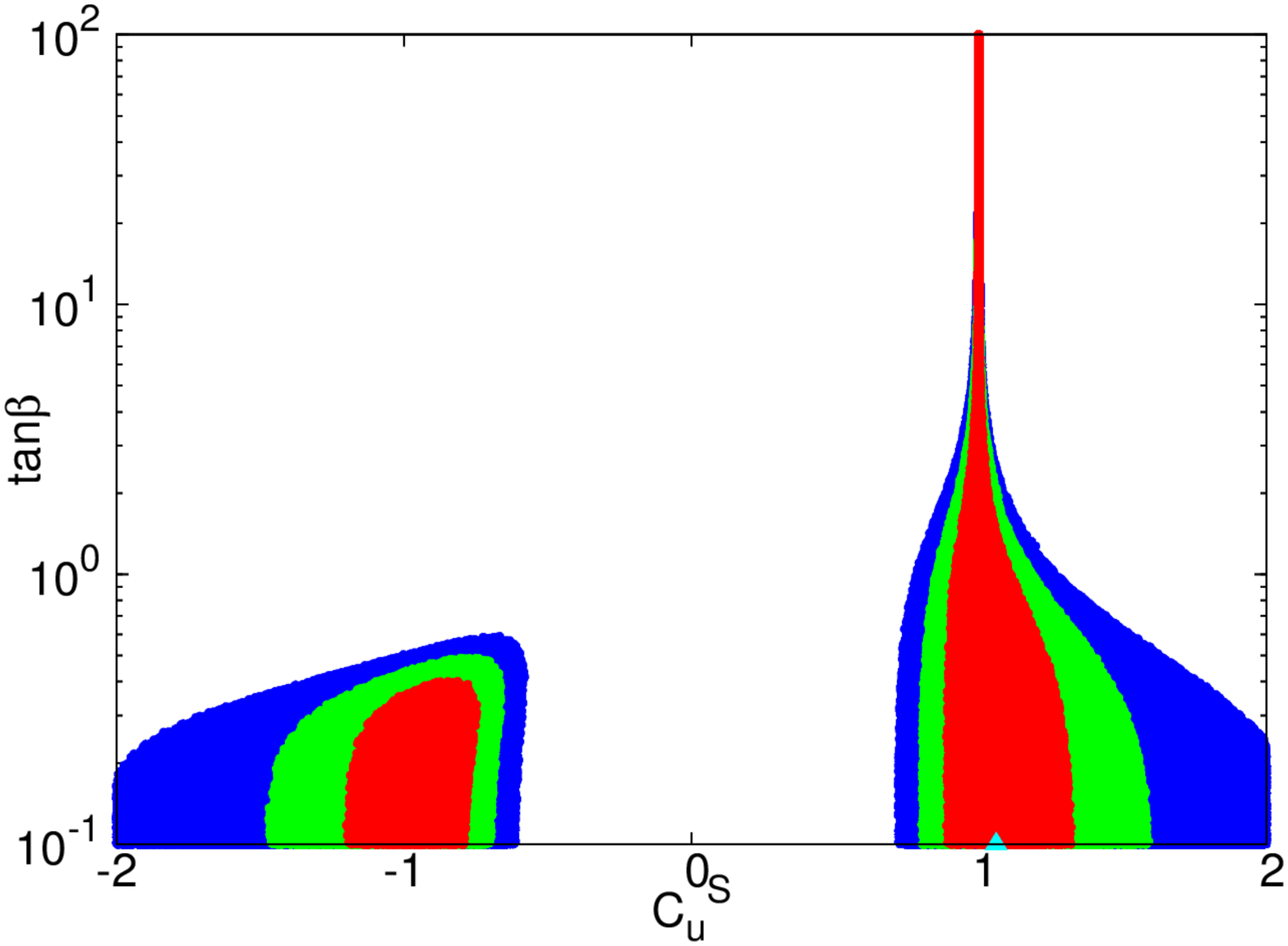}
\includegraphics[width=2.0in]{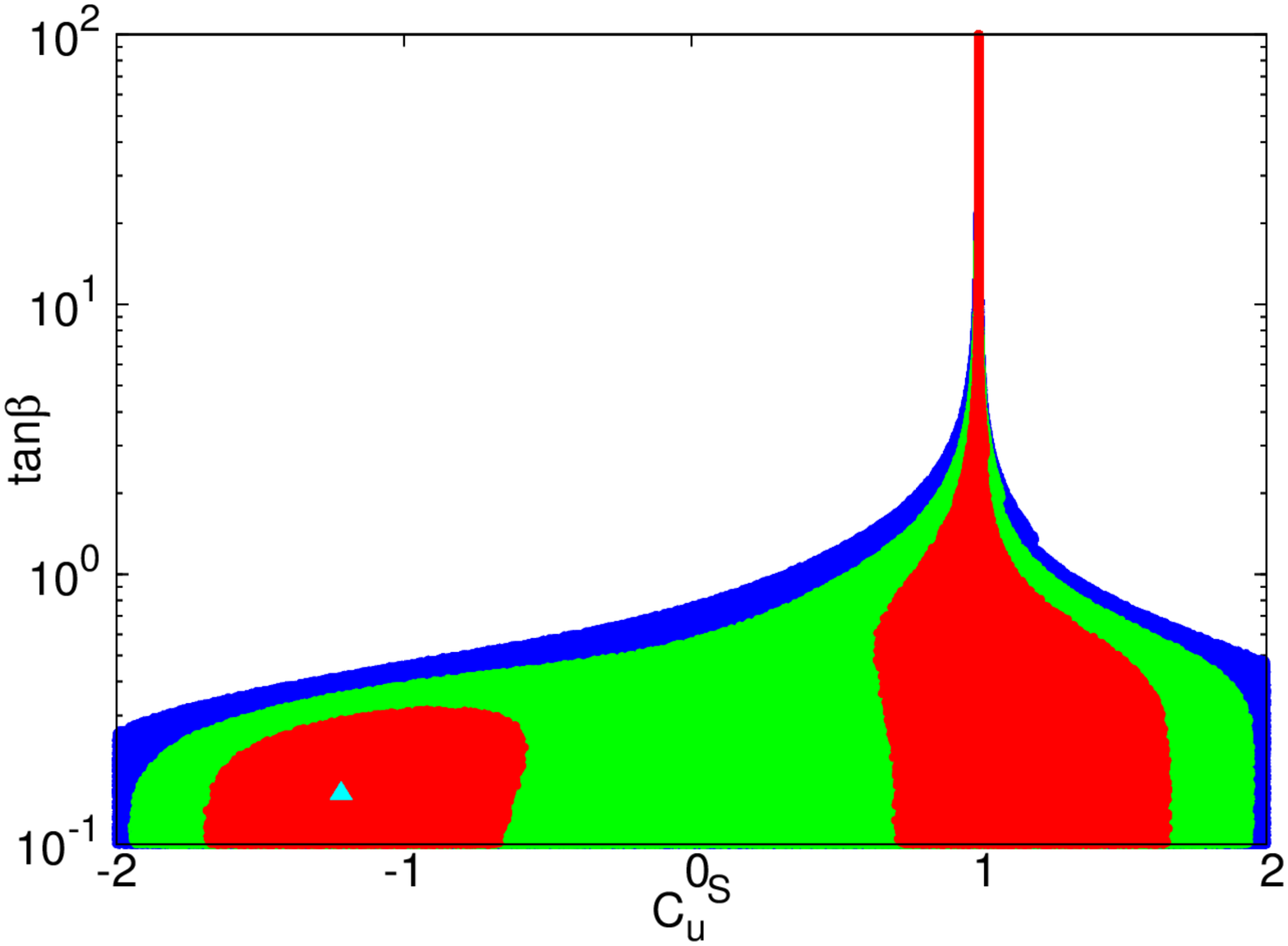}
\includegraphics[width=2.0in]{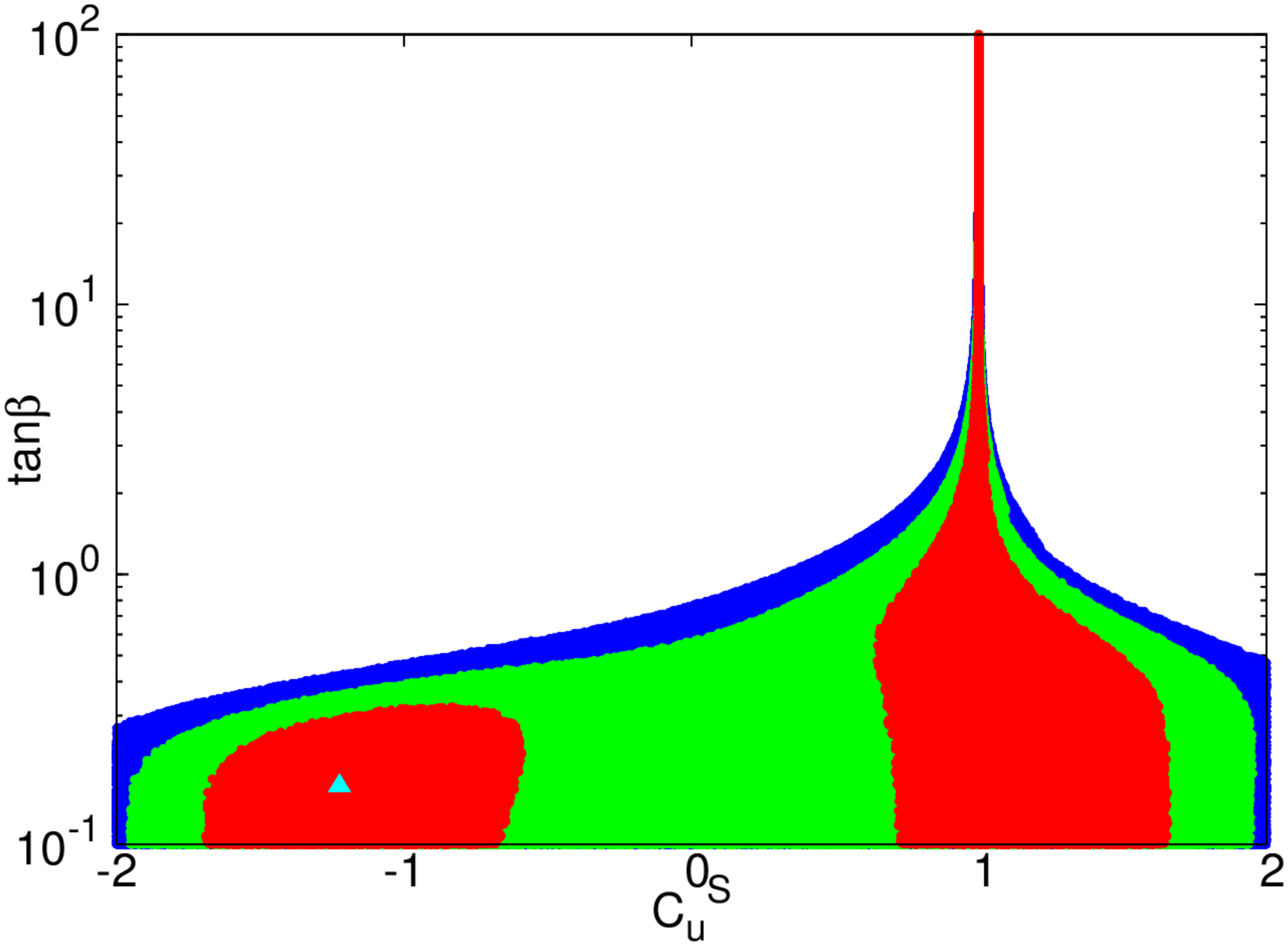}
\includegraphics[width=2.0in]{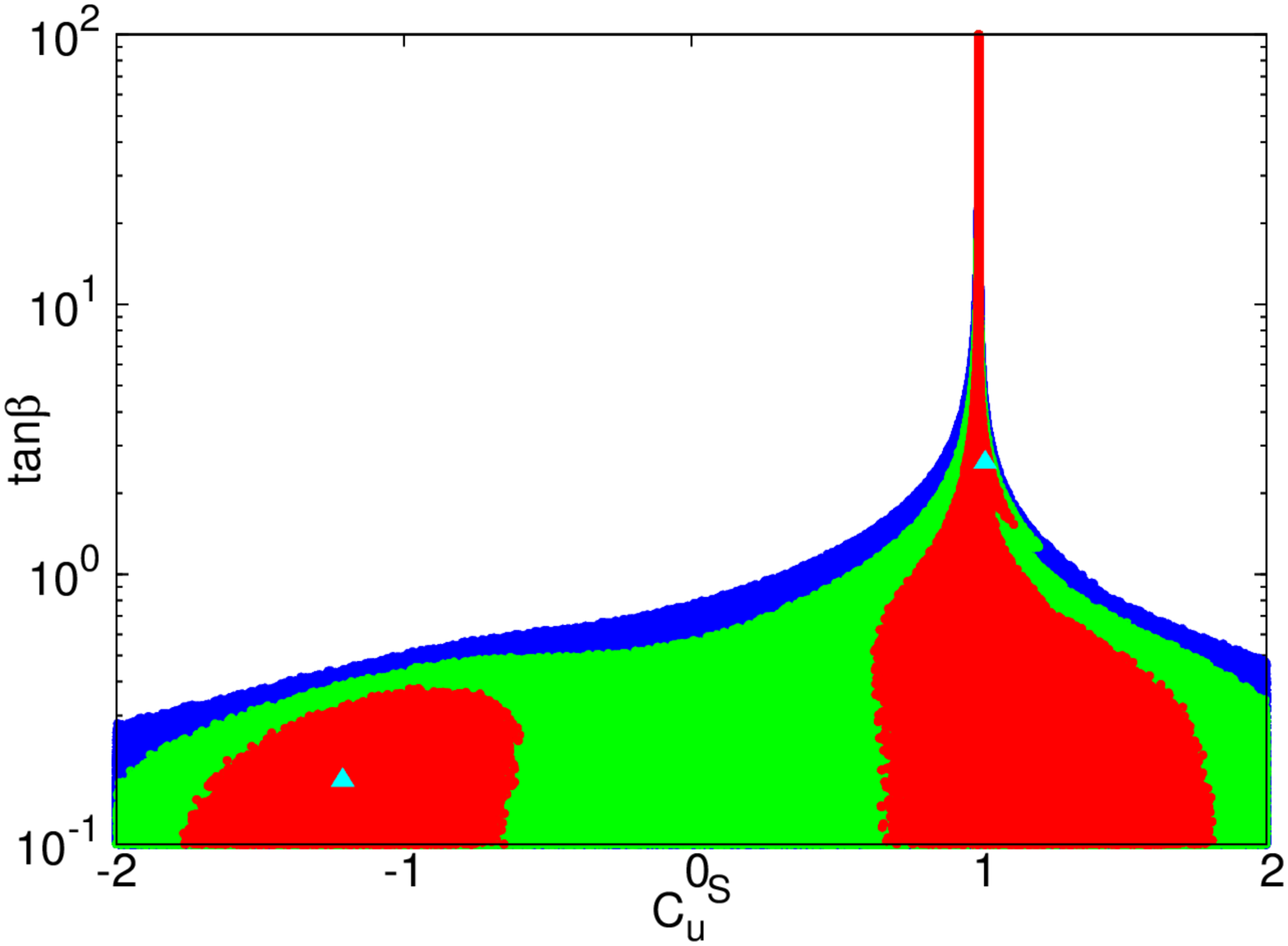}
\caption{\small \label{fig:tanbeta}
The confidence-level regions on the $(C_u^S, \tan\beta)$ plane
for three categories of CPC fits:
{\bf CPC.II} (upper row), {\bf CPC.III} (middle row), and
{\bf CPC.IV} (lower row) fits.
The left frames show the cases of
{\bf CPC.II.2} (varying $C_u^S$, $\tan\beta$), 
{\bf CPC.III.3} (varying $C_u^S$, $\tan\beta$, $\Delta S^\gamma$), and 
{\bf CPC.IV.4} (varying $C_u^S$, $\tan\beta$, $\Delta S^\gamma$, $\Delta S^g$).
In the middle frames, 
the cases {\bf CPC.II.3, CPC.III.4}, {\bf CPC.IV.5} are shown by adding
$\kappa_d$ to the corresponding set of varying parameters. 
The right frames are for the cases of
{\bf CPC.II.4, CPC.III.5}, and {\bf CPC.IV.6} in which
$\Delta \Gamma_{\rm tot}$ is further varied.
The confidence regions shown are for
$\Delta \chi^2 \le 2.3$ (red), $5.99$ (green), and $11.83$ (blue)
above the minimum, which correspond to confidence levels of
$68.3\%$, $95\%$, and $99.7\%$, respectively.  The best-fit point
is denoted by the triangle.
}
\end{figure}

We show the confidence-level regions on the $(C_u^S, \tan\beta)$ plane
for three categories of {\bf CPC} fits:
{\bf CPC.II} (upper row), {\bf CPC.III} (middle row), and
{\bf CPC.IV} (lower row) in Fig.~\ref{fig:tanbeta}.
The confidence level (CL) regions shown are for
$\Delta \chi^2 \le 2.3$ (red), $5.99$ (green), and $11.83$ (blue)
above the minimum, which correspond to CLs of
$68.3\%$, $95\%$, and $99.7\%$, respectively.  The best-fit point
is denoted by the triangle.
We observe that the plots are very close to those of the
Type II of the 2HDM \cite{Cheung:2013rva},
though the regions in general shrink by small amounts.
First of all,
the vertical $68.3\%$ confidence (red) regions around $C_u^S=1$ can be understood from
Eq.~(\ref{eq:tbsq}) by observing
that the value of $t_\beta$ changes from $\infty$ to $1$ when $(C_u^S-1)$ deviates from
$-(1-C_v)$ by the amount of $\pm\sqrt{1-C_v^2}$ and there are 
generally many points around $C_v=1$ as shown in 
Fig.~\ref{fig:cv}.

In each category of fits, Fig.~\ref{fig:chi2} is helpful to understand
the basic behavior of the CL regions as $C_u^S$ is varied.
In the {\bf CPC.II} fits, the region around $C_u^S=1$ is much more preferred. 
The negative $C_u^S$ values are not allowed at $68\%$ CL. 
In the {\bf CPC.III} fits, the region around $C_u^S=-1$ falls into
the stronger 68.3\% CL 
but $C_u^S=0$ is not allowed even at $99.7\%$ CL.
On the other hand, 
the whole range of $-2 < C_u^S < 2$ is allowed at 95\% CL 
for the {\bf CPC.IV} fits though not at $68.3\%$ CL.
In all the fits, the negative values of $C_u^S$ are not allowed at $95\%$ CL
when $\tan\beta \gsim 0.5$  is imposed, which is in general
required by the perturbativity of the top-quark Yukawa coupling.
The CL regions hardly change by varying  $\kappa_d$ additionally,
but the CL regions can extend to the regions of $|C_u^S|>1$
by further varying $\Delta\Gamma_{\rm tot}$.

The CL regions on the $(C_u^S, C_v)$ plane are shown in Fig.~\ref{fig:cv}
for the three categories of {\bf CPC} fits:
{\bf CPC.II} (upper row), {\bf CPC.III} (middle row), and
{\bf CPC.IV} (lower row).
The CL regions are labeled in the same way as in Fig.~\ref{fig:tanbeta}. 
We observe $C_v\gsim 0.75$ at $68.3\%$ CL except in the {\bf CPC.IV.6} fit.
Otherwise,
one may make similar observations as in Fig.~\ref{fig:tanbeta} for
the behavior of the CL regions as $C_u^S$ is varied.

Figure~\ref{fig:cd} shows the CL regions on the $(C_u^S, C_d^S)$ plane
in the same format as Fig.~\ref{fig:tanbeta}. $C_d^S\approx 1$ is 
preferred except for the {\bf CPC.IV.6} fit, in which the best-fit values 
 of $C_d^S$ are about $0.96$ and $-0.81$ when
$C_u^S\sim -1.2$ and $1.0$, respectively:
see Table~\ref{tab:best}. 
Nevertheless, the difference in $\Delta\chi^2$ between the 
true minima and the local minimum around the SM limit $(C_u^S,C_d^S)=(1,1)$ is 
small.
The CL regions, centered around
the best-fit values, significantly expand as the fit progresses 
from {\bf CPC.II} to {\bf CPC.III} and
from {\bf CPC.III} to {\bf CPC.IV}, 
as well as by adding
$\Delta\Gamma_{\rm tot}$ to the set of varying parameters.

\begin{figure}[t!]
\centering
\includegraphics[width=2.0in]{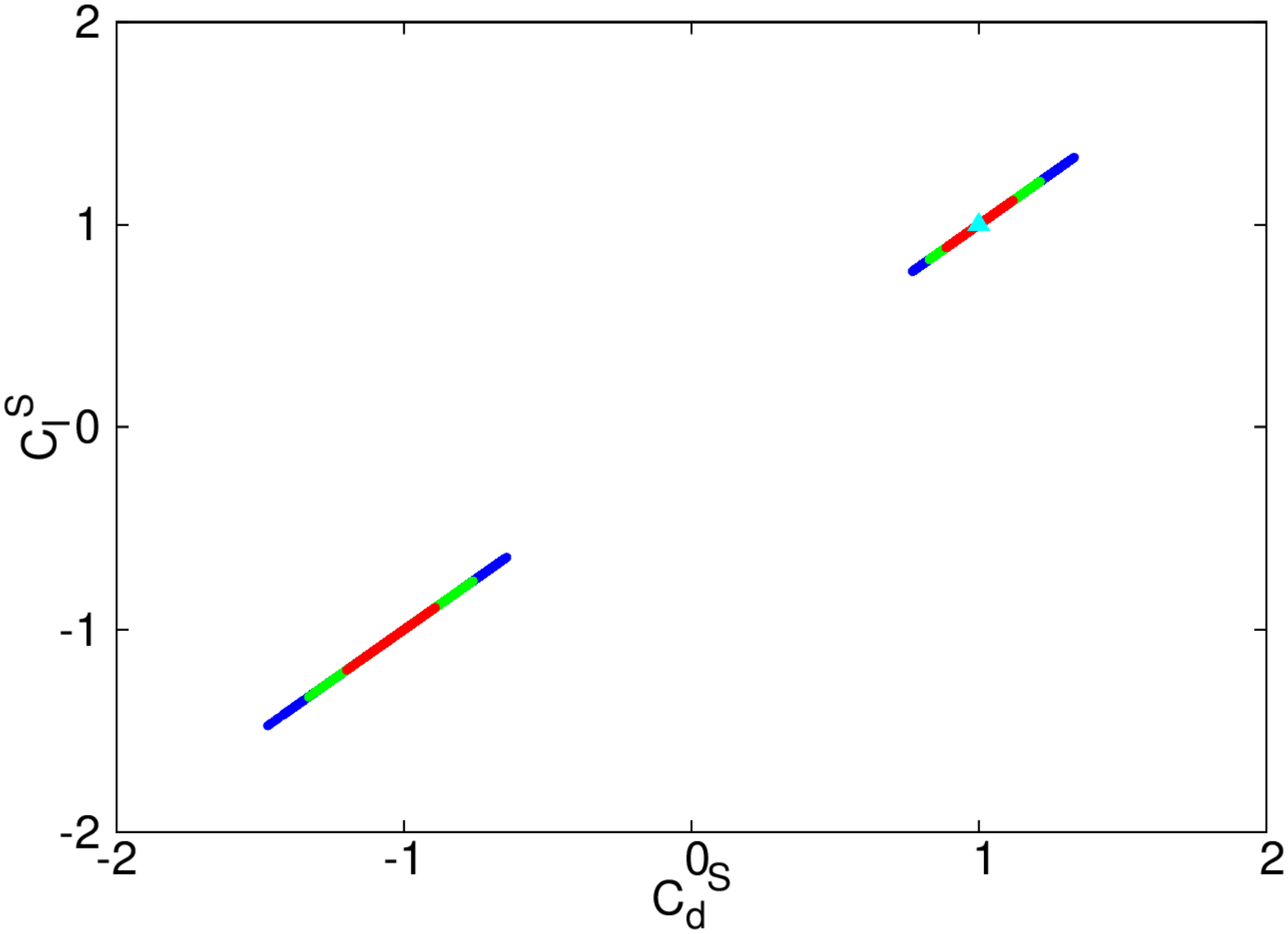}
\includegraphics[width=2.0in]{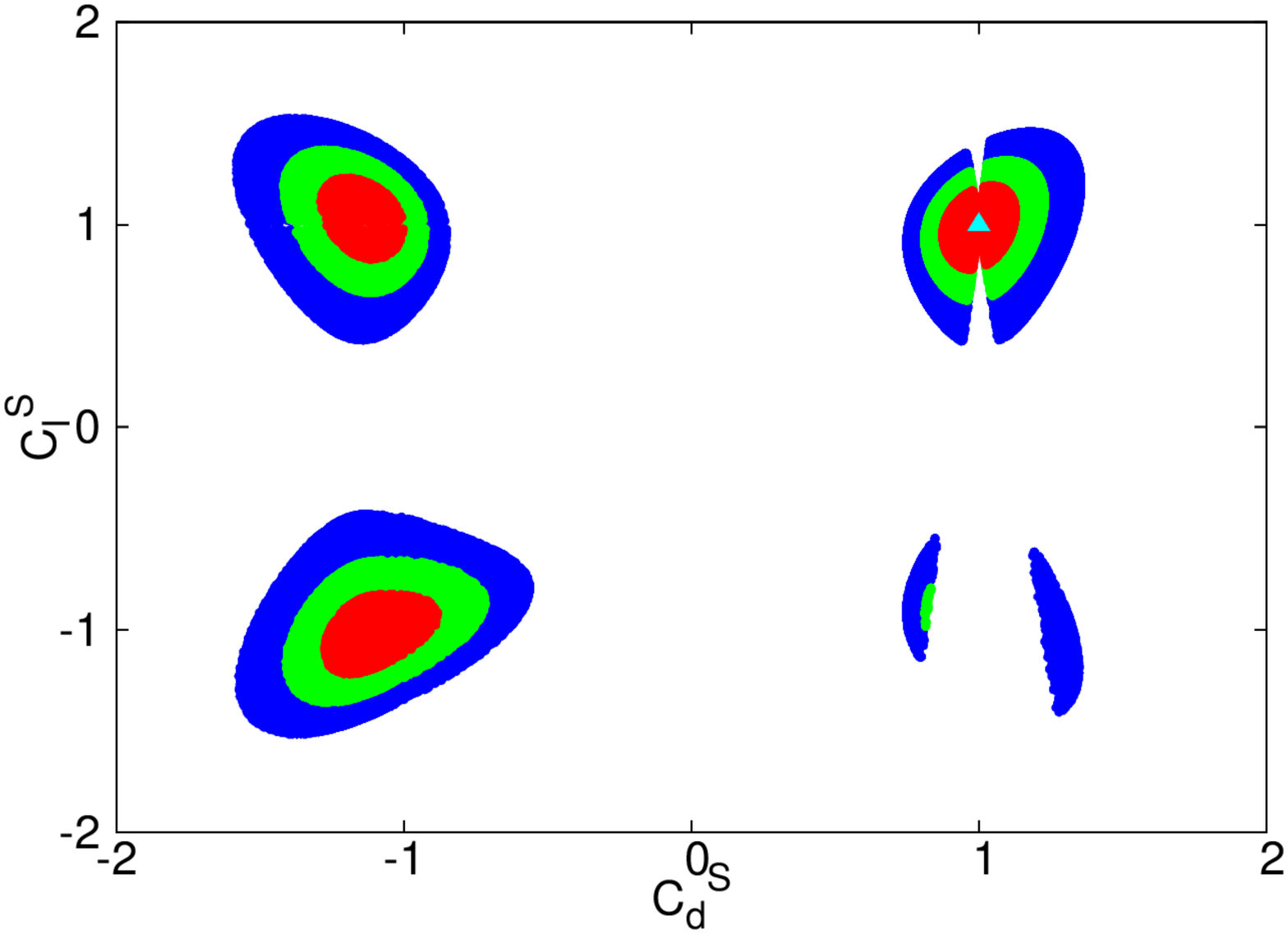}
\includegraphics[width=2.0in]{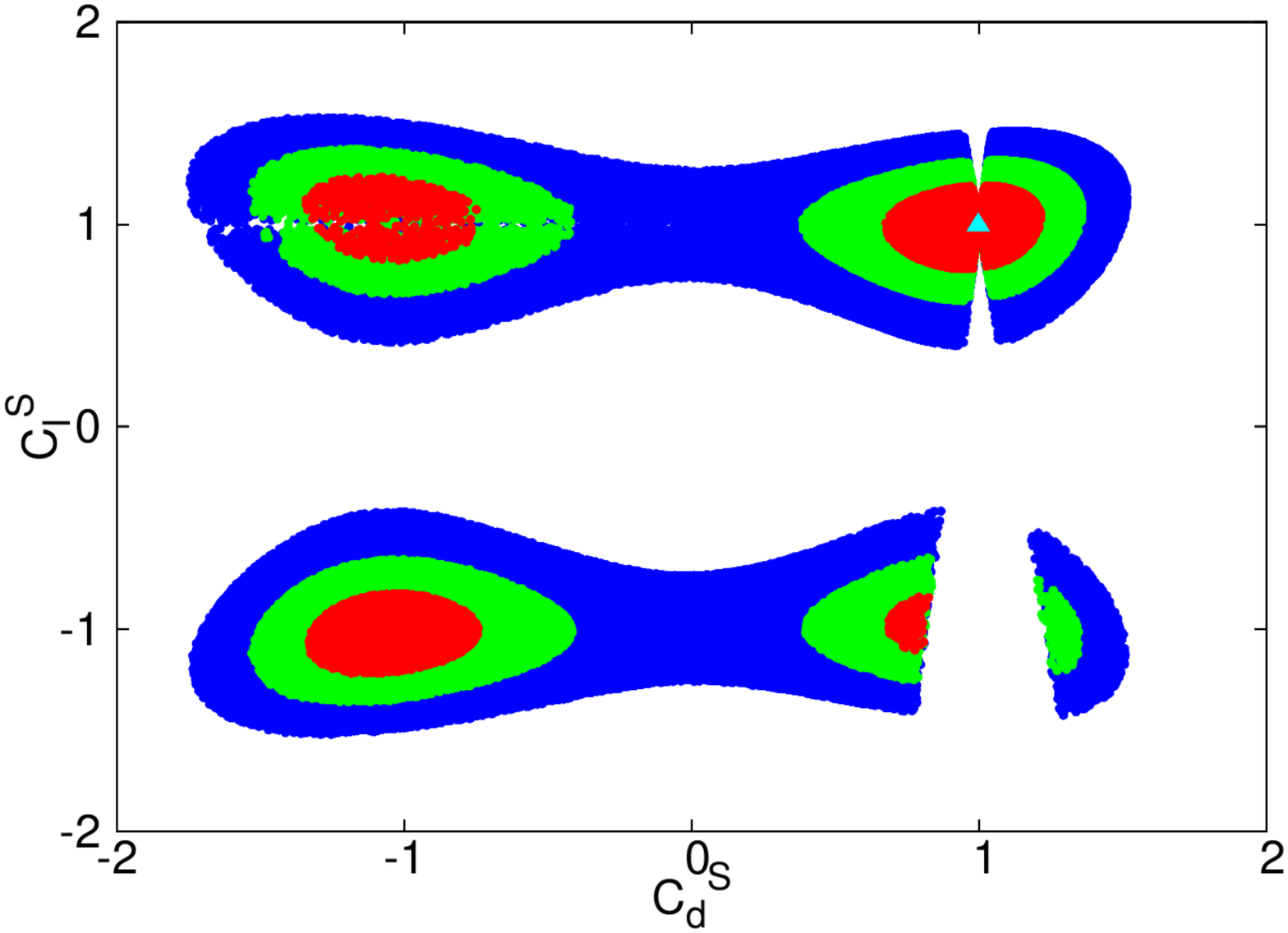}
\includegraphics[width=2.0in]{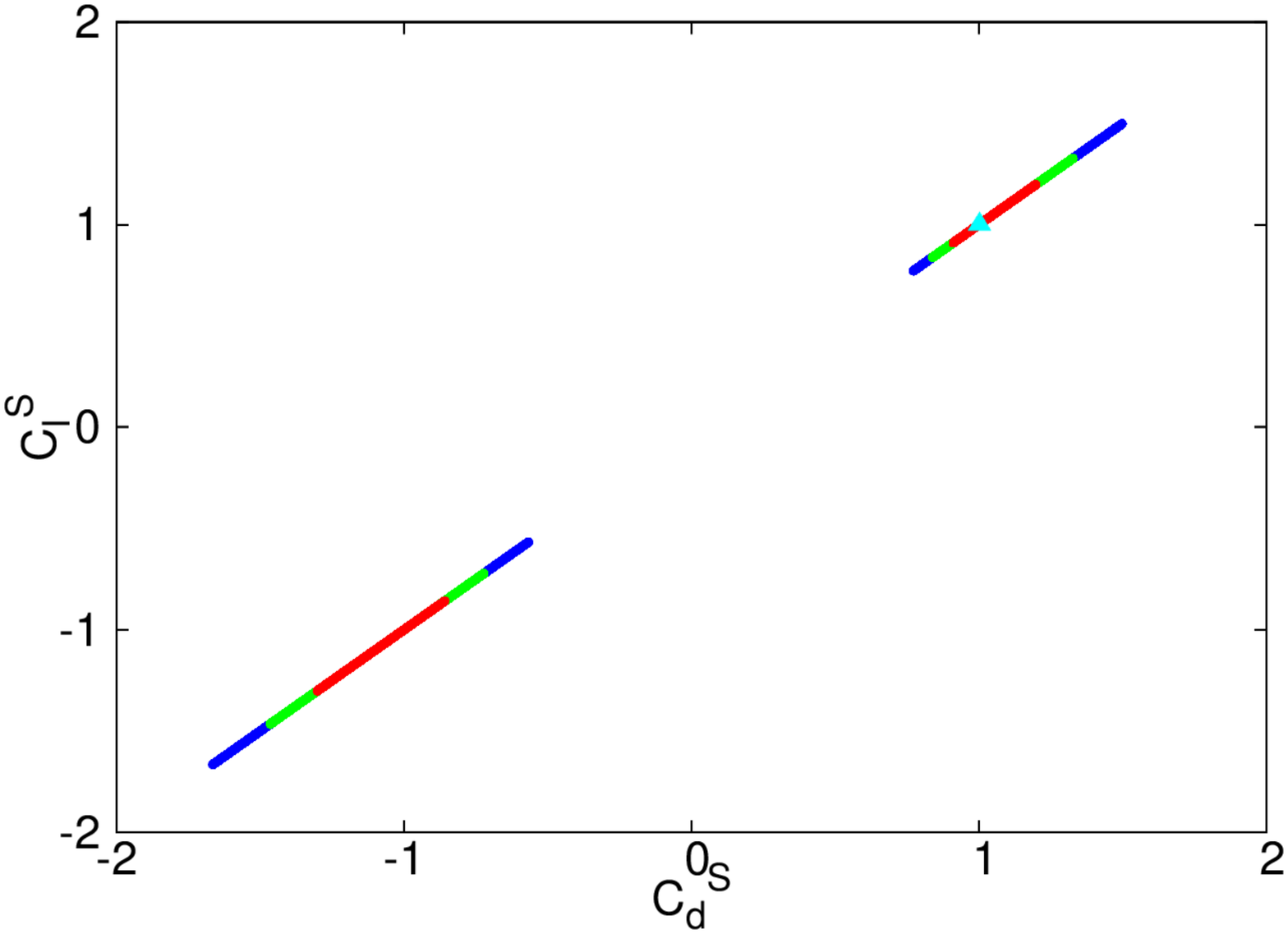}
\includegraphics[width=2.0in]{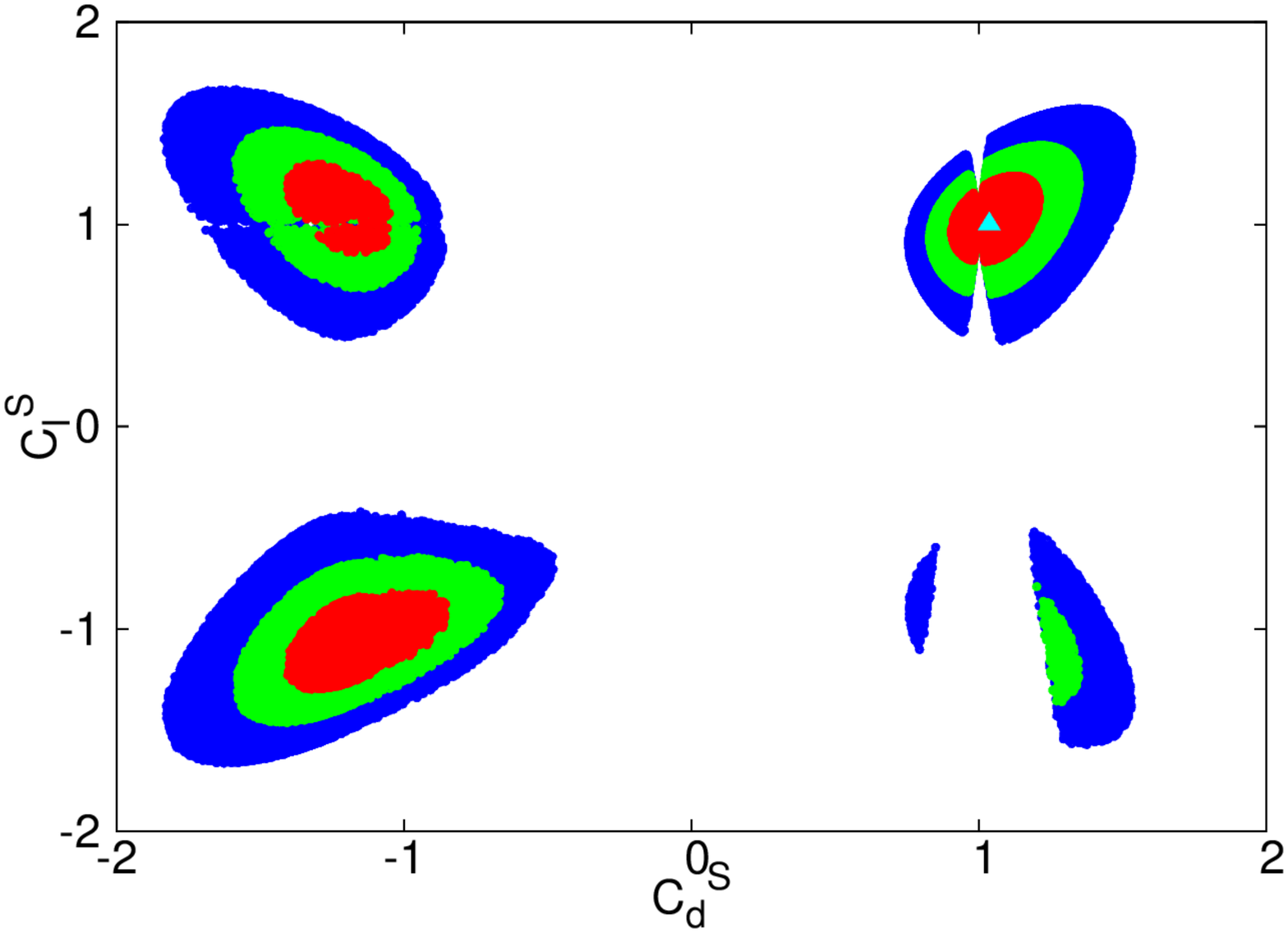}
\includegraphics[width=2.0in]{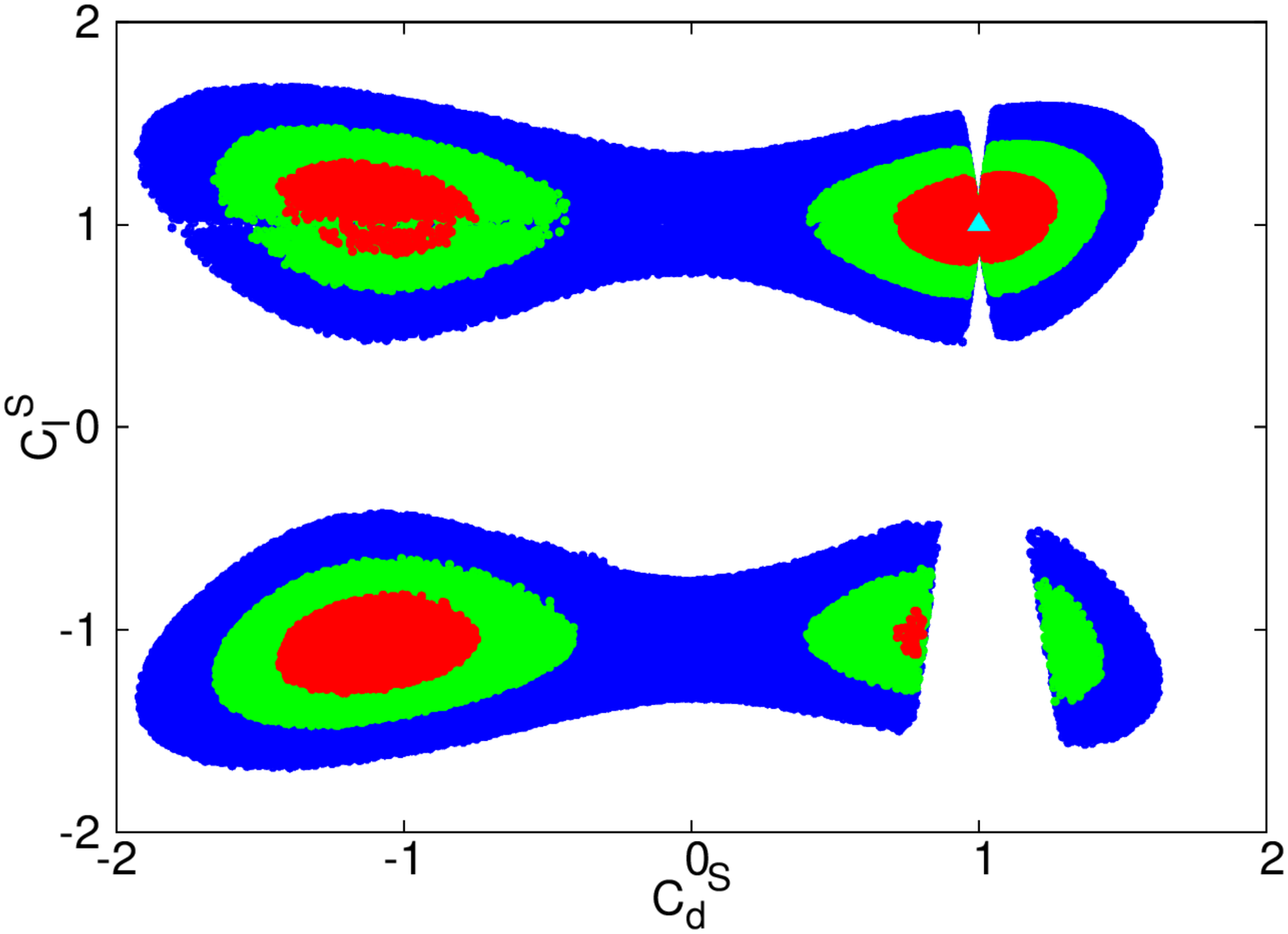}
\includegraphics[width=2.0in]{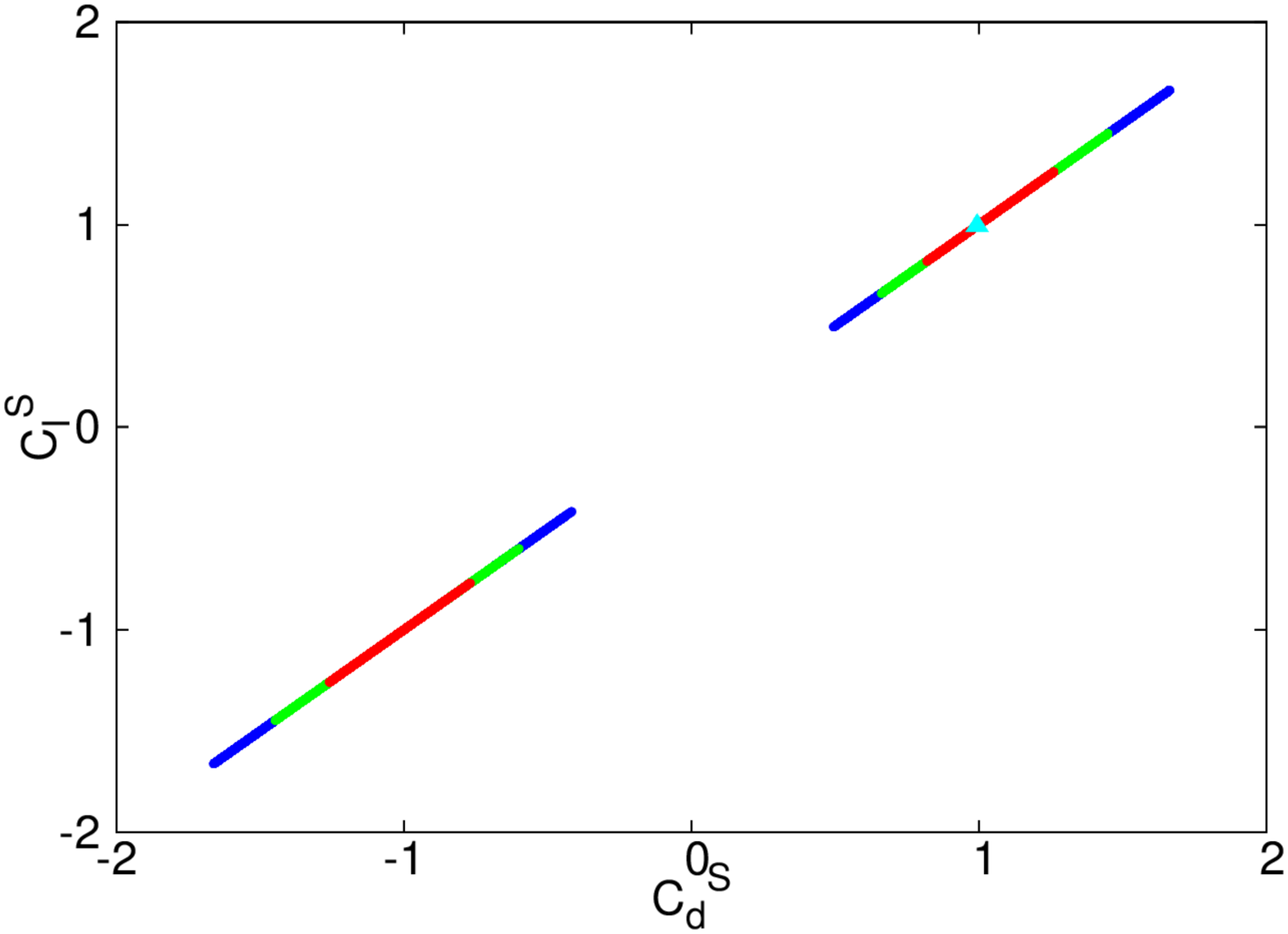}
\includegraphics[width=2.0in]{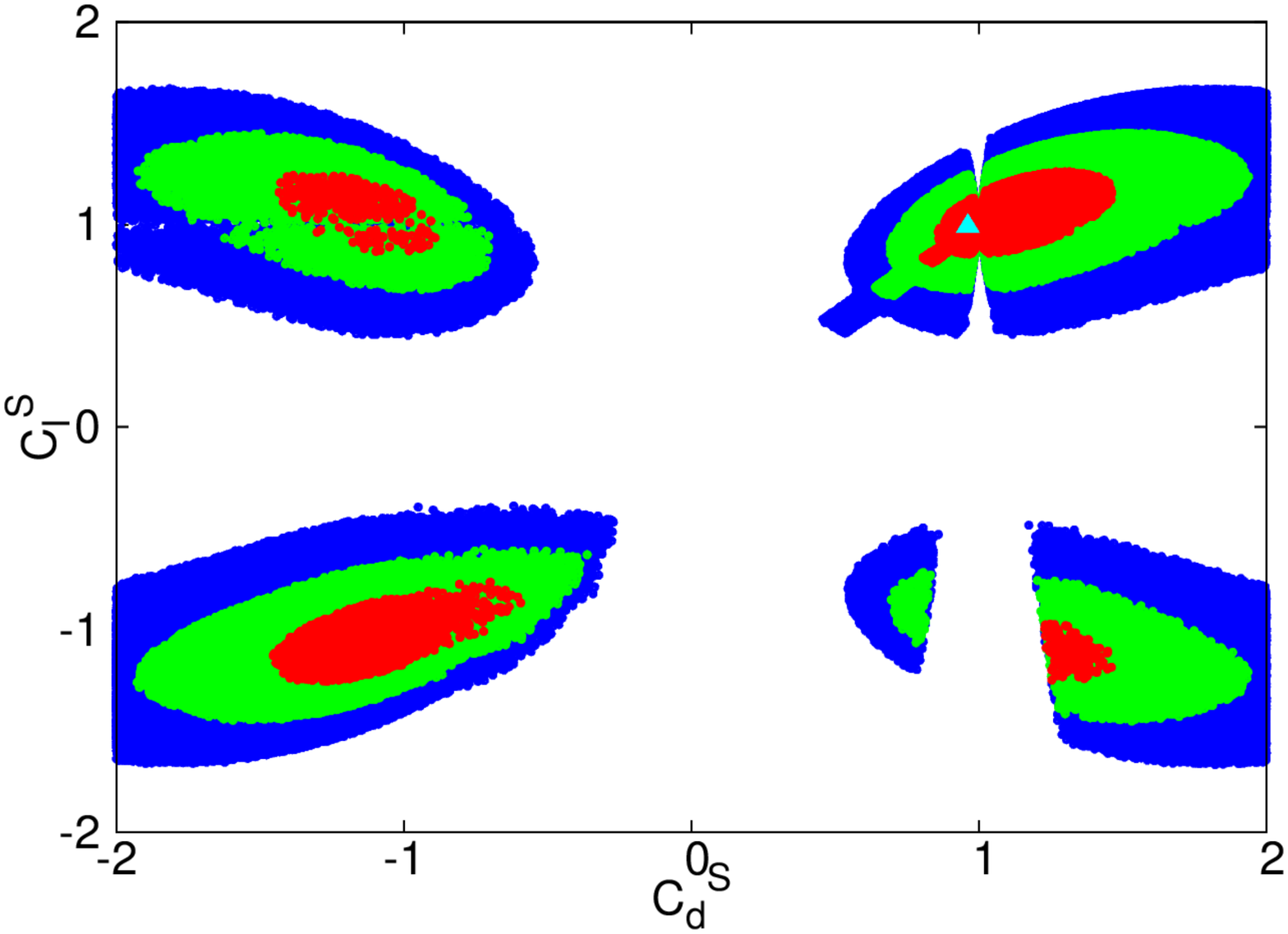}
\includegraphics[width=2.0in]{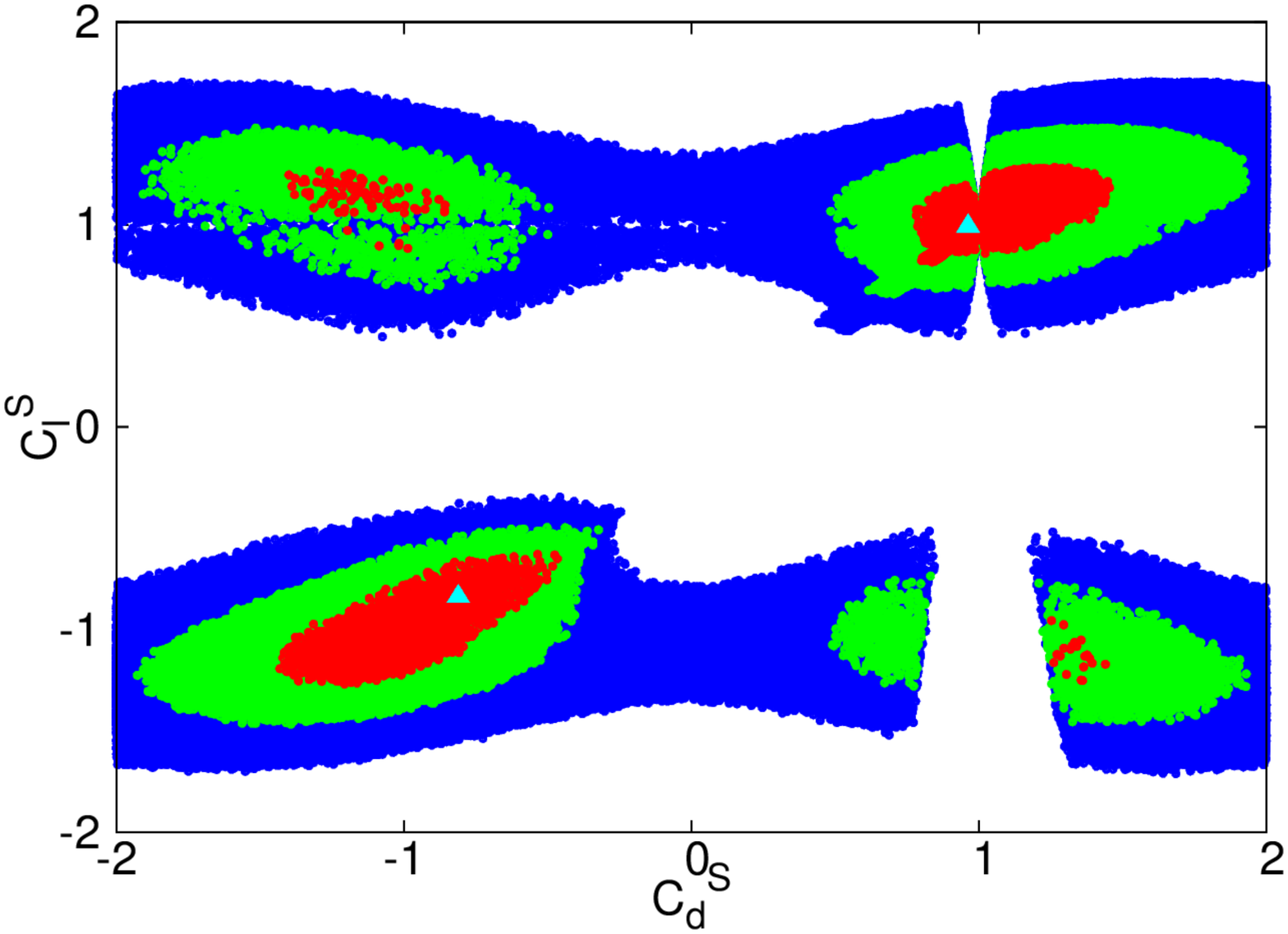}
\caption{\small \label{fig:cdcl}
The same as in Fig.~\ref{fig:tanbeta} but on
the $(C_d^S, C_\ell^S)$ plane.
}
\end{figure}

We show the CL regions on the $(C_d^S, C_\ell^S)$ plane in Fig.~\ref{fig:cdcl}. 
The format
is the same as in Fig.~\ref{fig:tanbeta}. At tree level without including
$\kappa_d$, $C_\ell^S=C_d^S=O_{\phi_1i}/\cos\beta$
as clearly seen in the left frames  and the true and local minima are located
at $(C_d^S, C_\ell^S)=(1,1)$ and $(-1,-1)$.
The tree-level relation is modified by introducing
$\kappa_d$ and the local minima around $(C_d^S, C_\ell^S)=(-1,1)$ are developed as shown in
the middle frames.
Further varying $\Delta \Gamma_{\rm tot}$, we observe that $C_d^S=0$ is allowed at the
$99.7\%$ CL but $|C_\ell^S|>0$ always: see the right frames.

\begin{figure}[t!]
\centering
\includegraphics[width=2.0in]{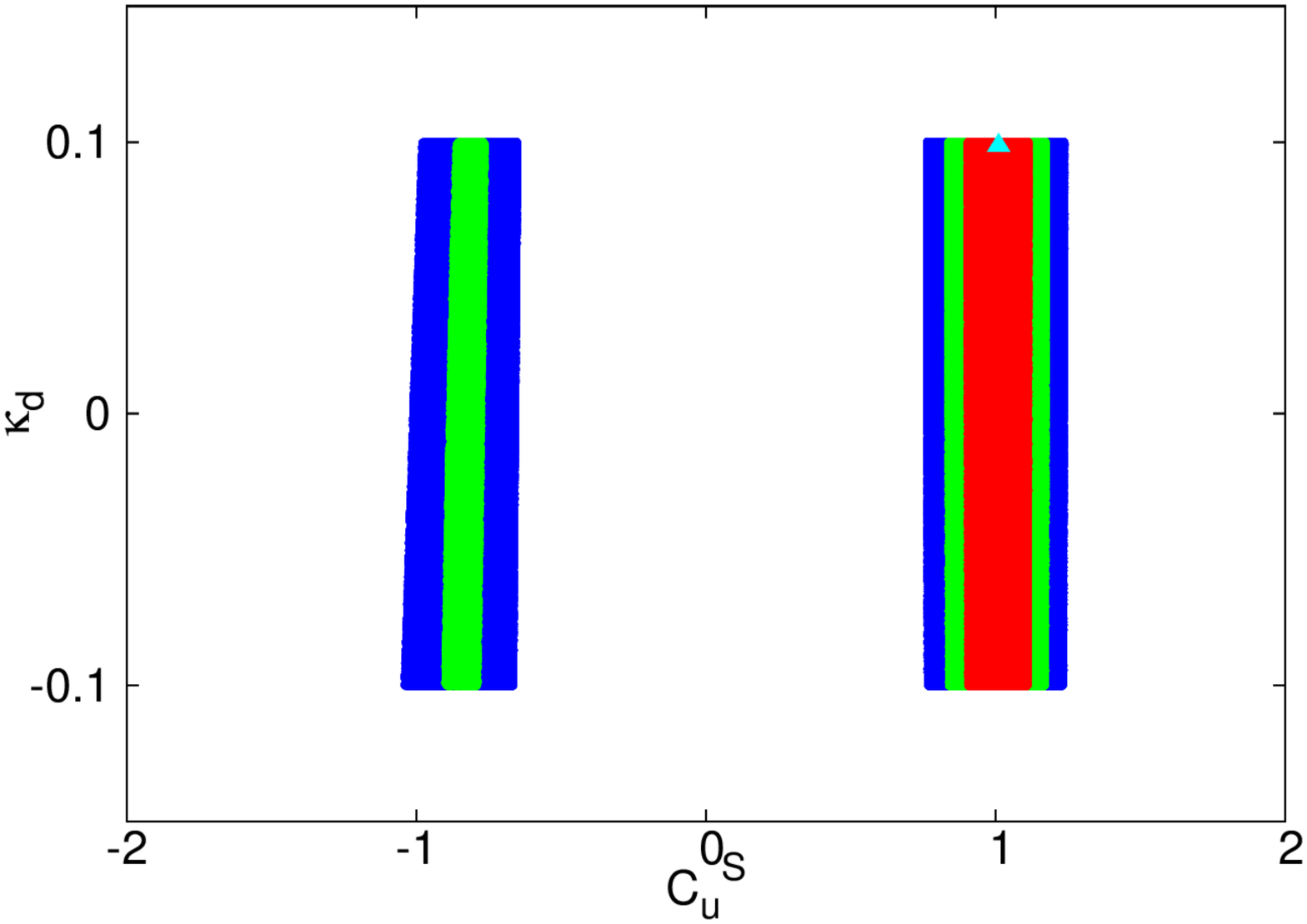}
\includegraphics[width=2.0in]{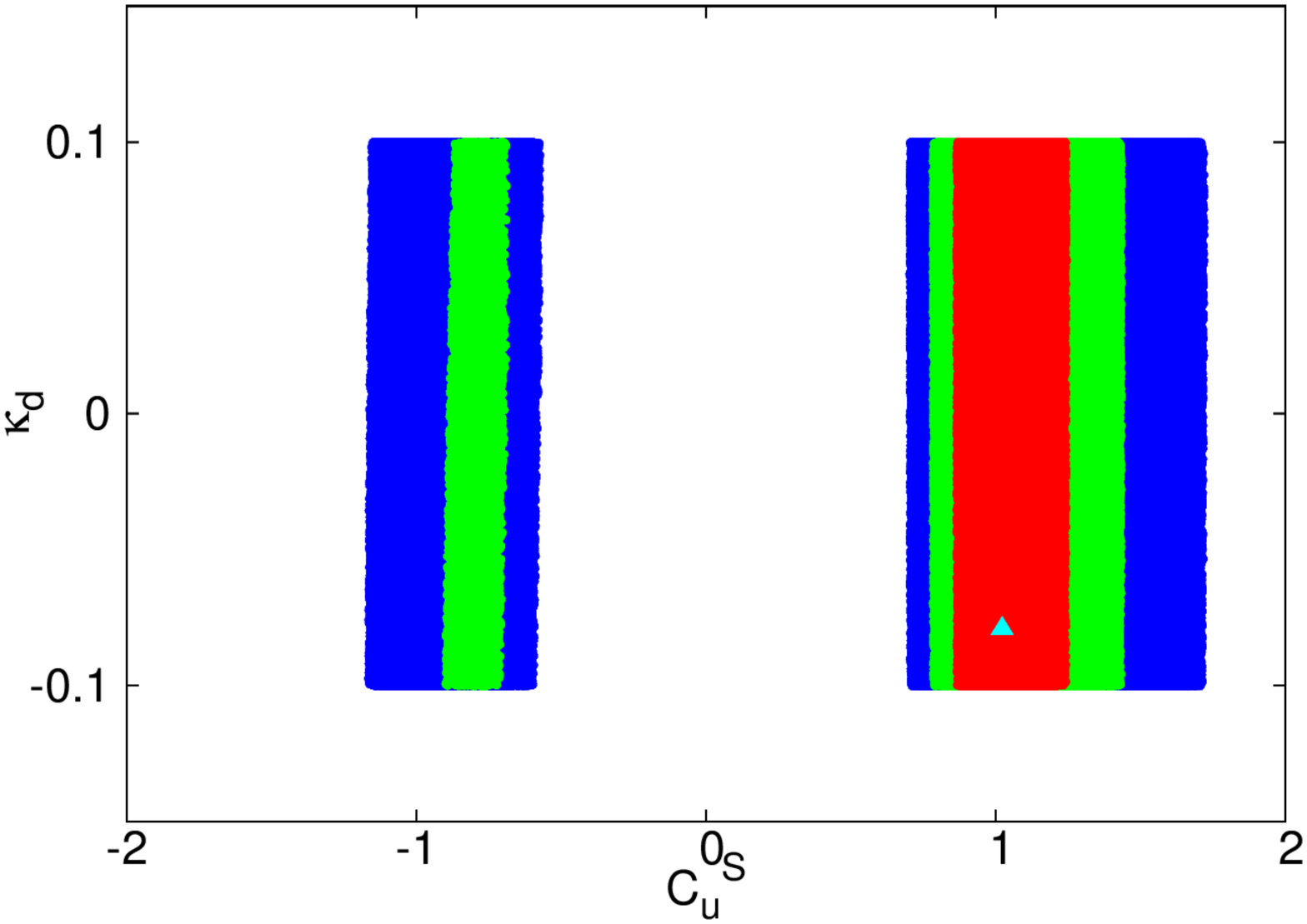}
\includegraphics[width=2.0in]{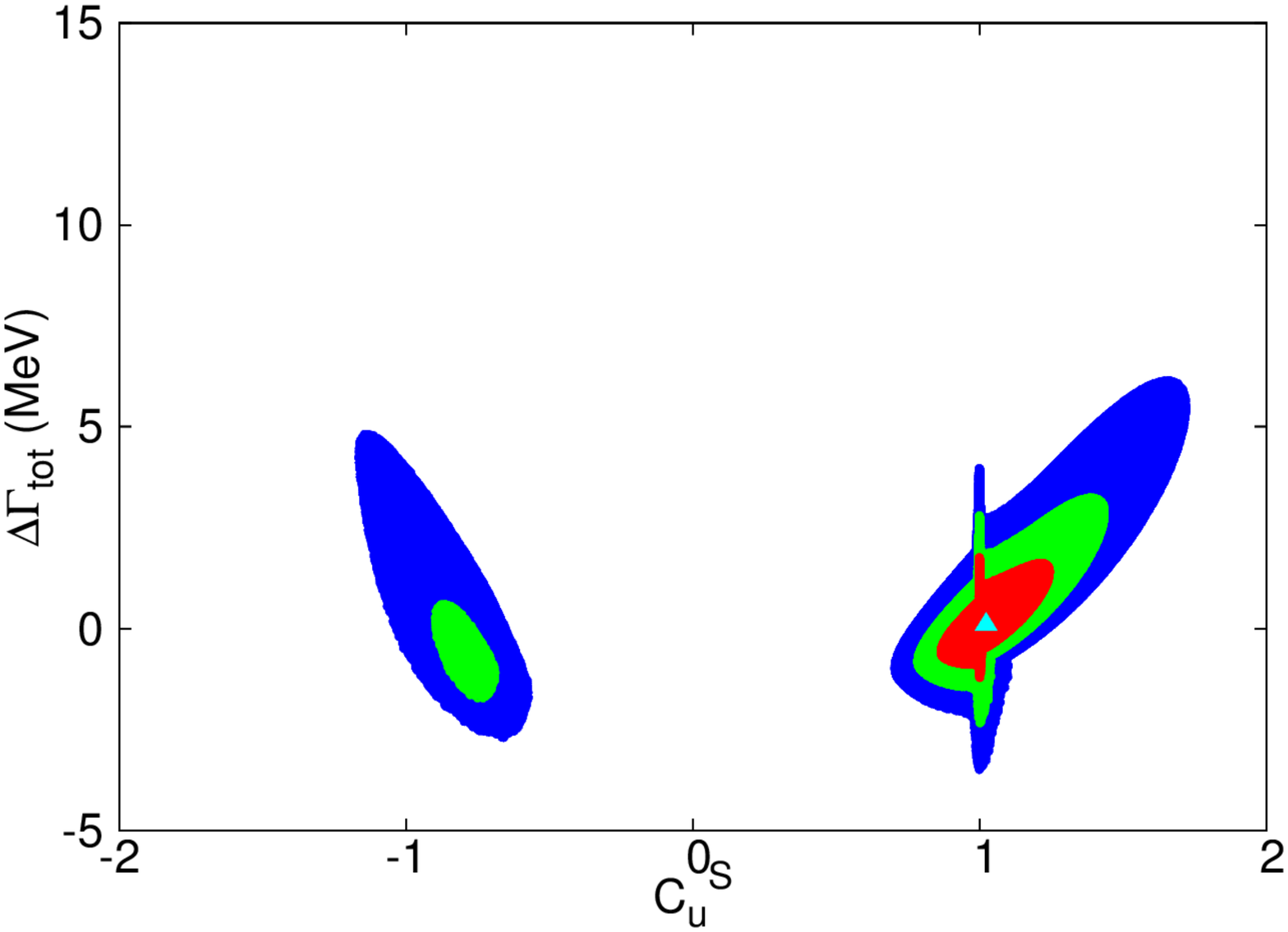}
\includegraphics[width=2.0in]{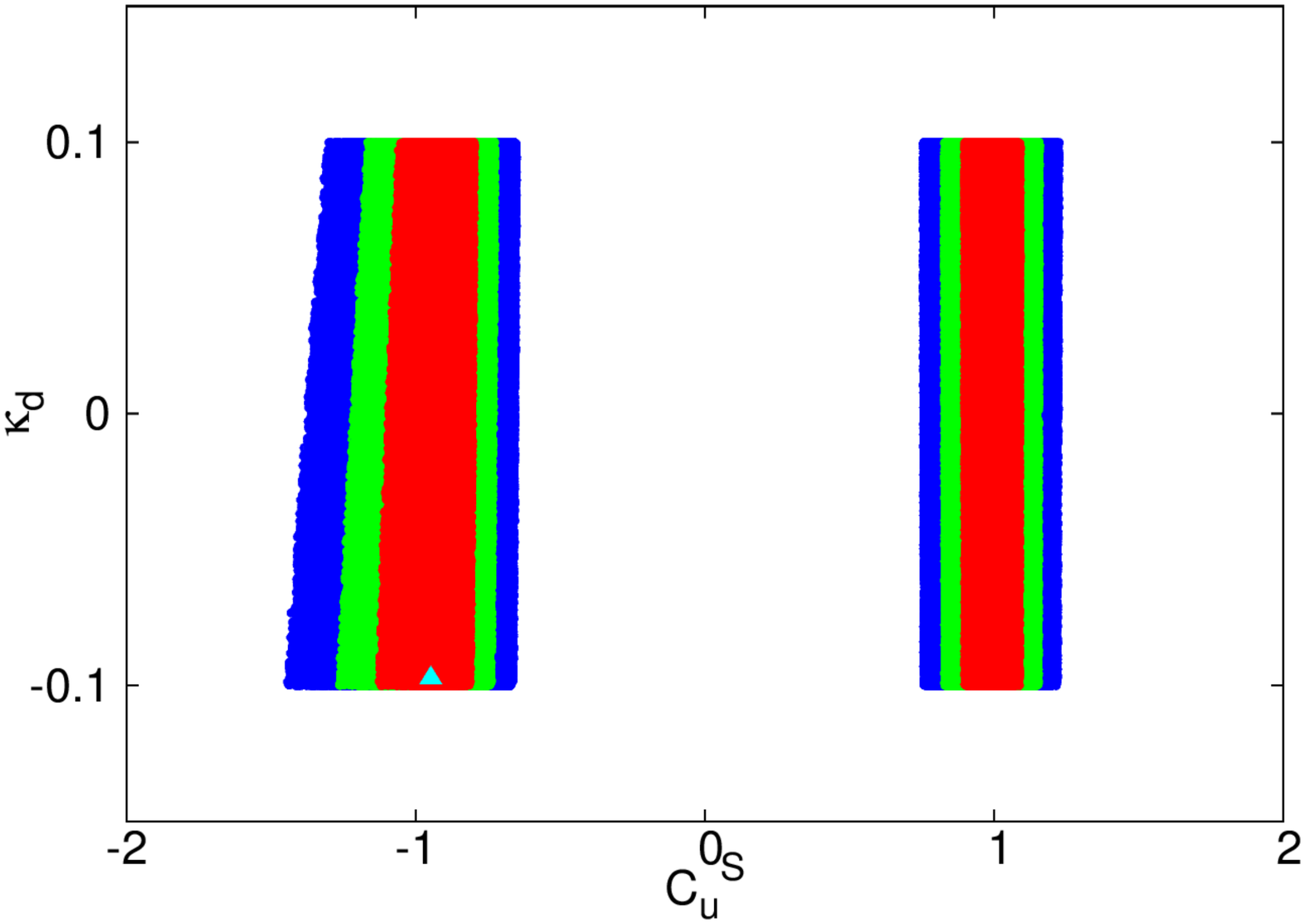}
\includegraphics[width=2.0in]{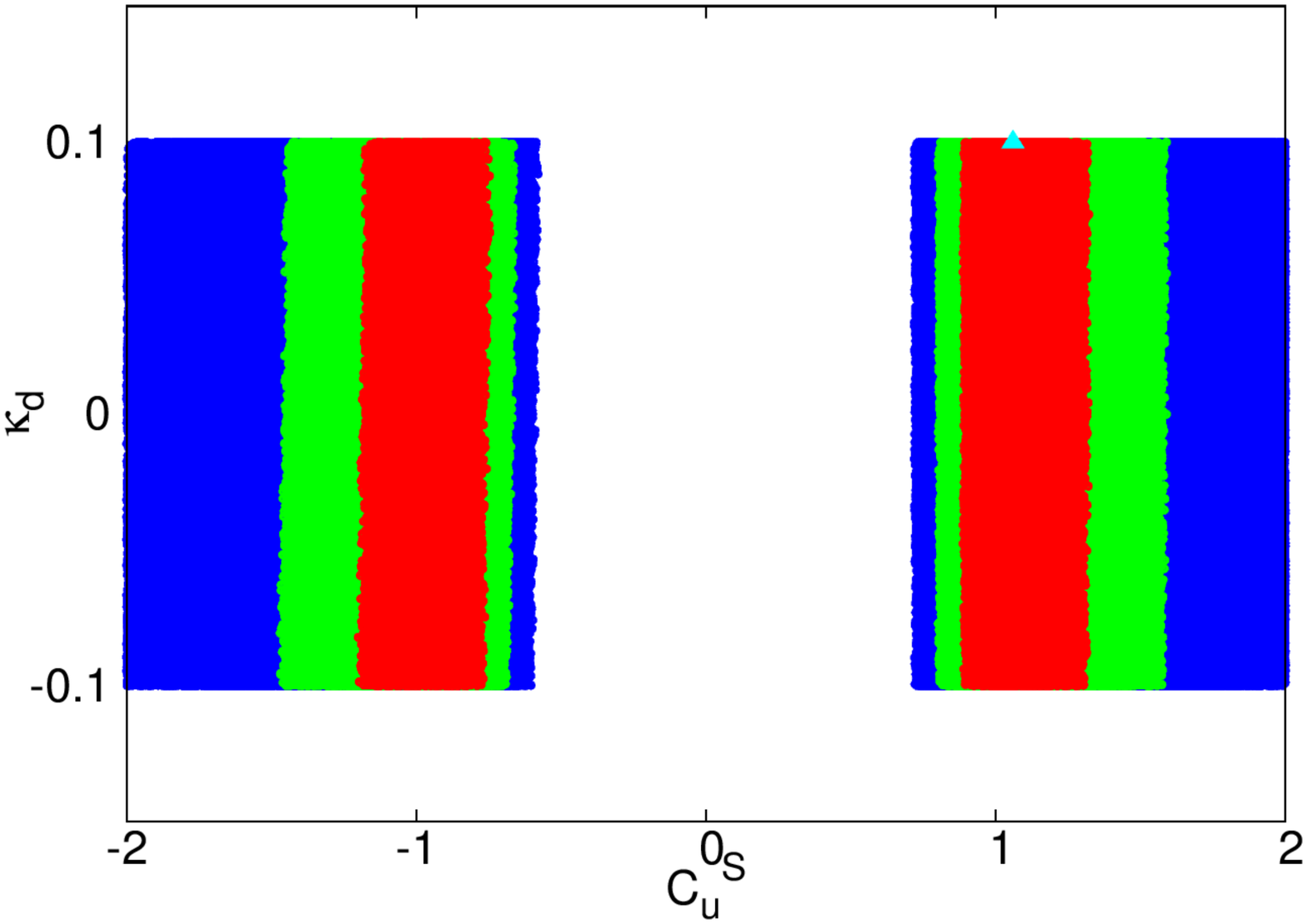}
\includegraphics[width=2.0in]{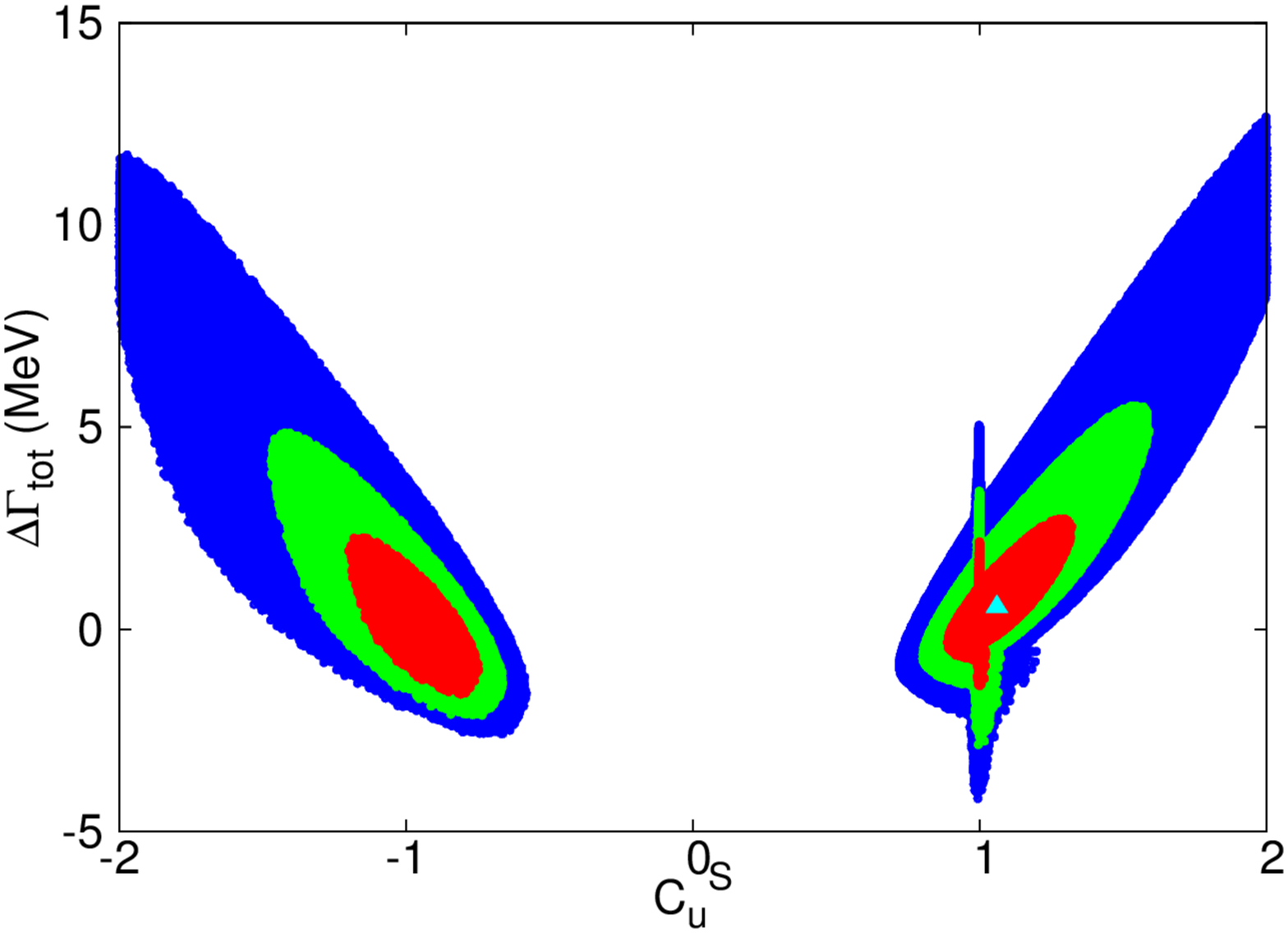}
\includegraphics[width=2.0in]{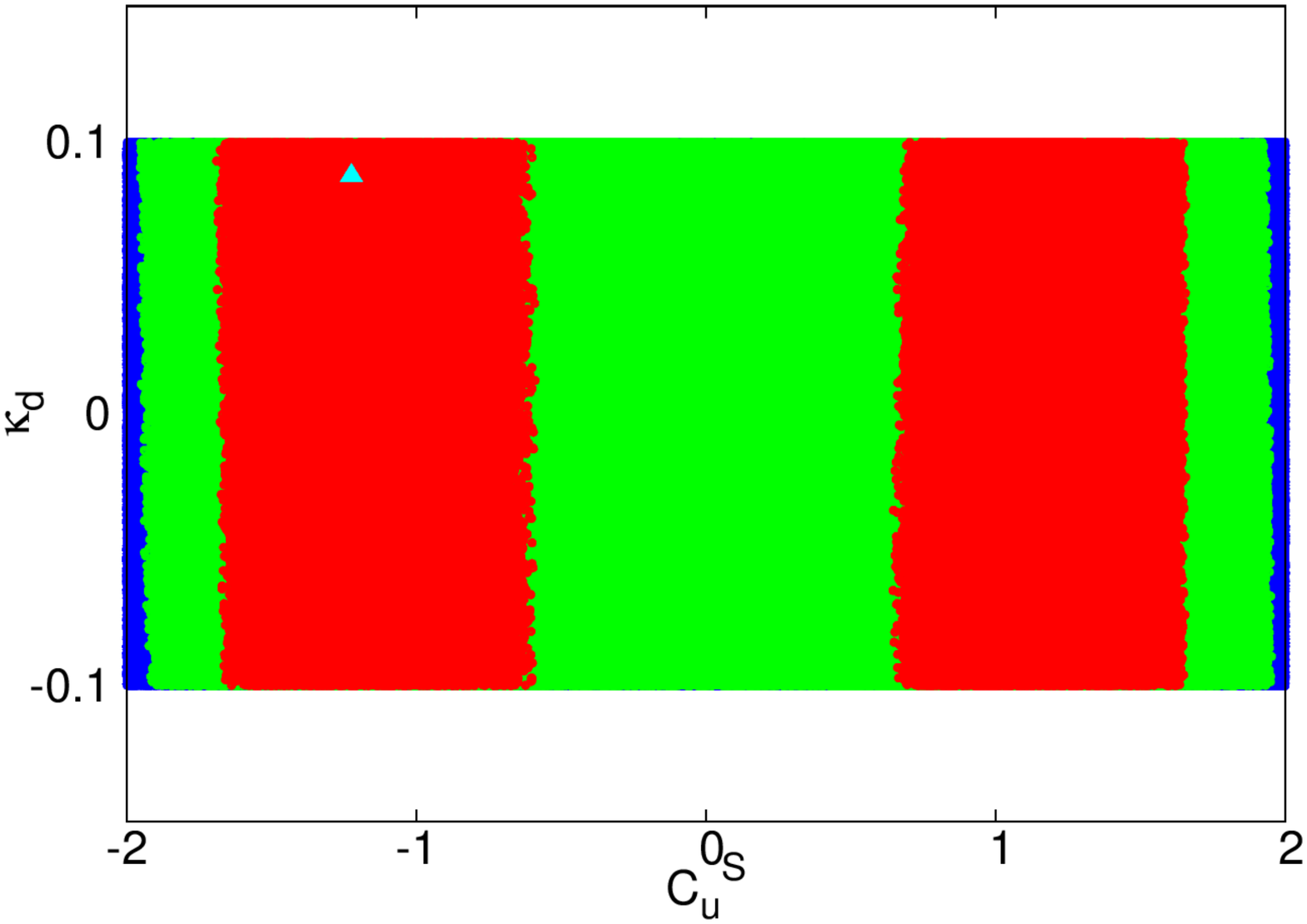}
\includegraphics[width=2.0in]{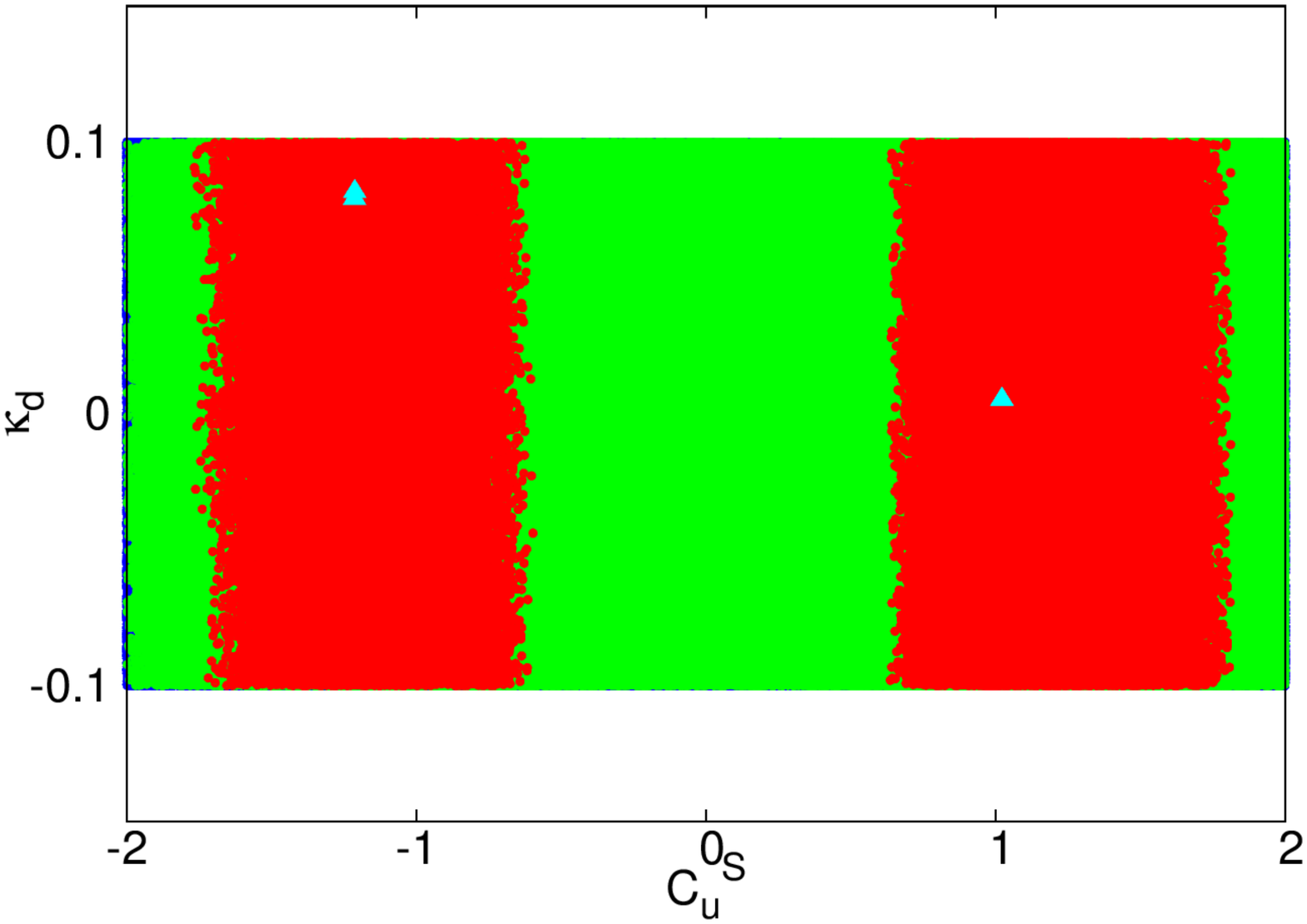}
\includegraphics[width=2.0in]{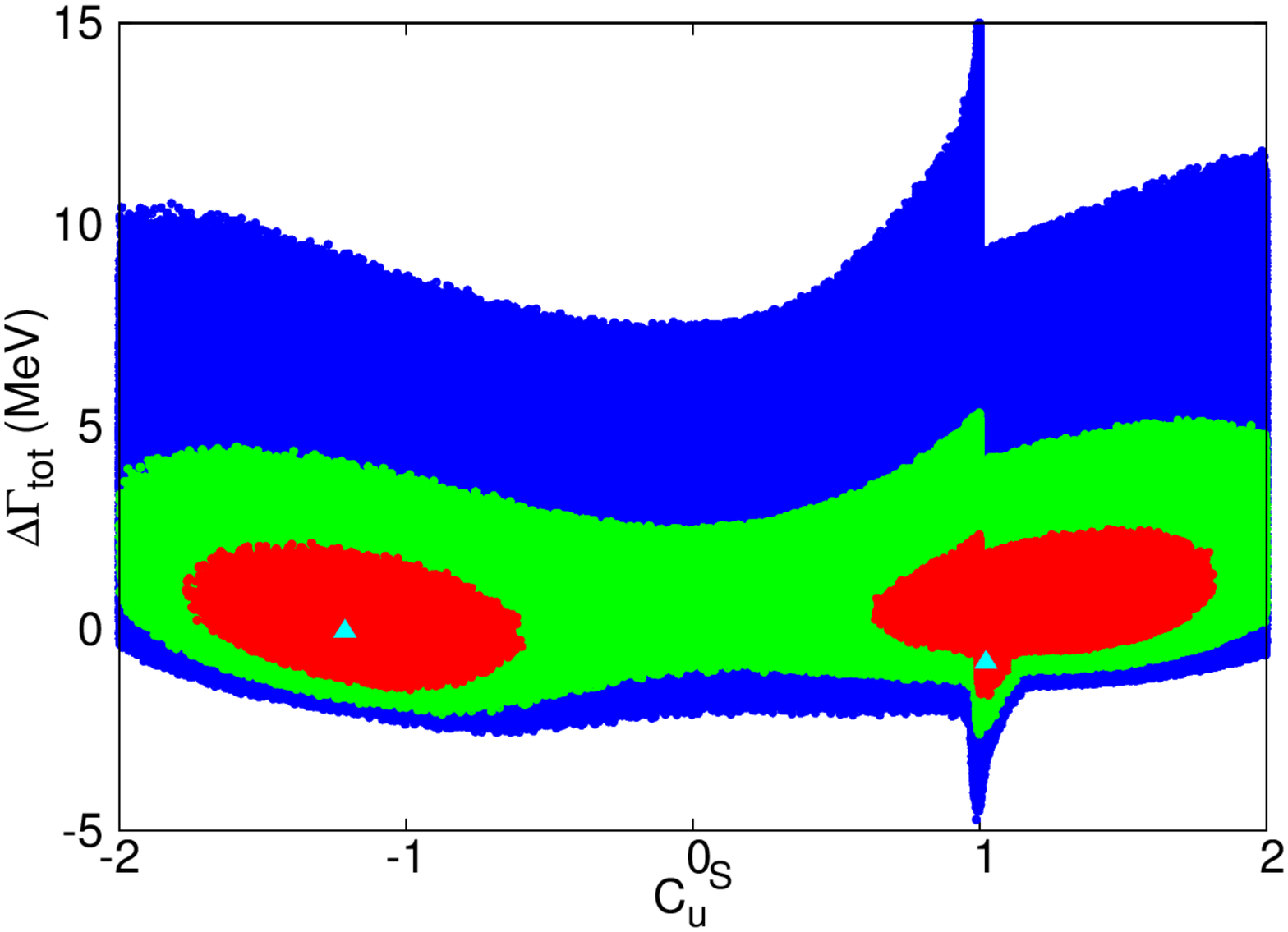}
\caption{\small \label{fig:kddgam}
The confidence-level regions on the $(C_u^S, \kappa_d)$  (left and middle)
and the $(C_u^S, \Delta\Gamma_{\rm tot})$ (right) planes.
The left frames show the cases of
{\bf CPC.II.3, CPC.III.4}, {\bf CPC.IV.5} and 
the middle and right frames are for the cases of
{\bf CPC.II.4, CPC.III.5}, and {\bf CPC.IV.6}.
The labeling of confidence regions is the same as in Fig.~\ref{fig:tanbeta}.
}
\end{figure}

The CL regions involved with $\kappa_d$ are shown in the left and middle 
frames of
Fig.~\ref{fig:kddgam} for the {\bf CPC.II} (upper), {\bf CPC.III} (middle), 
and {\bf CPC.IV} (lower) fits. We see any value of $\kappa_d$ between 
$-0.1$ and $0.1$ is allowed.

Note that in the most recent update
\cite{Cheung:2014noa} when $\Delta \Gamma_{\rm tot}$ is the only
parameter allowed to vary, the fitted value of $\Delta \Gamma_{\rm
  tot}$ is consistent with zero and is constrained by
$\Delta \Gamma_{\rm tot} < 0.97 \;{\rm MeV}$ at 95\% CL.
From the right frames of Fig.~\ref{fig:kddgam}, we observe that 
the range of $\Delta \Gamma_{\rm tot}$ at 95\% CL (green region) varies from 
$-2.4$ MeV to $3.3$ MeV ({\bf CPC.II.4}) and
$-2.9$ MeV to $5.6$ MeV ({\bf CPC.III.5} and {\bf CPC.IV.6}).
Such a large range is not very useful in constraining the exotic decay branching
ratio of the Higgs boson.  Usually we have to limit the number of
varying parameters
to be small enough to draw a useful constraint on $\Delta \Gamma_{\rm tot}$.

\begin{figure}[t!]
\centering
\includegraphics[width=2.0in]{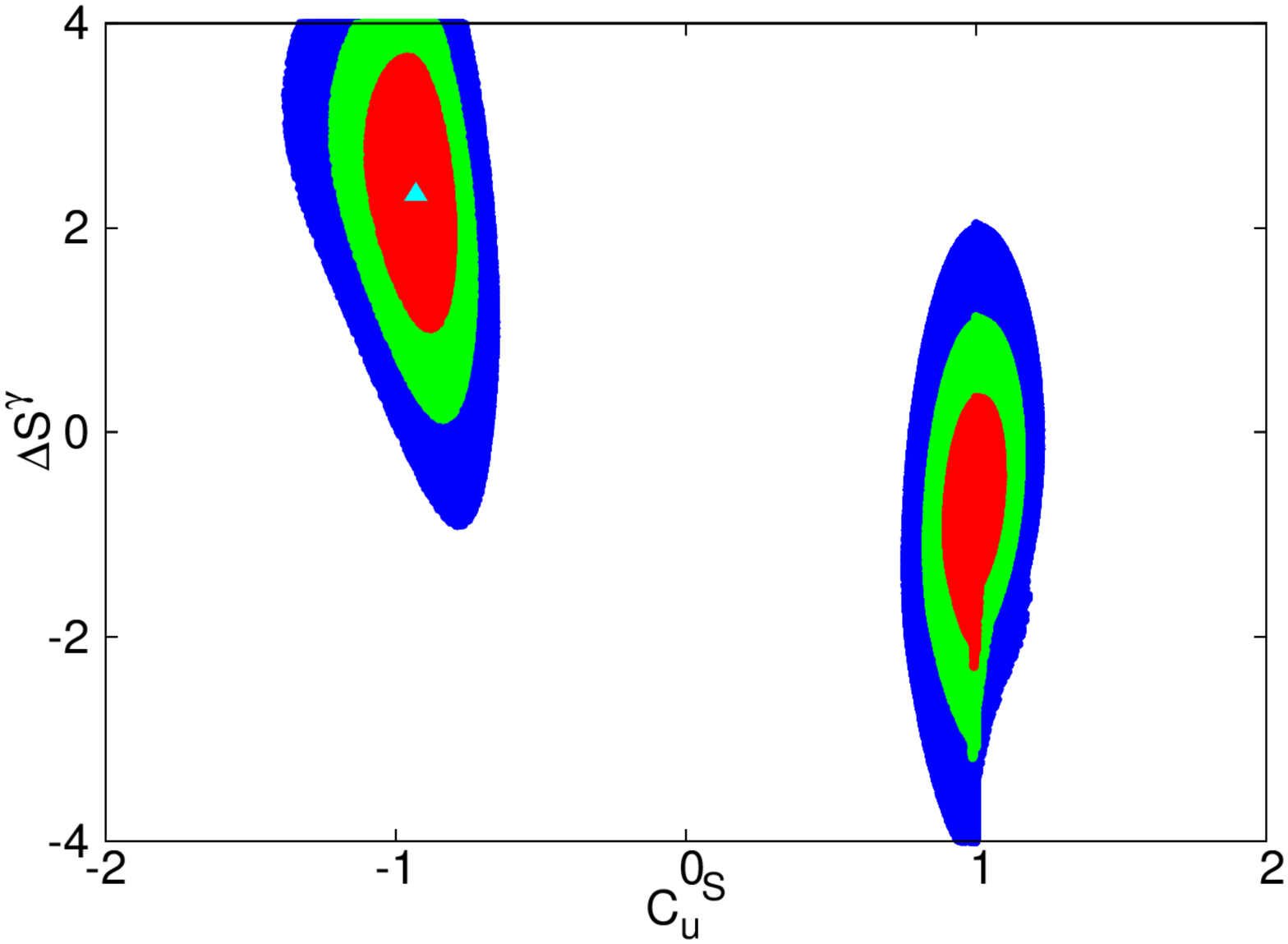}
\includegraphics[width=2.0in]{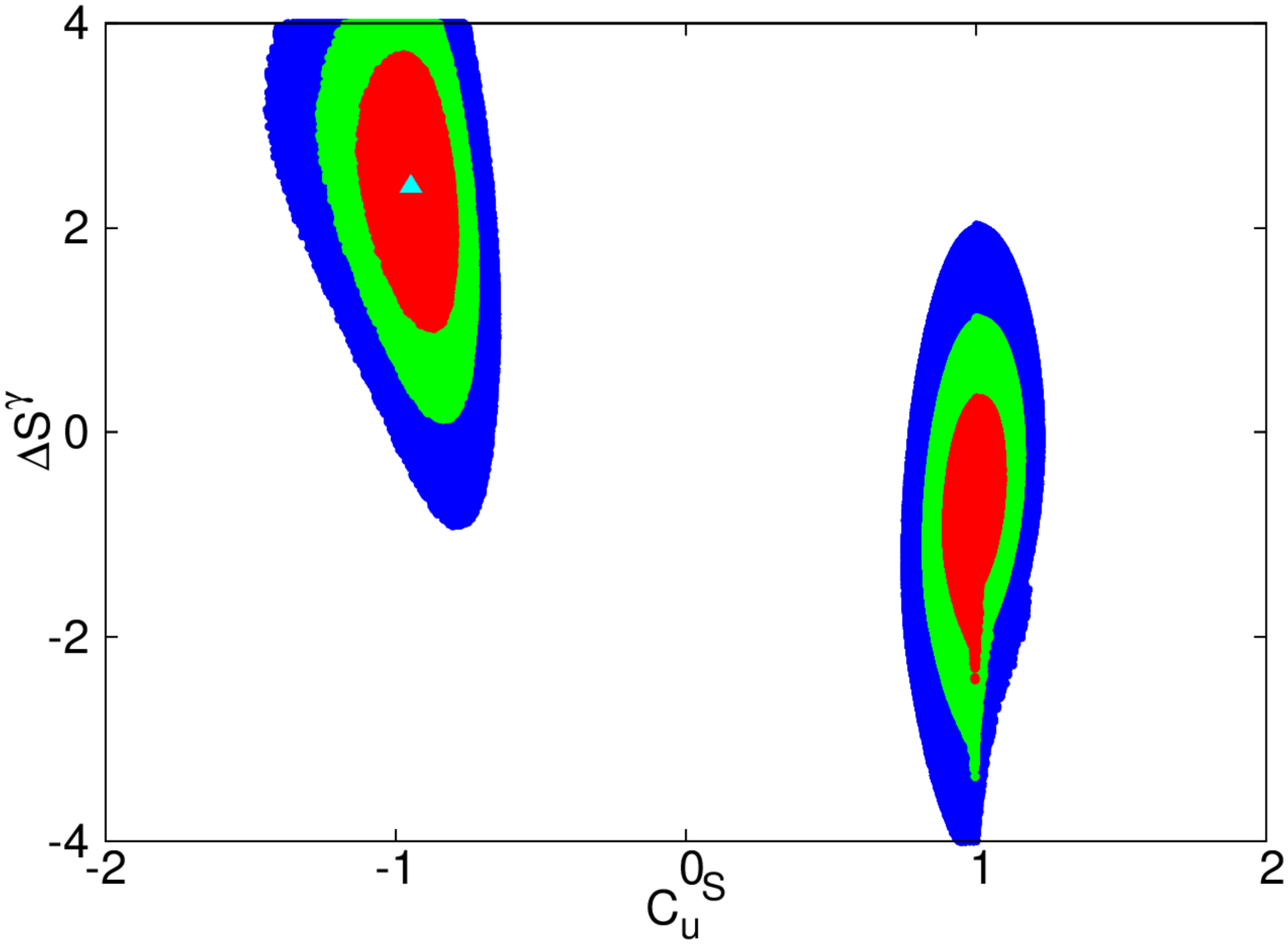}
\includegraphics[width=2.0in]{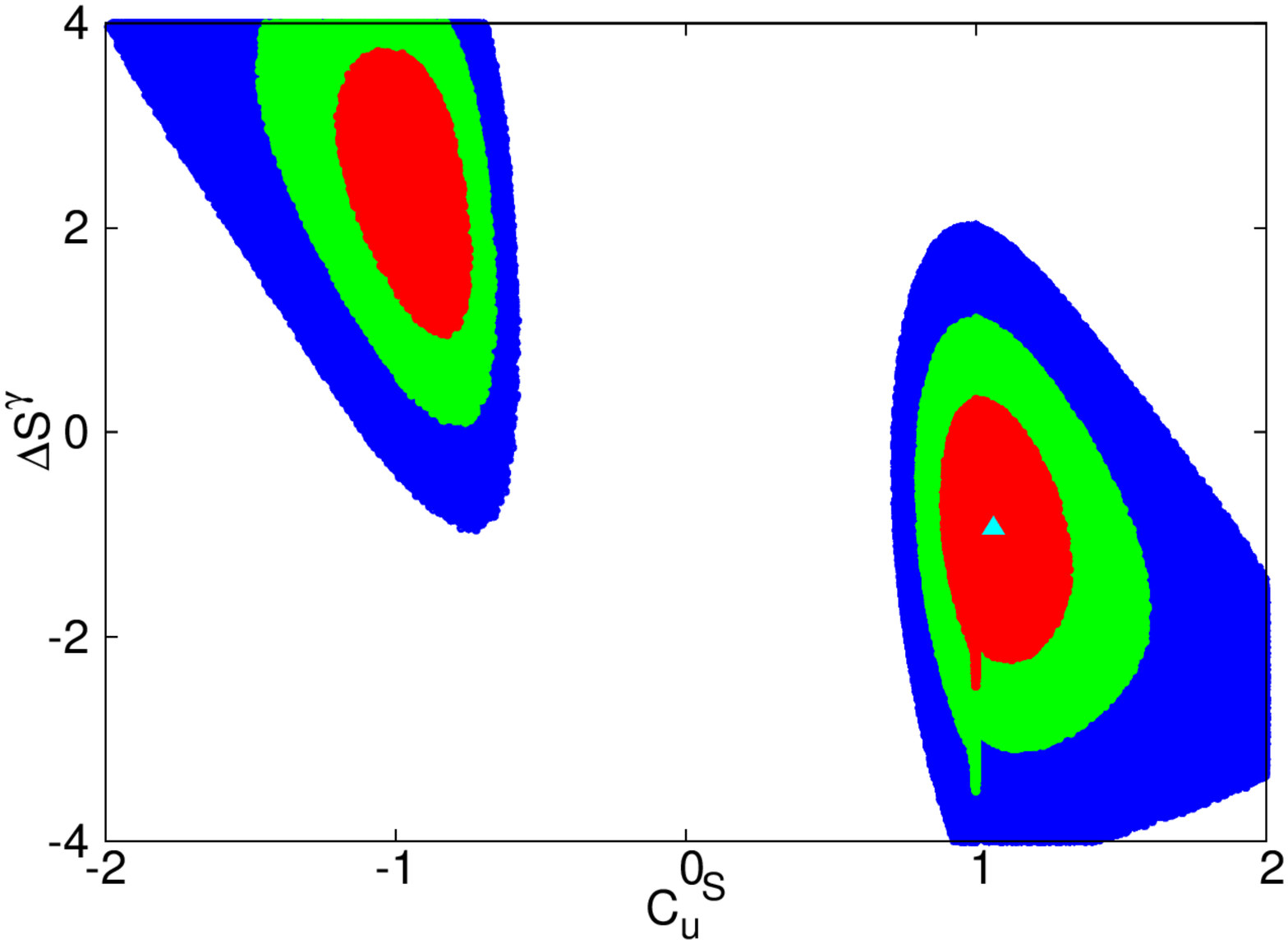}
\includegraphics[width=2.0in]{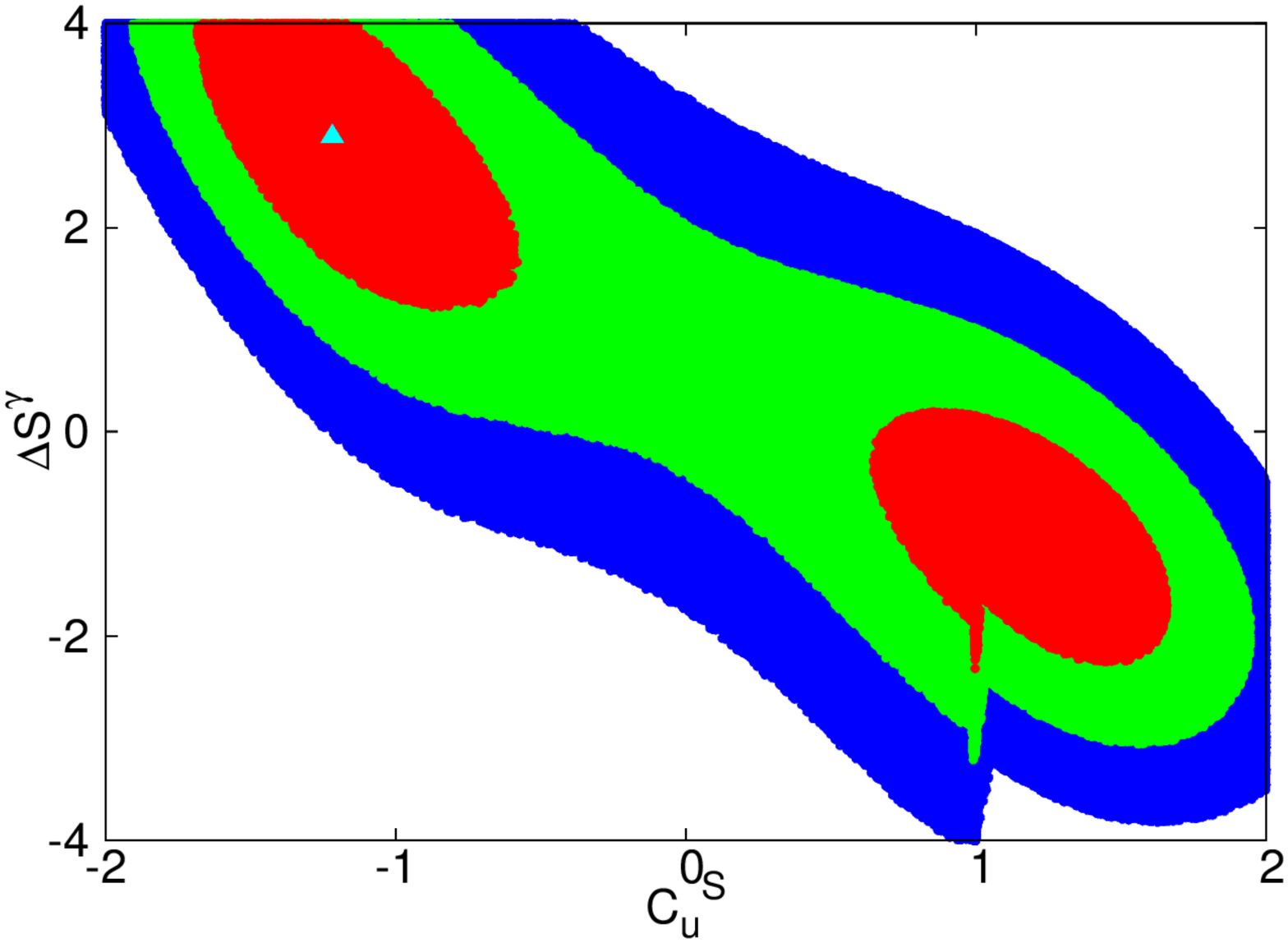}
\includegraphics[width=2.0in]{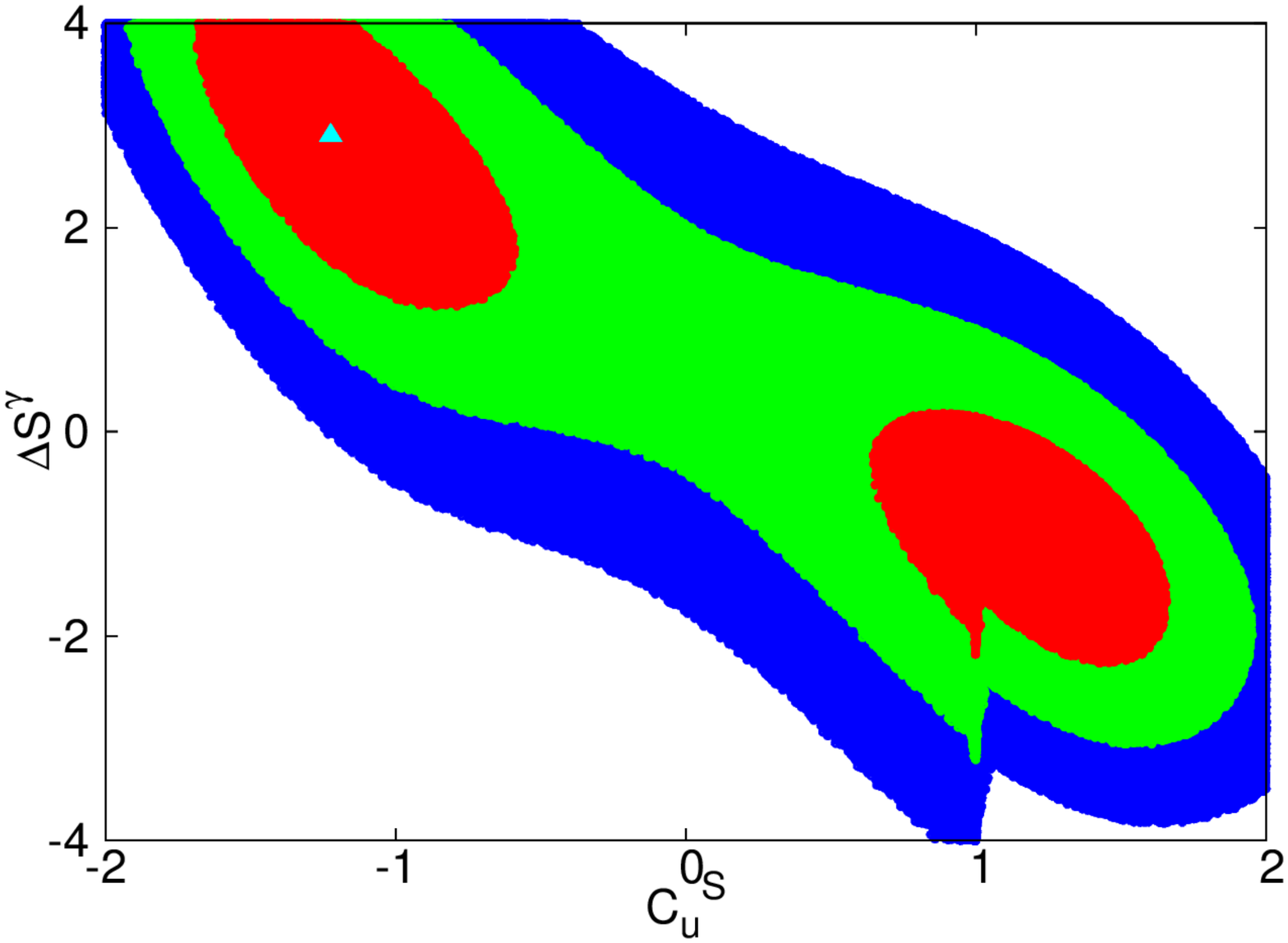}
\includegraphics[width=2.0in]{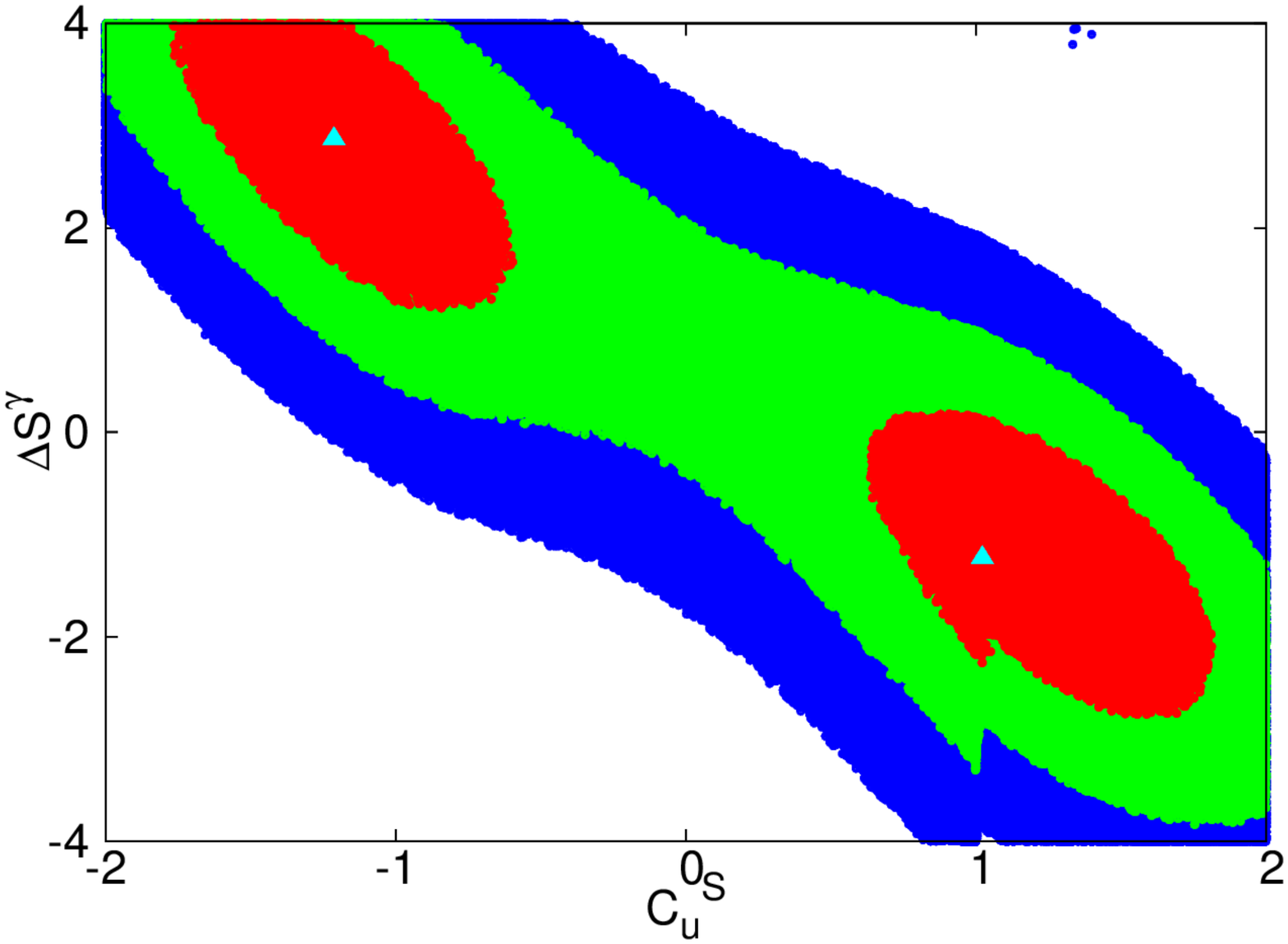}
\caption{\small \label{fig:dcp}
The upper frames show
the confidence-level regions on the $(C_u^S, \Delta S^\gamma)$  plane
for the {\bf CPC.III.3} (left), {\bf CPC.III.4} (middle), 
and {\bf CPC.III.5} (right) fits.
The lower frames are for
the {\bf CPC.IV.4} (left), {\bf CPC.IV.5} (middle), and {\bf CPC.IV.6} (right) fits.
The labeling of confidence regions is the same as in Fig.~\ref{fig:tanbeta}.
}
\end{figure}

We show the CL regions on the $(C_u^S, \Delta S^\gamma)$ plane
in Fig.~\ref{fig:dcp}
for the {\bf CPC.III} (upper) and {\bf CPC.IV} (lower) fits.
In the {\bf CPC.III} fits, the range of $\Delta S^\gamma$ is from
$-2.5\,(1)$ to $0.3\,(3.7)$ at $68.3\%$ CL for the positive (negative) $C_u^S$.
In the {\bf CPC.IV} fits, the range is a bit widened.

\begin{figure}[t!]
\centering
\includegraphics[width=2.0in]{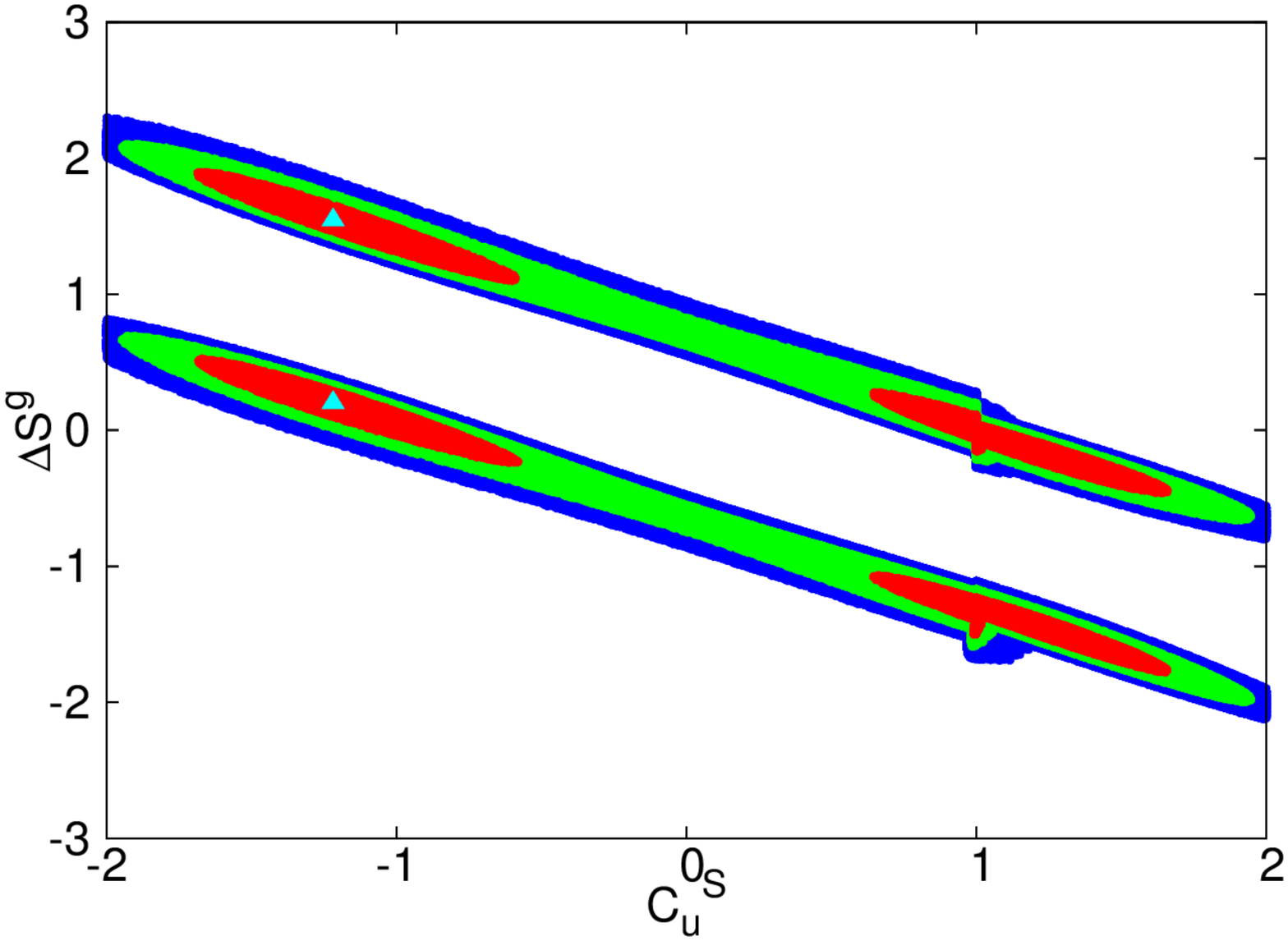}
\includegraphics[width=2.0in]{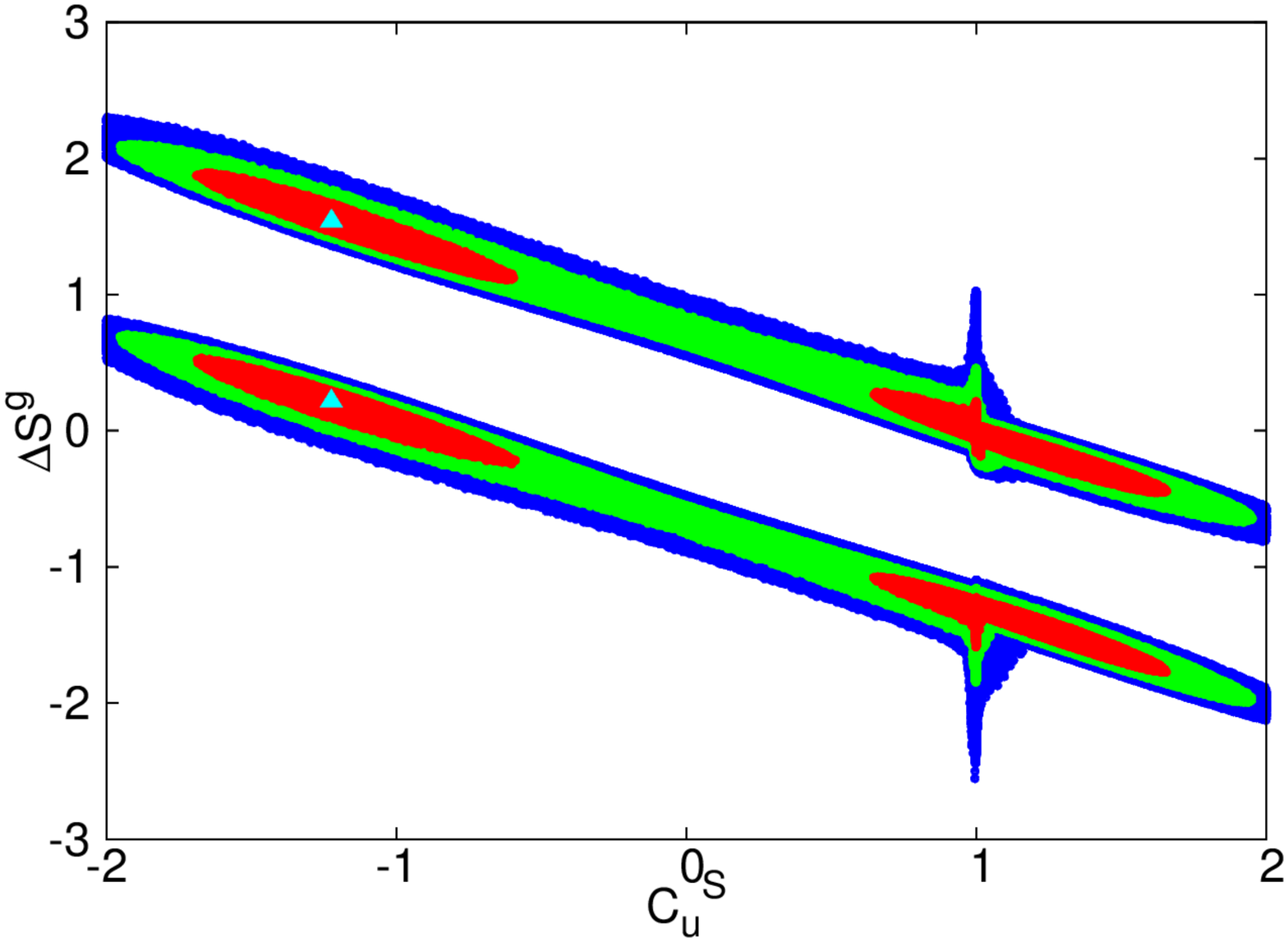}
\includegraphics[width=2.0in]{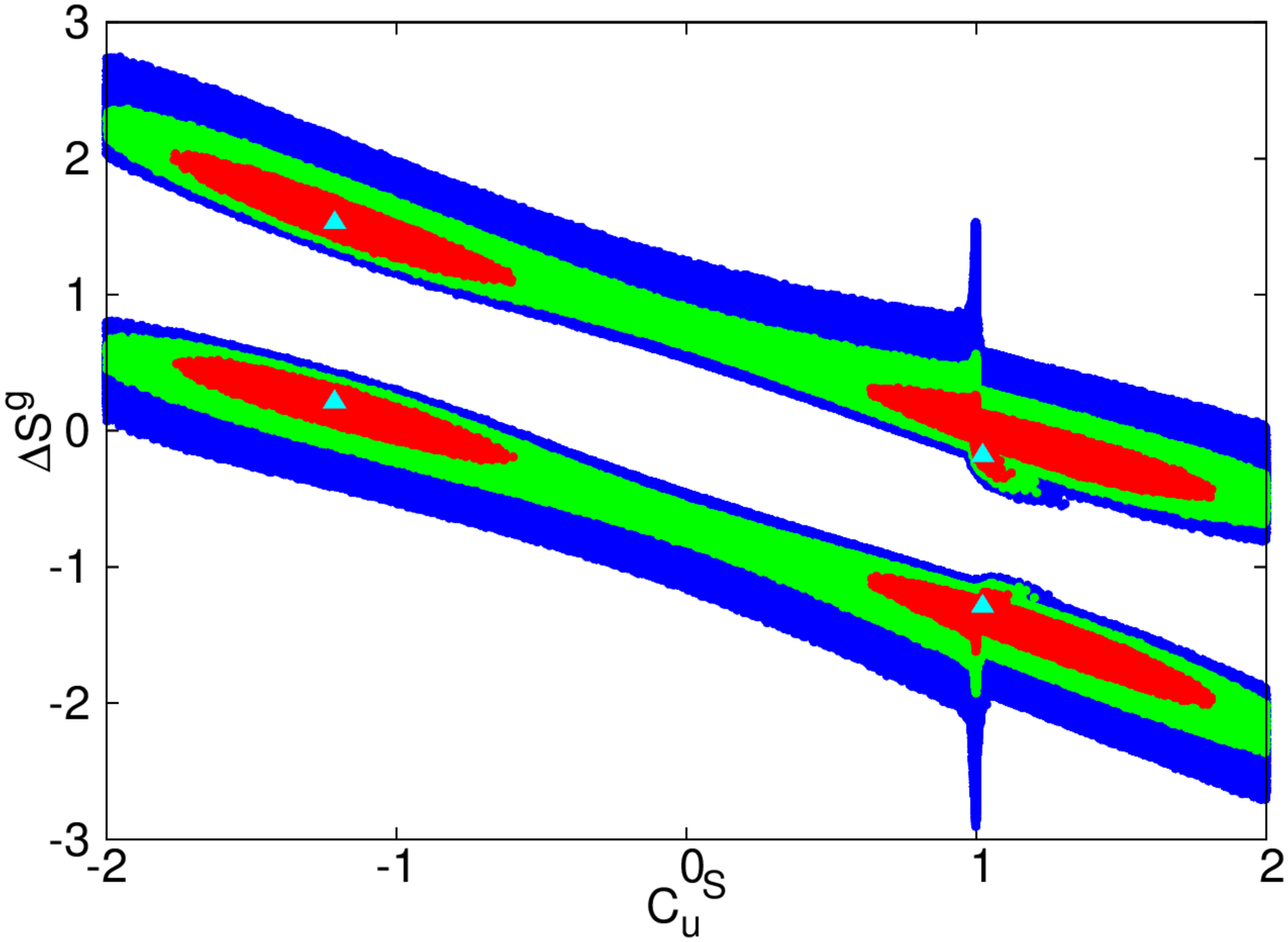}
\includegraphics[width=2.0in]{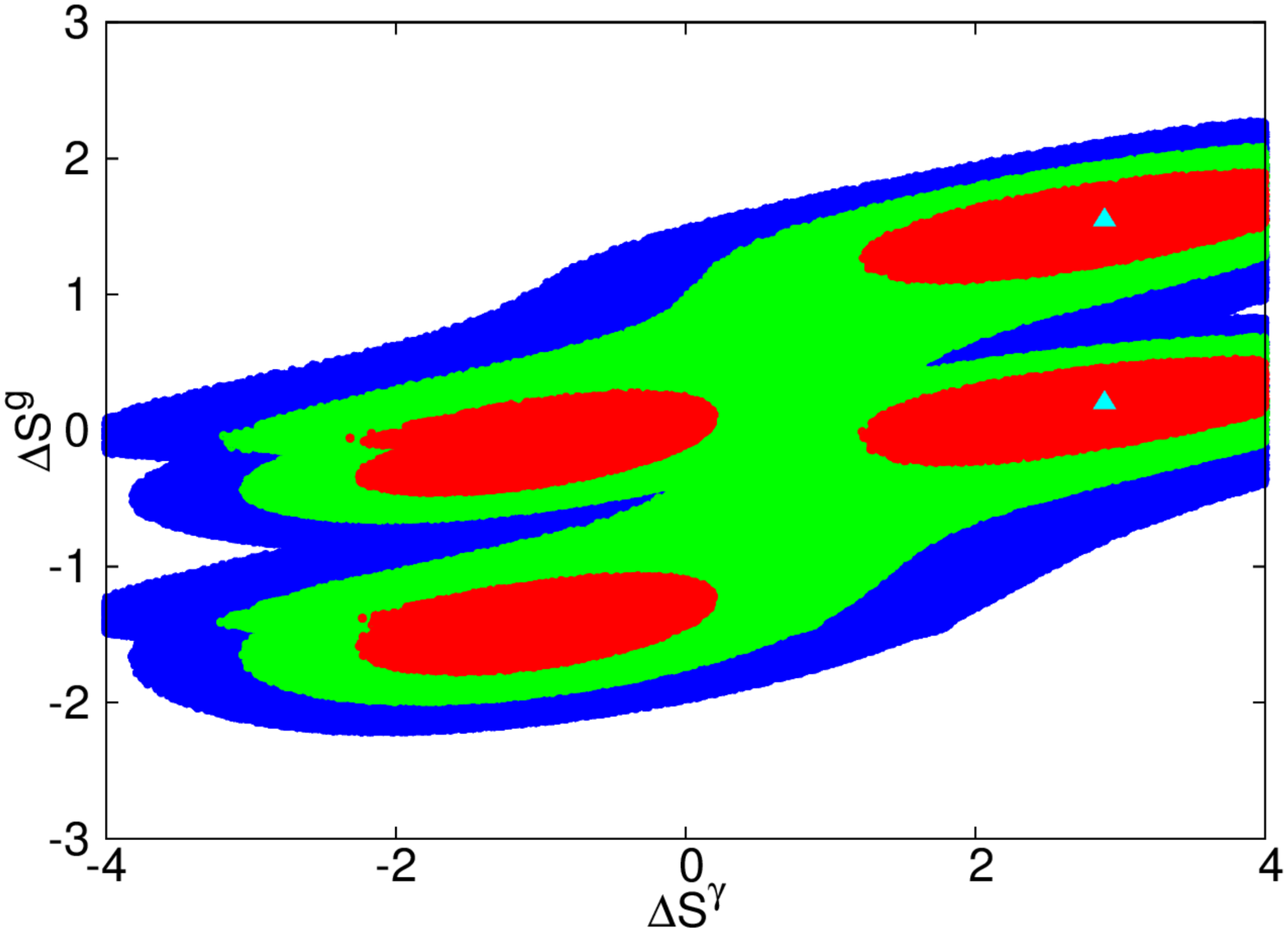}
\includegraphics[width=2.0in]{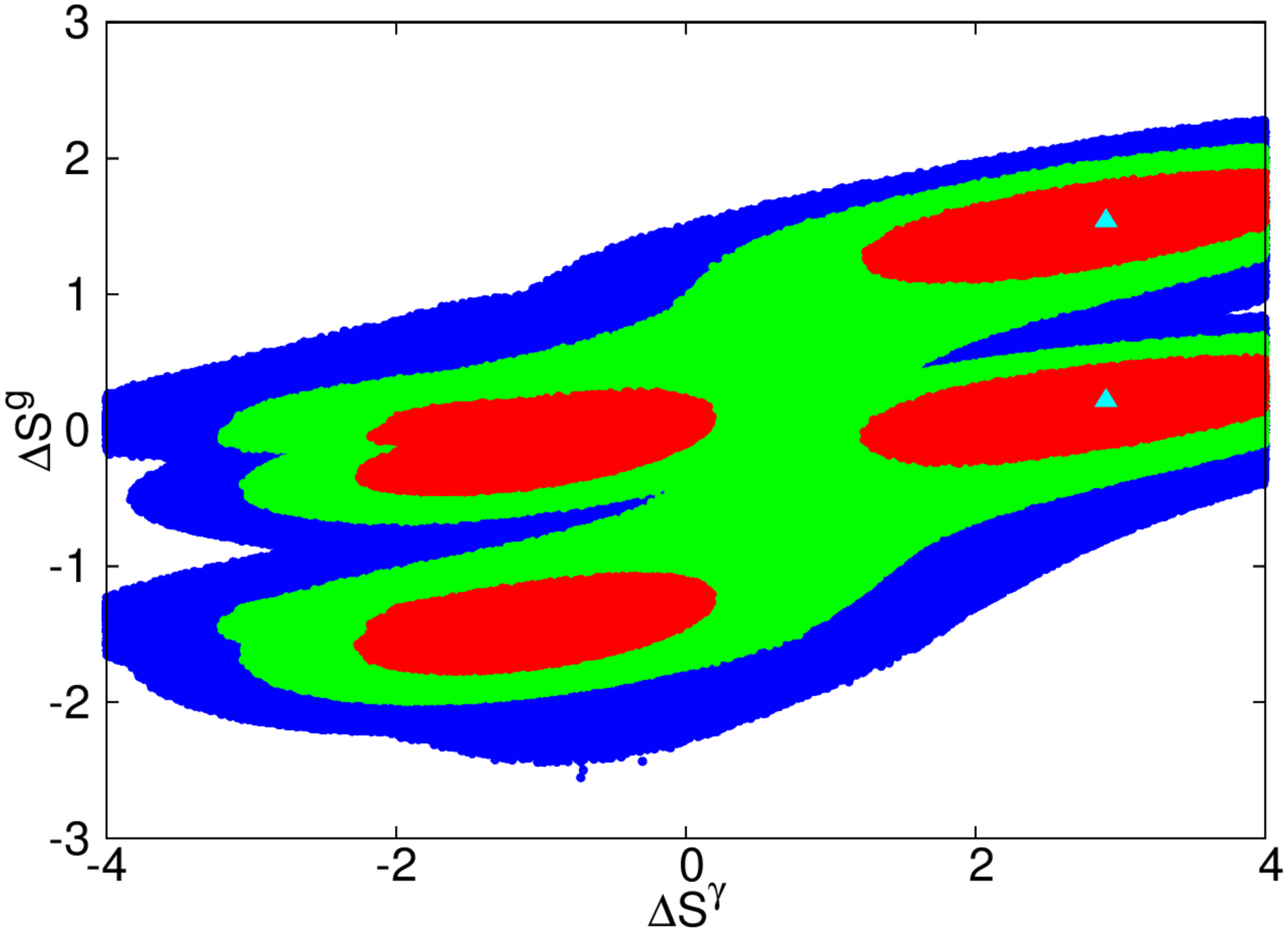}
\includegraphics[width=2.0in]{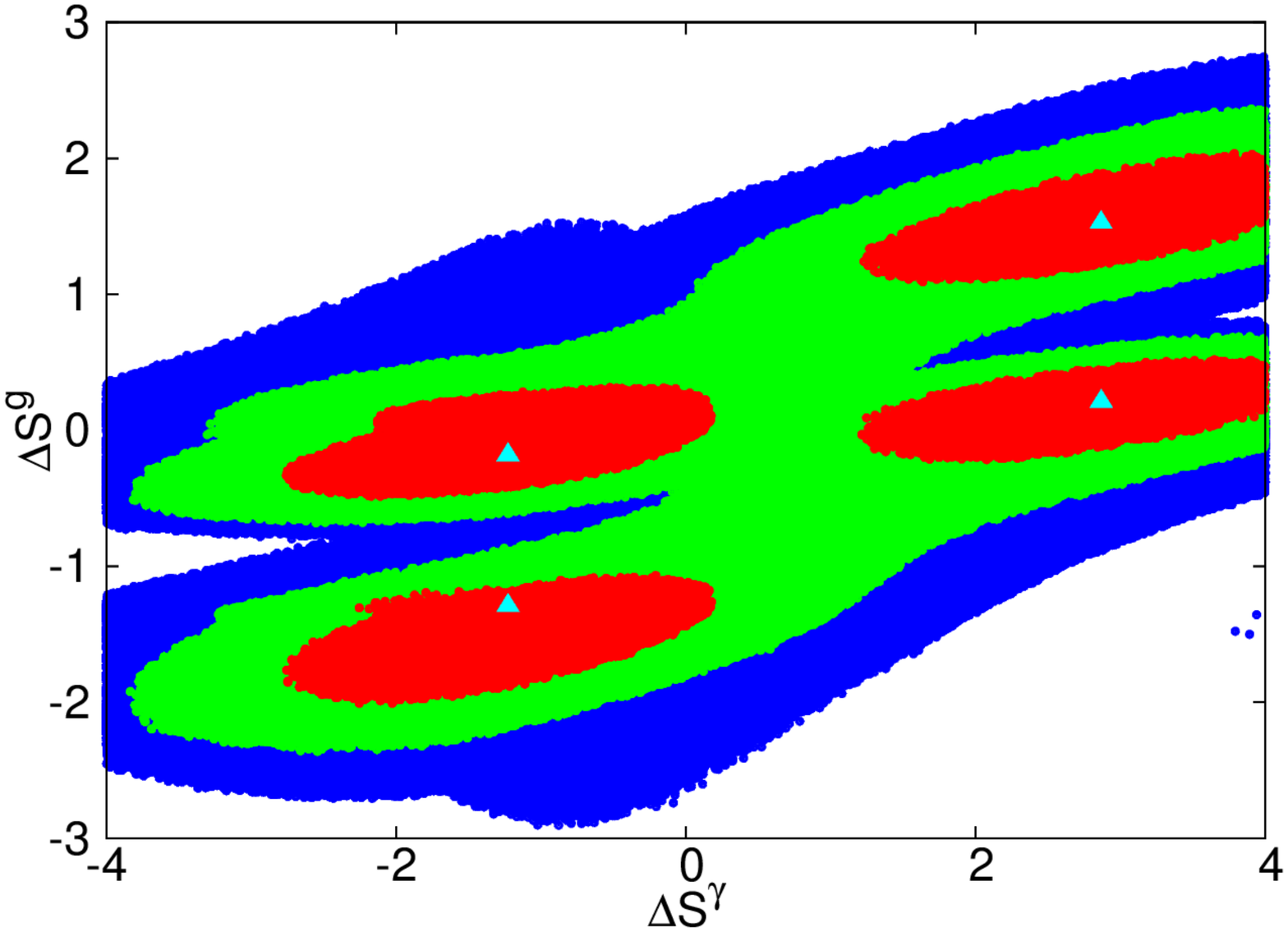}
\caption{\small \label{fig:dcpdcg}
The confidence-level regions on the $(C_u^S, \Delta S^g)$  (upper)
and the $(\Delta S^\gamma, \Delta S^g)$ (lower) planes for
the {\bf CPC.IV.4} (left), {\bf CPC.IV.5} (middle), and {\bf CPC.IV.6} (right) fits.
The labeling of confidence regions is the same as in Fig.~\ref{fig:tanbeta}.
}
\end{figure}

In Fig.~\ref{fig:dcpdcg}, we show the  CL regions of the {\bf CPC.IV} fits
on the $(C_u^S, \Delta S^g)$  (upper) and $(\Delta S^\gamma, \Delta S^g)$ (lower)
planes.
We found that there are two bands of $\Delta S^g$ allowed
by data, which are consistent with the results in the model-independent fits
\cite{Cheung:2013kla}.
In the plots of $\Delta S^\gamma$ vs $\Delta S^g$
there are four almost degenerate solutions to the local minimum of $\chi^2$,
which only differ from one another by a very small amount. It happens because
$\Delta S^\gamma$ and $\Delta S^g$ satisfy a set of elliptical-type equations,
which imply two solutions for each of $\Delta S^\gamma$ and $\Delta S^g$
\cite{Cheung:2013kla}.

A quick summary of the {\bf CPC} fits is in order here. The confidence
regions in various fits are similar to the Type II of the 2HDM. When
$\kappa_d$ and $\Delta \Gamma_{\rm tot}$ (not investigated in the
previous 2HDM fits) are allowed to vary, the confidence regions are
slightly and progressively enlarged due to more varying
parameters. Especially the linear relation between $C_d^S$ and $C_\ell^S$
are ``diffused'' when $\kappa_d$ varies between $\pm 0.1$ as shown in
Eq.~(\ref{eq5}).  The two possible solutions for $\Delta S^\gamma$ in
the  {\bf CPC.III} and {\bf CPC.IV}
cases are consistent with what we have
found in previous works \cite{Cheung:2013kla,Cheung:2013rva}.
The best-fit point of each fit is shown in Table~\ref{tab:best} with
the corresponding $p$-value.  It is clear that the SM fit provides the
best $p$-value in consistence with our previous works
\cite{Cheung:2013kla,Cheung:2013rva,Cheung:2014noa}.  
Among the fits other than the
SM one, the {\bf CPC.III.3} fit 
gives the smallest $\chi^2$ per degree of freedom
and thus the largest $p$-value. It demonstrates that the set of
parameters consisting of the top-Yukawa coupling $C_u^S$, $\tan\beta$
or equivalently the gauge-Higgs coupling $C_v$, and $\Delta S^\gamma$
is the minimal set of parameters that gives the best description of
the data, other than the SM.
In this fit, the $C_v =0.93$ being very close
to the SM value while $C_u^S$ takes on a negative value $-0.93$,
which is then compensated by a relatively large $\Delta S^\gamma = 2.3$.
The derived $C_d^S$ and $C_\ell^S$ are very close to the SM values.
On the other hand, we show in Table~\ref{tab:best3} the other local
minima for various {\bf CPC} fits.  We can see that 
the {\bf CPC.III.3} fit indeed
has another local minimum, which has a $\chi^2$ very close to the
true minimum, at which $C_u^S$, $C_v$, $C_d^S$, and $C_\ell^S$ are
extremely close to their SM values while $\Delta S^\gamma = -0.85$.

\section{Implications on the MSSM spectrum}

In this section, we shall try to find the implications of
the current Higgs signal strength data 
on the masses of charginos, sleptons,  sbottoms, and stops, as
well as the $A$ parameters -- SUSY spectrum -- through the virtual effects.
Supersymmetric particles can enter into the picture of the observed Higgs boson
via (i) exotic decays, e.g., into neutralinos, (ii) contributions to
$\Delta S^\gamma$ by charginos, sleptons, squarks, and (iii) contributions
to $\Delta S^g$ by squarks. Note that virtual effects are also present in
$\kappa_d$.

Being different from the fits considered in the previous section,
we restrict $\tan\beta$ to be larger than $1/2$ 
so that the top-quark Yukawa coupling is supposed to be perturbative and 
the one-loop contributions of the SUSY particles to the $H\gamma\gamma$
and $Hgg$ vertices remain reliable.
Furthermore, as we shall see, the  best-fit values of the couplings
are close to the SM ones and, accordingly,  we take the lightest 
Higgs state ($H_1$) for the observed Higgs boson with 
$M_{H_1}\sim 125.5$ GeV.

A comprehensive survey over the full parameter space of the MSSM is 
a demanding task requiring a large amount of computing time.
Since we are in pursuit of the implications of the current Higgs data on
SUSY spectrum, we consider the following three representative fits
instead of carrying out the comprehensive study:
\begin{itemize}
\item{\bf MSSM-1}: Only with chargino contributions.
\item{\bf MSSM-2}: Only with scalar-tau contributions.
\item{\bf MSSM-3}: With all chargino, scalar-tau, sbottom, and 
stop contributions.
\end{itemize}

In the {\bf MSSM-1} fit, we assume all the scalar
fermions are too heavy to affect the Higgs signal strengths,
and the heavy scalar fermions can easily generate the 
lightest Higgs boson weighing 125.5 GeV through the large 
renormalization group running effects, such as in 
Split SUSY~\cite{Giudice:2011cg}.
In this case, the lightest supersymmetric stable particle (LSP) is 
in general a mixed state of bino, wino, and higgsinos. 

In the {\bf MSSM-2} fit, except for the neutral LSP, we assume only the scalar
taus are light enough to affect the Higgs signal strengths.
Similar to the {\bf MSSM-1} case, the heavy stop and sbottoms 
can easily give $M_{H_1}\sim 125.5$ GeV. In this fit, we are
assuming the charginos are heavy and, therefore, the LSP is bino-like and
its mass is fixed by the bino mass parameter $M_1$.

In the {\bf MSSM-3} fit, we consider all the chargino, scalar-tau, sbottom, 
and stop contributions. Being different from the previous two fits, 
the mass spectrum of the Higgs sector is closely 
correlated with the SUSY contributions to Higgs signal strengths.
To calculate the lightest Higgs mass, we adopt the the approximated
two-loop level analytical expression~\cite{Heinemeyer:1999be,Espinosa:2000df} 
which is precise enough for the purpose of the current study.
For the heavier Higgses, we assume  that they are decoupled or heavier 
than $\sim 300$ GeV. 
To be more specific, we are taking $M_A=300$ GeV and require
$|M_{H_1}-125.5\,{\rm GeV}|\leq 6$ GeV, 
taking account of the $\sim 3$ GeV theoretical error of the lightest Higgs 
mass.

Note that the charginos and sleptons have negligible effects on
the Higgs boson mass and thus we do not impose Higgs boson mass
constraints in the {\bf MSSM-1} and {\bf MSSM-2} fits.  

\subsection{MSSM-1: Charginos only}
\begin{figure}[t!]
\centering
\includegraphics[height=3.0in,angle=-90]{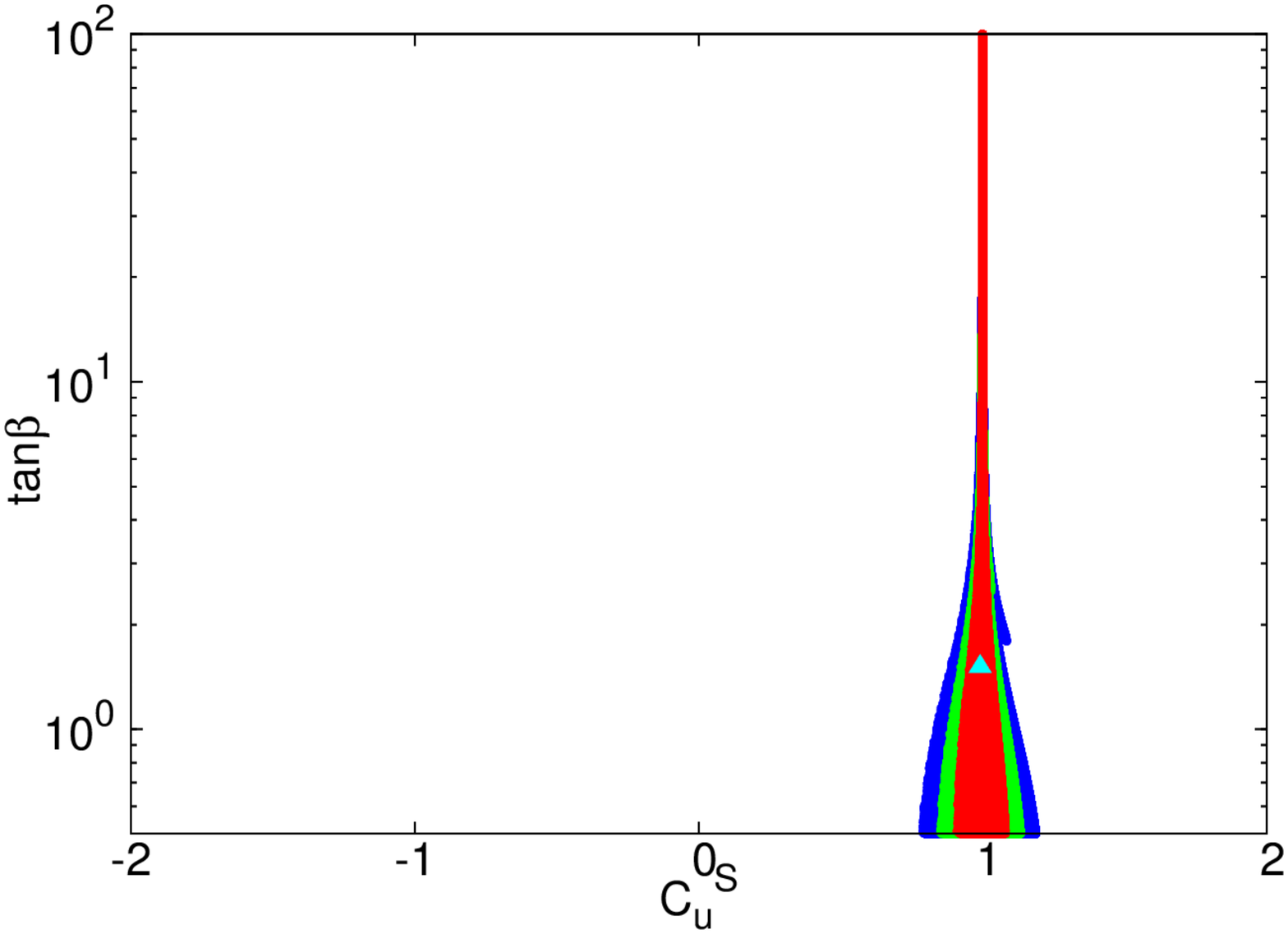}
\includegraphics[height=3.0in,angle=-90]{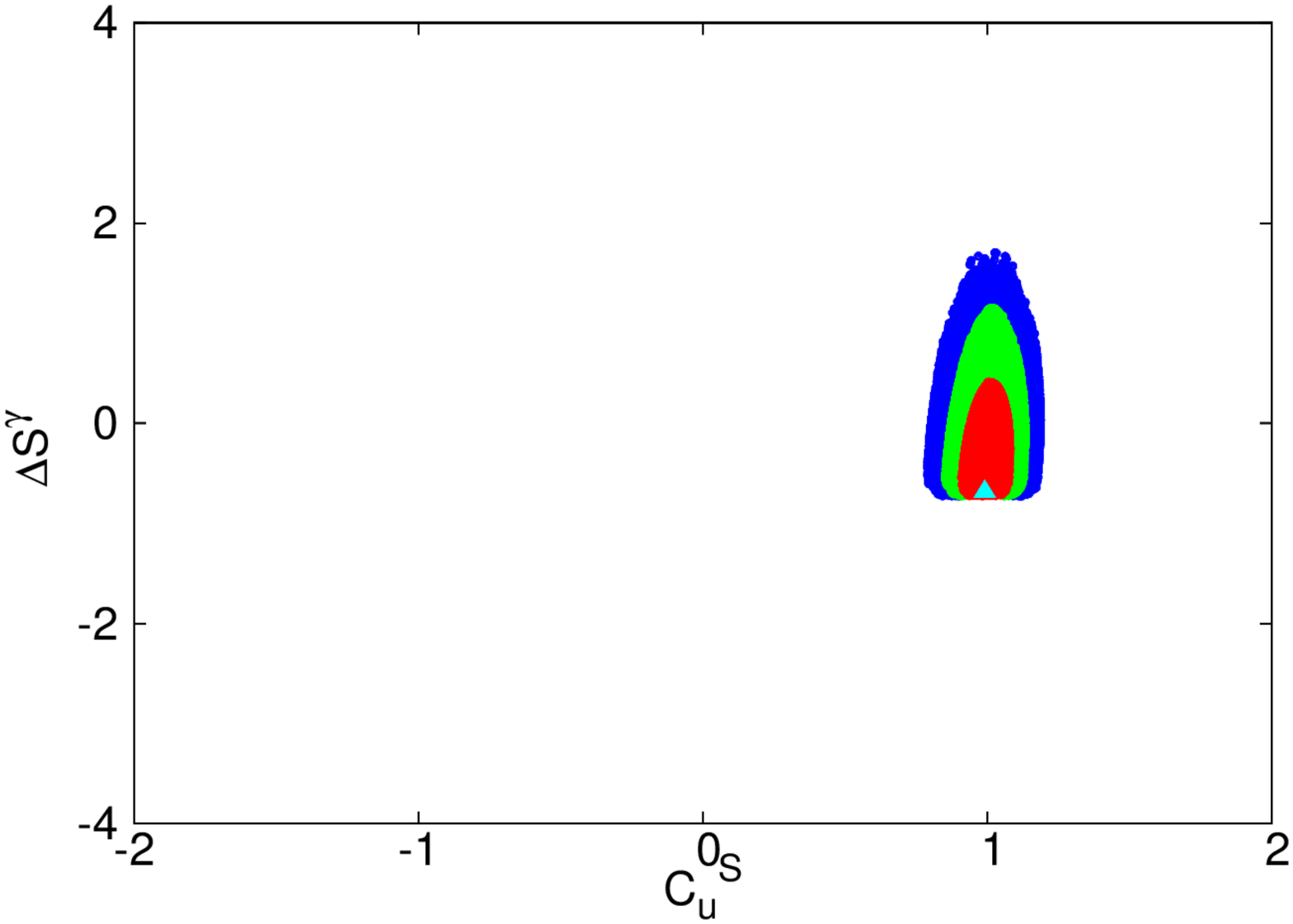}
\includegraphics[height=3.0in,angle=-90]{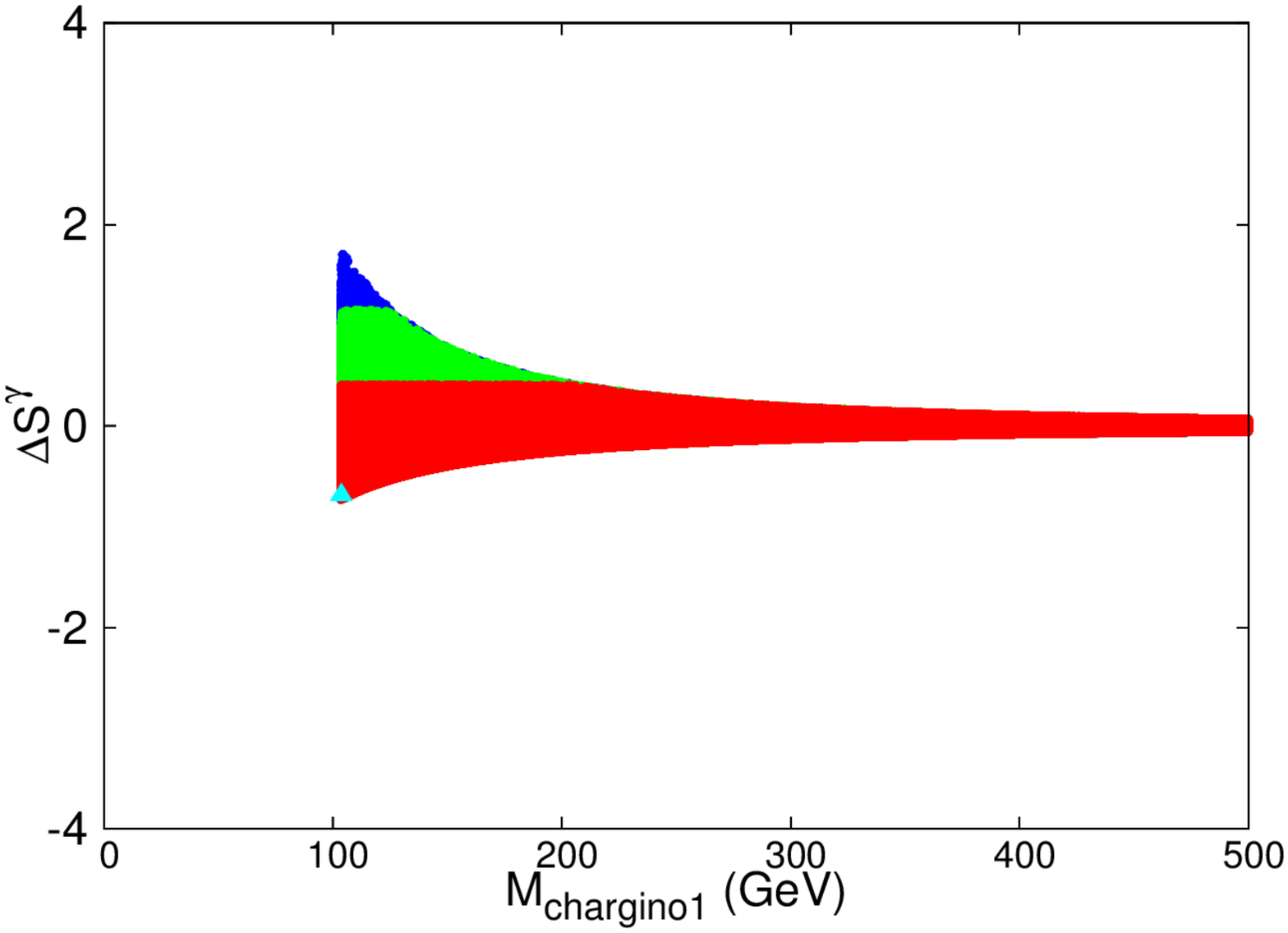}
\includegraphics[height=3.0in,angle=-90]{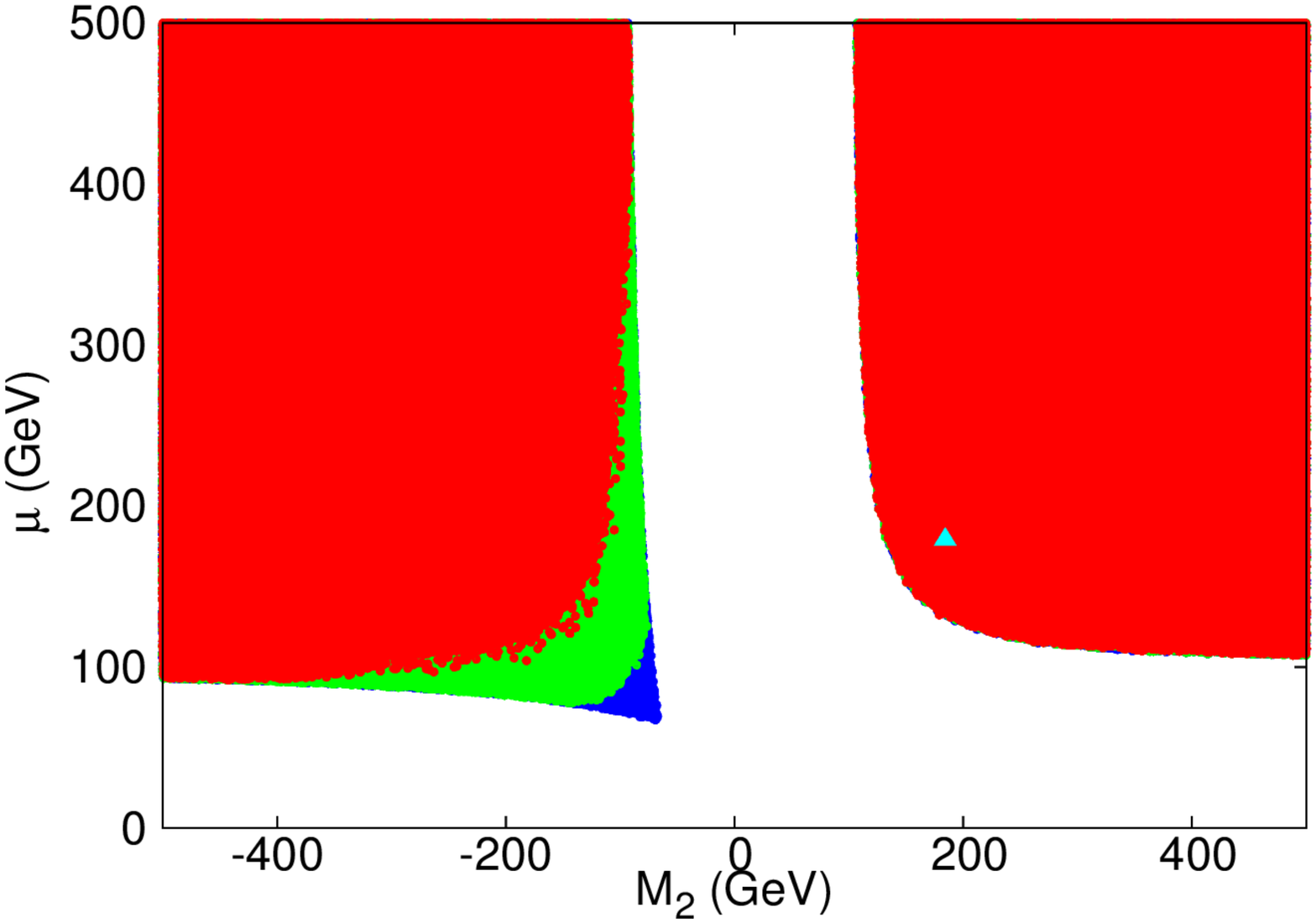}
\caption{\small \label{fig:char}
{\bf MSSM-1} (Charginos):
The confidence-level regions of the fit by varying $C_u^S$, $\tan\beta$, $M_2$,
and $\mu$ with $\tan\beta > 1/2$ and $M_{\tilde{\chi}^\pm_1} > 103.5$
GeV.
The description of the confidence regions is the same as in Fig.~\ref{fig:tanbeta}.
}
\end{figure}

%
%\begin{table}[thb!]
\begin{table}[t!]
  \caption{\small \label{tab:char}
The best-fit values for chargino contributions to $\Delta
S^{\gamma}(\tilde{\chi}^\pm_1,\tilde{\chi}^\pm_2)$. We imposed
$M_{\tilde{\chi}^\pm_1}>103.5$ GeV and $\tan\beta > 1/2$.
The parameters: $C_u^S$, $\tan\beta$, $M_2\subset [-1$TeV$,1$TeV$]$,
$\mu\subset [0,1$TeV$]$ are scanned.
  }
\begin{ruledtabular}
\begin{tabular}{ l|ccc|rcrrrr}
Fits  & $\chi^2$ & $\chi^2$/dof & $p$-value &
\multicolumn{6}{c}{Best-fit values} \\
& & & & $C_u^S$ & $\tan\beta$ & $\kappa_d$ &
$\Delta S^\gamma$ & $\Delta S^g$ & $\Delta{\Gamma}_{\rm tot}$ \\
\hline\hline
%CPC3-2 chargino
{\bf Charginos}
& $15.78$ & $0.631$ & $0.921$ & $0.992$ &
$1.513$ & $-$ & $-0.683$ & $-$ & $-$
\end{tabular}
\begin{tabular}{ lcccrrrrrrr}
 & & &  & \multicolumn{7}{c}{Best-fit values} \\
& & & & $C_v$ & $C_{d}^{S}$ & $C_{\ell}^{S}$  & $M_2$(GeV) & $\mu$(GeV) &
     $M_{\tilde{\chi}^\pm_1}$(GeV) & $M_{\tilde{\chi}^\pm_2}$(GeV)    \\
\hline\hline
& & & & $1.000$ & $1.019$ & $1.019$ & $184$ & $179$ & $103.7$ & $261.3$  \\
\end{tabular}
\end{ruledtabular}
\end{table}
We first investigate the effects of charginos. The lower mass limit of
chargino is
103.5 GeV, so that the only place that it can affect the Higgs boson is in the
loop factor $\Delta S^\gamma$. The MSSM parameters that affect the chargino mass
and the interactions with the Higgs boson are: $M_2$, $\mu$, and $\tan\beta$,
shown in Eqs.~(\ref{eq13}) and (\ref{eq14}).
We show in Fig.~\ref{fig:char} the confidence regions when we vary
$C_u^S$, $\tan\beta$, $M_2$, and $\mu$ with the additional constraint on the
chargino mass:
\[
  M_{\tilde{\chi}^\pm} > 103.5 \;{\rm GeV} \;.
\]
The results are analogous to those of the {\bf CPC.III.3} case if we do not
impose the chargino mass constraint and 
the restriction of $\tan\beta>1/2$.  
In the {\bf CPC.III.3} fit, 
$\Delta S^\gamma$ is free to vary both negatively and positively,
while here 
the sign of the chargino contribution correlates with
$C_u^S$ in the parameter space of $M_2$ and $\mu$.
From the upper frames, we note that
$C_u^S$ is always positive under the requirement of $\tan\beta>1/2$ and
$\Delta S^\gamma$ tends to be positive taking its value in the range between
$-0.75$ and $1.7$ at $99.7\%$ CL. 
In the lower-left frame, we show the $M_{\tilde\chi_1^\pm}$ dependence of the
CL regions of $\Delta S^\gamma$.  We observe that all the points fall into the
$68.3\%$ CL region of $-0.25\lsim\Delta S^\gamma\lsim 0.43$
when $M_{\tilde\chi_1^\pm}\gsim 200$ GeV. We also observe that 
the $\mu$  parameter can be as low
as $70$ GeV when $M_2<0$ from the lower-right frame.

We show the best-fit point for the chargino contribution in
Table~\ref{tab:char}.
The best-fit point gives
$M_2 =184$ GeV and $\mu = 179$ GeV, which give the lightest
chargino mass $M_{\tilde{\chi}^\pm_1} = 103.7$ GeV, just above the current limit.
The corresponding $\Delta S^\gamma \approx -0.68$.
The $p$-value is slightly worse than
the {\bf CPC.III.3} case.

\subsection{MSSM-2: Scalar taus}
%
%\begin{table}[thb!]
\begin{table}[t!]
  \caption{\small \label{tab:stau}
The best-fit values for stau contributions to $\Delta
S^{\gamma}(\tilde{\tau}_1,\tilde{\tau}_2)$. We set $M_{E_3}=M_{L_3}$ and imposed
$\tan\beta > 1/2$, $\mu > 1$ TeV, and $M_{\tilde{\tau}_1}>81.9$ GeV.
The scanning parameters are
 $C_u^S$, $\tan\beta$, $M_{L_3}\subset [0,1$TeV$]$, $\mu\subset
[1,2$TeV$]$, $A_{\tau}\subset [-1$TeV$,1$TeV$]$.
  }
\begin{ruledtabular}
\begin{tabular}{ l|ccc|rcrrrr}
Fits  & $\chi^2$ & $\chi^2$/dof & $p$-value &
\multicolumn{6}{c}{Best-fit values} \\
& & & & $C_u^S$ & $\tan\beta$ & $\kappa_d$ &
$\Delta S^\gamma$ & $\Delta S^g$ & $\Delta{\Gamma}_{tot}$ \\
\hline\hline
%CPC3-2 stau
{\bf Scalar taus}
& $15.68$ & $0.653$ & $0.899$ & $1.000$ &
$47.14$ & $-$ & $-0.854$ & $-$ & $-$
\end{tabular}
\begin{tabular}{ lcccrrrrrrrr}
&&& & \multicolumn{8}{c}{Best-fit values} \\
&&&& $C_v$ & $C_{d}^{S}$ & $C_{\ell}^{S}$
& $M_{L_3}$(GeV) & $\mu$(GeV) & $A_{\tau}$ (GeV) & $M_{\tilde{\tau}_1}$(GeV) &
$M_{\tilde{\tau}_2}$(GeV)    \\
\hline\hline
&&&& $1.000$ & $1.040$ & $1.040$ & $323$ & $1075$ & $-43.2$ & $132.3$ & $442.4$
\\
\end{tabular}
\end{ruledtabular}
\end{table}

The staus contribute to $\Delta S^\gamma$ in a way similar to charginos.
The SUSY soft parameters that affect the stau contributions are
the left- and right-handed slepton masses $M_{L_3}$ and $M_{E_3}$, the
$A$ parameter $A_\tau$, and the $\mu$ parameter. 
We are taking $\mu>1$ TeV to avoid possibly large chargino
contributions to $\Delta S^\gamma$.
The $2\times 2$ stau
mass matrix is diagonalized to give two mass eigenstates $\tilde{\tau}_1$
and $\tilde{\tau}_2$, shown in ~(\ref{eq16}) and (\ref{eq18}).
The current mass limit on stau is
$M_{\tilde{\tau}_1} > 81.9\; {\rm GeV}$ \cite{pdg}.

\begin{figure}[t!]
\centering
\includegraphics[height=3.0in,angle=-90]{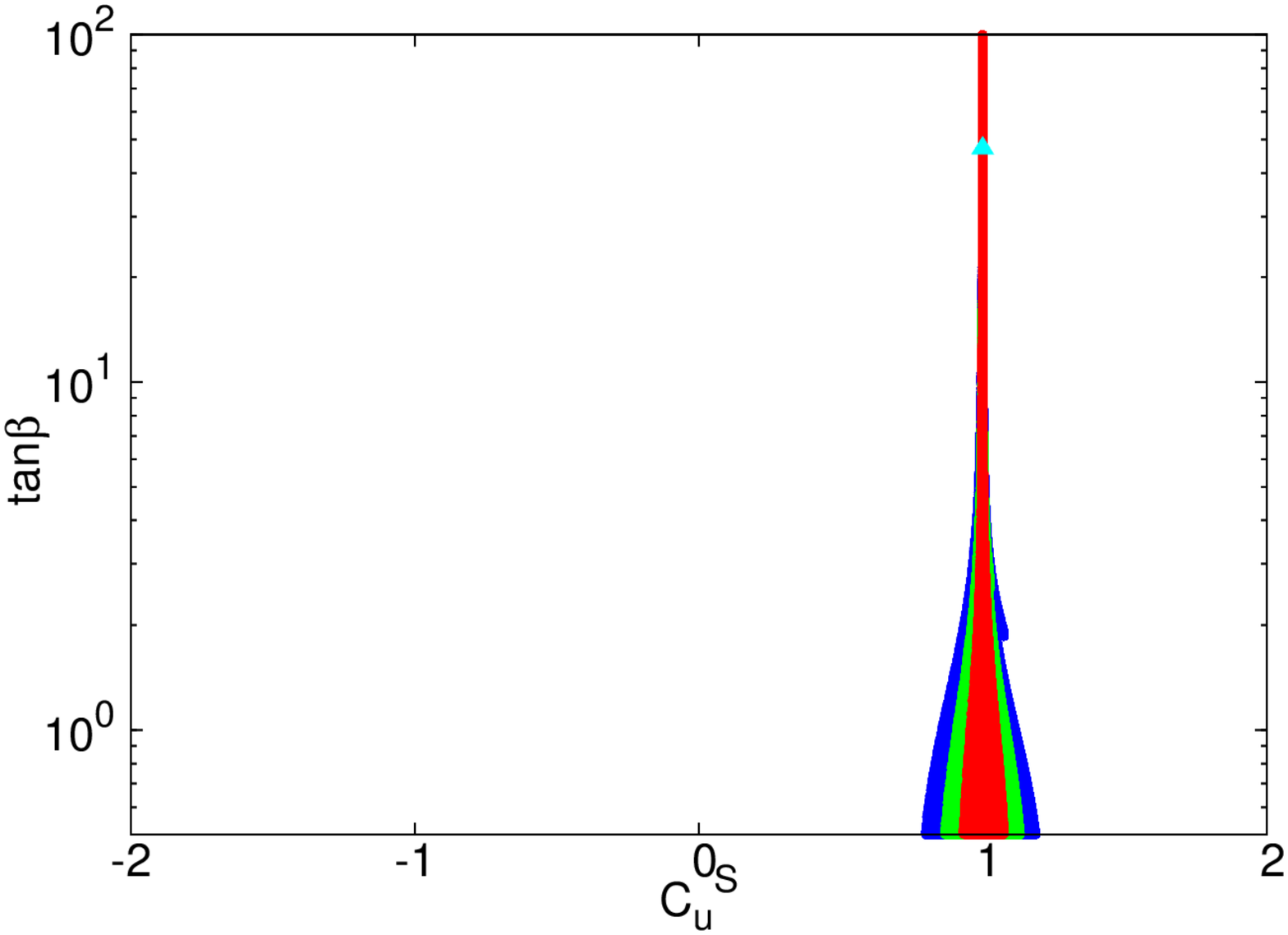}
\includegraphics[height=3.0in,angle=-90]{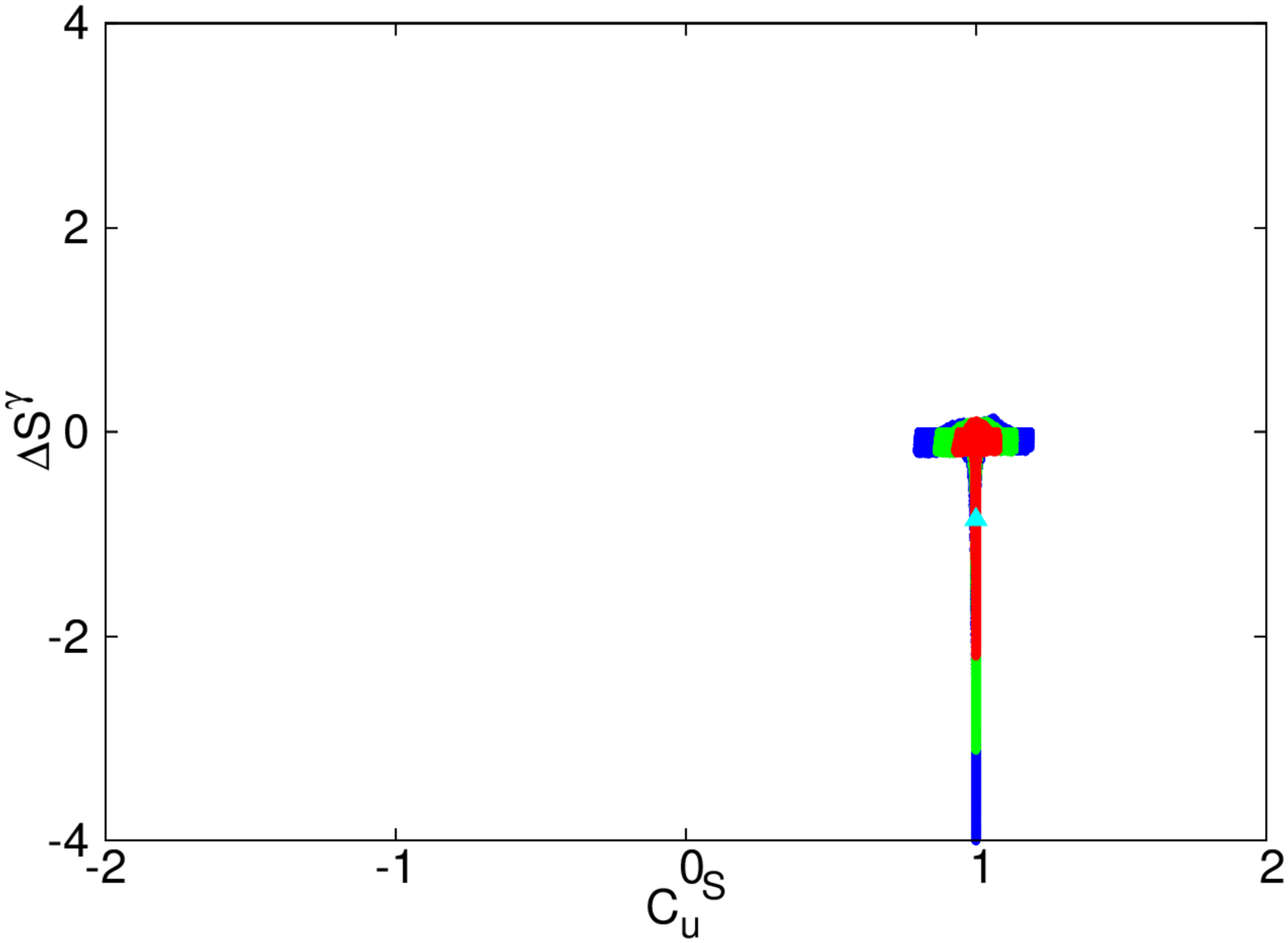}
\includegraphics[height=3.0in,angle=-90]{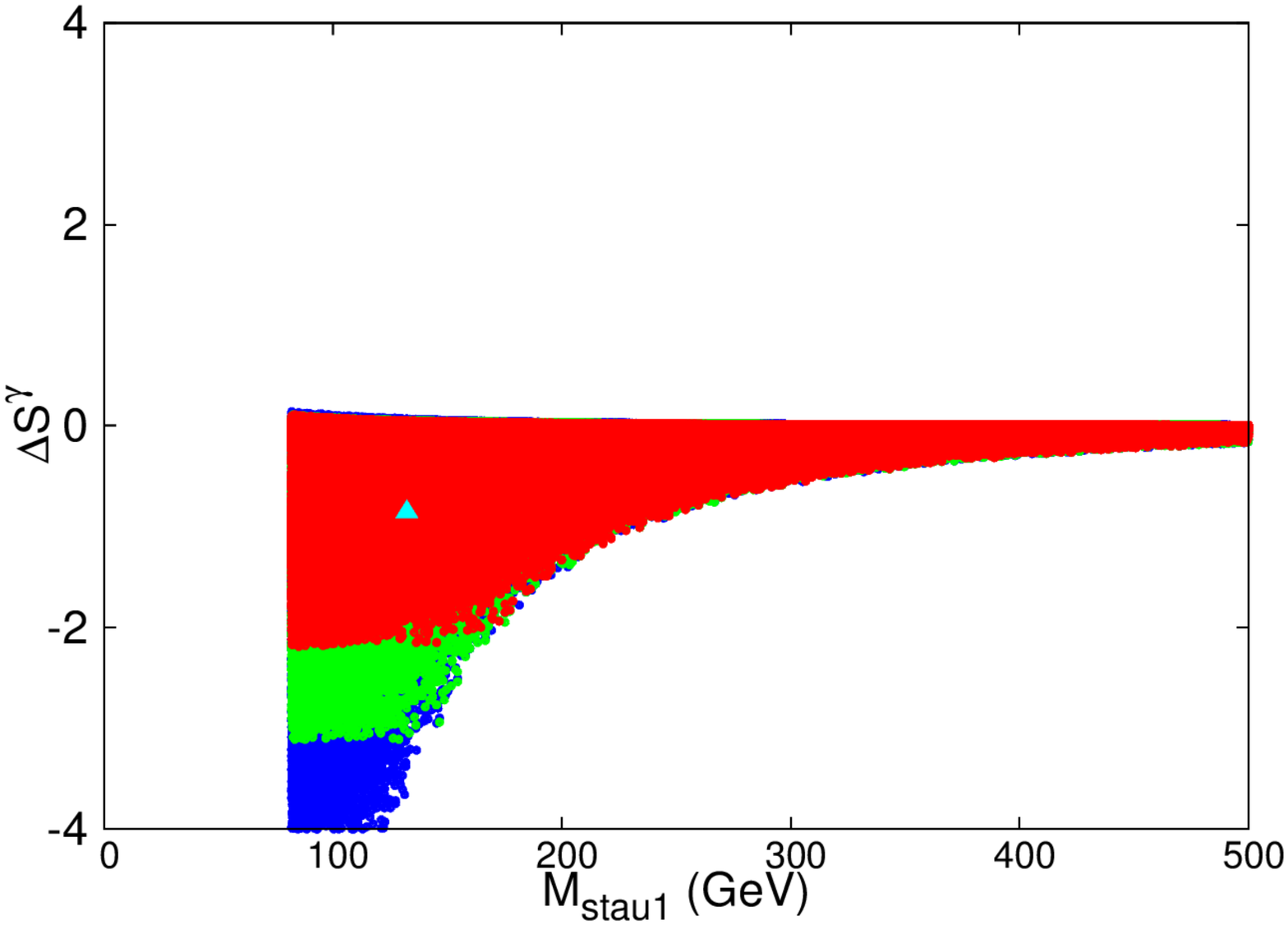}
\caption{\small \label{fig:stau}
{\bf MSSM-2} (staus):
The confidence-level regions of the fit by varying $C_u^S$, $\tan\beta$,
$M_{L_3}=M_{E_3}$, $\mu$, and $A_\tau$
with the restrictions: $\tan\beta > 1/2$, $\mu > 1$ TeV, and
$M_{\tilde{\tau}_1} > 81.9$ GeV.
The description of the confidence regions is the same as in Fig.~\ref{fig:tanbeta}.
}
\end{figure}

We show in Fig.~\ref{fig:stau} the confidence regions when we vary
$C_u^S$, $\tan\beta$, $M_{L_3}= M_{E_3}$, $\mu$, and $A_\tau$.
Requiring $\tan\beta>1/2$,
$C_u^S > 0$ and most allowed regions
are concentrated at $C_u^S\approx 1$
and $\Delta S^\gamma < 0$. 
Similar to the chargino case, $C_u^S$ and $\Delta S^\gamma$ 
correlate with each other in the parameter space.
The ``T" shape of the
CL regions of $\Delta S^\gamma$ (upper-right) can be understood by 
observing that $C_v$ is constrained to be very close to $1$
unless $C_u^S \approx 1$ when $C_u^S > 0$: see
the {\bf CPC.III} (middle) frames of Fig.~\ref{fig:cv}.
We observe that all the points fall into the
$68.3\%$ CL region of $-1.8\lsim\Delta S^\gamma\lsim 0$
when $M_{\tilde\tau_1}\gsim 180$ GeV. 

The best-fit values are shown in Table~\ref{tab:stau}.
The $\chi^2$ is just slightly worse than that 
of the {\bf CPC.III.3} case
and the $p$ value is lowered because of
more varying parameters.
The values for $C_u^S$, $C_v$, $C_\ell^S$ and $C_d^S$ are very close
to their SM values.
The lightest stau has a mass of 132.3 GeV.

%In order to further understand the behavior of $\Delta S^\gamma$ when
%we relax the $M_{L_3} = M_{E_3}$ condition, we show in Fig.~\ref{ml3-me3} the
%regions of $\Delta S^\gamma$ in the plane of $(M_{L_3},\;M_{E_3})$
%for some choices of $\tan\beta$ and $A_\tau$ and with
%$C_u^S =1$ and $\mu = 1$ TeV.
%We also show the contour of $M_{\tilde{\tau}_1} =  81.9$ GeV, and
%above the contour stau mass is heavier than 81.9 GeV.
%We can see how $\Delta S^\gamma$ changes from positive
%to negative values. With the $M_{\tilde{\tau}_1} >  81.9$ GeV,
%the large positive $\Delta S^\gamma$ are ruled out leaving behind
%only the small positive and negative values for $\Delta S^\gamma$.

\subsection{MSSM-3: With all chargino, scalar tau, sbottom, and stop contributions}
%
% chargino, scalar taus, stop, sbottom case
\begin{table}[thb!]
  \caption{\small \label{tab:all}
The chargino, scalar tau, sbottom, and stop contributions to $\Delta
S^{\gamma}(\tilde{\chi}^{\pm}_1,\tilde{\chi}^{\pm}_2,
\tilde{\tau}_1,\tilde{\tau}_2,\tilde{b}_1,\tilde{b}_2,\tilde{t}_1,\tilde{t}_2)$,
$\Delta S^g(\tilde{b}_1,\tilde{b}_2,\tilde{t}_1,\tilde{t}_2)$, $\kappa_d$.
We are taking
$M_{L_3}=M_{E_3}$, $M_{Q_3}=M_{U_3}=M_{D_3}$, $A_t=A_b=A_{\tau}$, $M_3=1$TeV,
$M_A=300$GeV, $M_2=\pm \mu$, and imposing mass limits
$|M_{H_1}-125.5\,{\rm GeV}|\leq 6$ GeV,
$M_{\tilde{\chi}^{\pm}_1}>103.5$GeV,
$M_{\tilde{\tau}_1}>81.9$GeV, $M_{\tilde{t}_1}>95.7$ GeV, and
$M_{\tilde{b}_1}>89$ GeV.
Scanning parameters: $C_u^S$, $\tan\beta \subset [1,100]$, 
$M_{L_3}\subset [0,2$TeV$]$, $M_{Q_3}\subset [0,2$TeV$]$, 
$\mu\subset [0,2$TeV$]$,
$A_t\subset [-6$TeV$,6$TeV$]$.}
\begin{ruledtabular}
\begin{tabular}{ l|ccc|rcrrrr}
Fits  & $\chi^2$ & $\chi^2$/dof & $p$-value &
\multicolumn{6}{c}{Best-fit values} \\
& & & & $C_u^S$ & $\tan\beta$ & $\kappa_d$ &
$\Delta S^\gamma$ & $\Delta S^g$ & $\Delta{\Gamma}_{tot}$ \\
\hline\hline
%CPC stop sbottom
{\bf All-SUSY}
& $15.68$ & $0.682$ & $0.869$ & $1.000$ &
$16.85$ & $0.002$ & $-0.846$ & $0.001$ & $-$ \\
\end{tabular}
\begin{tabular}{ lcccrrrrrrrrrrrrrrr}
&&&& \multicolumn{14}{c}{Best-fit values} \\
&&&& $C_v$ & $C_{d}^{S}$ & $C_{\ell}^{S}$
& $M_{L_3}$ & $M_{Q_3}$ & $M_2$ & $A_{t}$
& $M_{\tilde{\chi}^{\pm}_1}$ & $M_{\tilde{\chi}^{\pm}_2}$
& $M_{\tilde{\tau}_1}$& $M_{\tilde{\tau}_2}$
& $M_{\tilde{t}_1}$ & $M_{\tilde{t}_2}$ &
$M_{\tilde{b}_1}$ & $M_{\tilde{b}_2}$   \\
\hline\hline
&&&& $1.000$ & $1.040$ & $1.041$ & $220$ & $1732$ & $-1255$ & $-2218$ & $1203$ &
$1310$ &
$94.5$ & $303$ & $1640$ & $1829$ & $1717$ & $1748$ \\
\end{tabular}
\end{ruledtabular}
\end{table}
Here we include all contributions from charginos, scalar taus, sbottoms,
and stops. 
The relevant SUSY soft parameters are $M_{Q_3}$, $M_{U_3}$,
$M_{D_3}$, $M_{L_3}$, $M_{E_3}$, $A_t$, $A_b$, $A_\tau$, $M_3$, $M_2$, and $M_A$.
In addition to $C_u^S$ and $\tan\beta$,
we are varying $M_{Q_3}$, $M_{L_3}$, $A_t$, $\mu$ while taking 
$M_{Q_3}=M_{U_3}=M_{D_3}$, $M_{L_3}=M_{E_3}$,
$A_t=A_b=A_\tau$, and $M_2=\pm\mu$. We fix the other parameters
as $M_3=1$ TeV and $M_A=300$ GeV.
Furthermore, we impose the following constraints
on the masses:
\begin{eqnarray}
&& M_{\tilde{\chi}^{\pm}_1}>103.5~{\rm GeV},~~~
   M_{\tilde{\tau}_1}>81.9~{\rm GeV}, \nonumber \\
&& M_{\tilde{t}_1}>95.7 ~{\rm GeV}, ~~~
   M_{\tilde{b}_1}>89 ~{\rm GeV}, \nonumber \\
&& |M_{H_1}-125.5\,{\rm GeV}|\leq 6 ~{\rm GeV}. ~~~ \nonumber
\end{eqnarray}
Note that we adopt rather loose mass limits quoted in PDG \cite{pdg}
and impose the Higgs-boson mass constraint.

\begin{figure}[t!]
\centering
\includegraphics[height=3.0in,angle=-90]{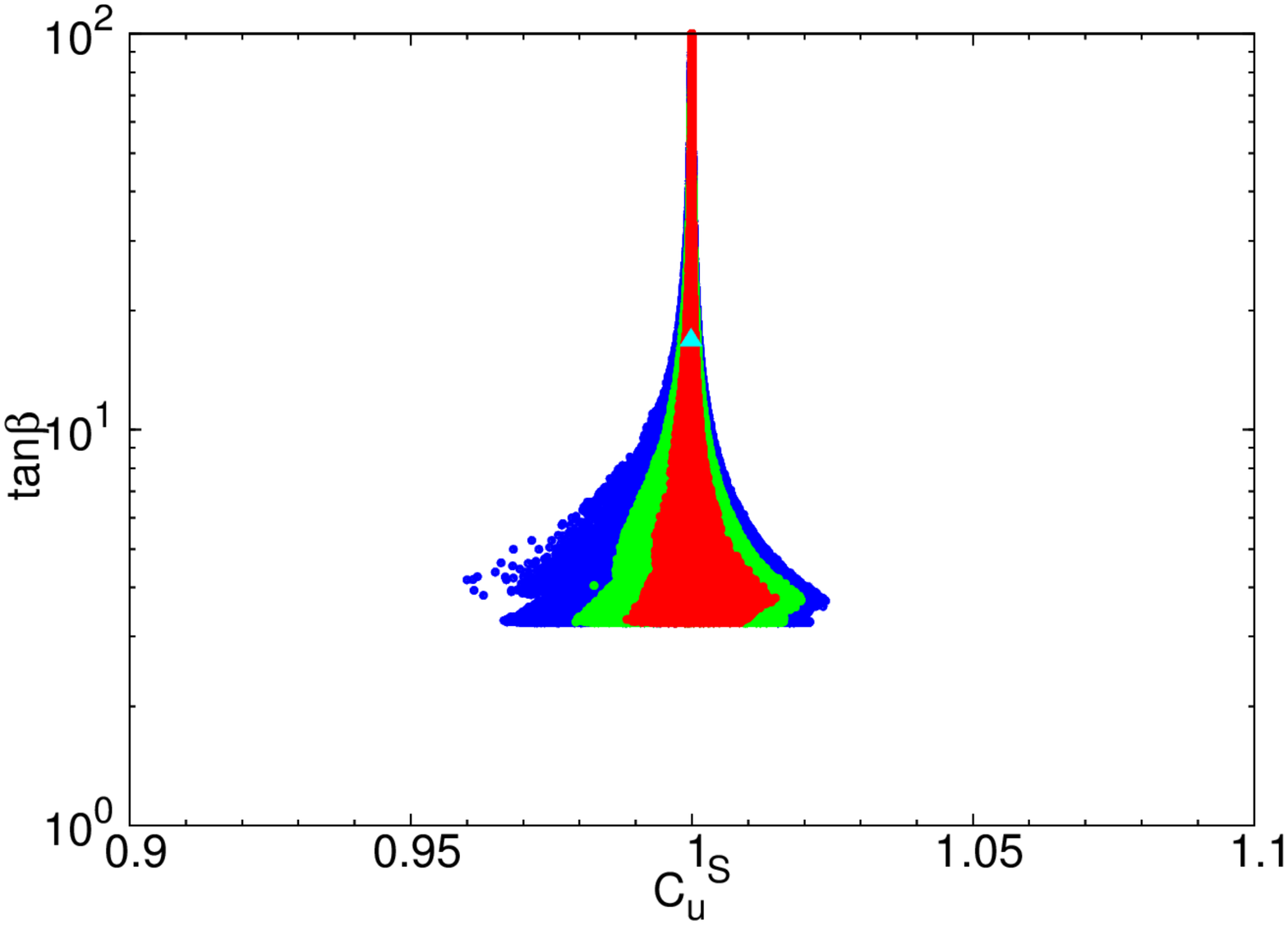}
\includegraphics[height=3.0in,angle=-90]{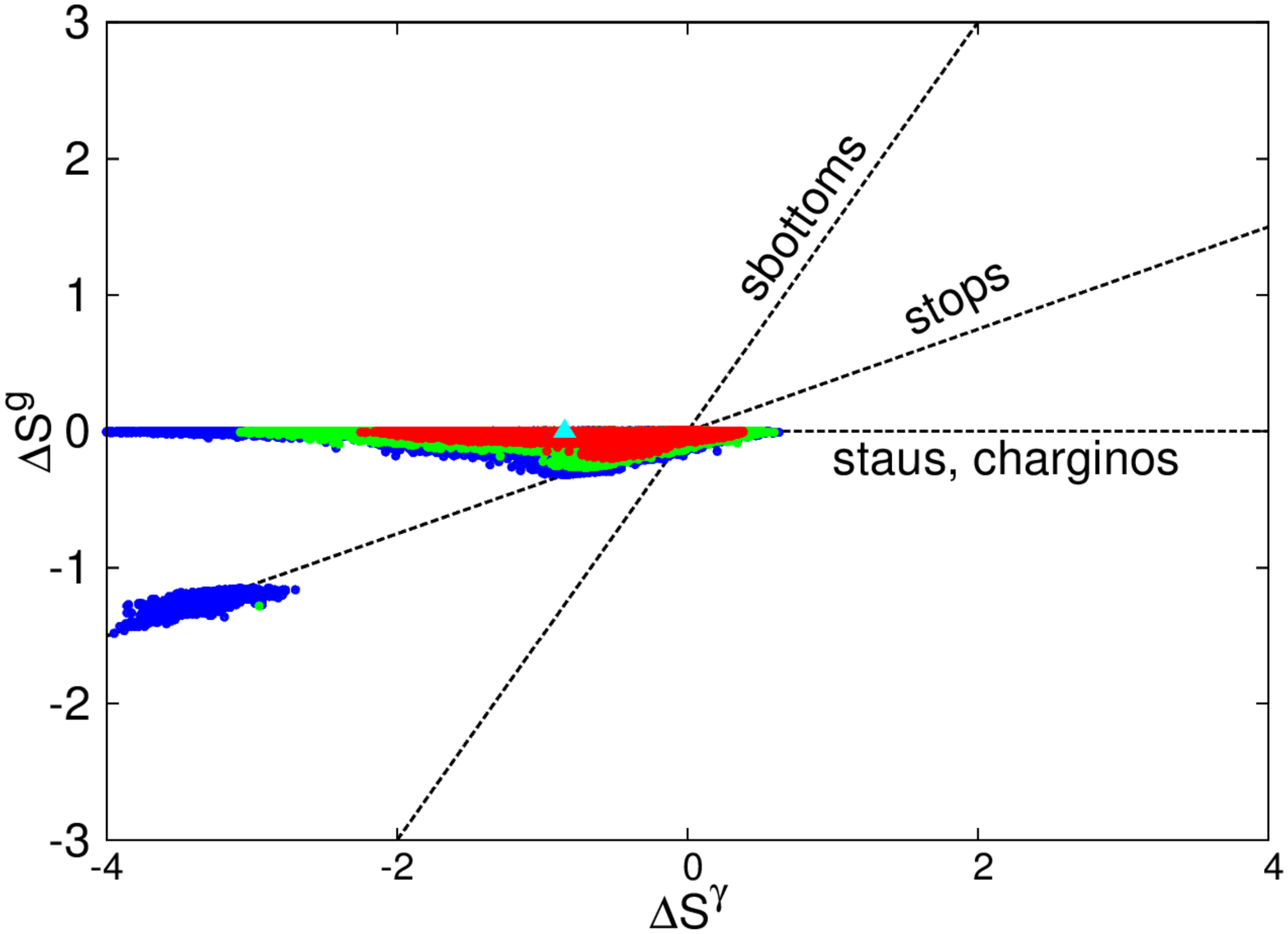}
\includegraphics[height=3.0in,angle=-90]{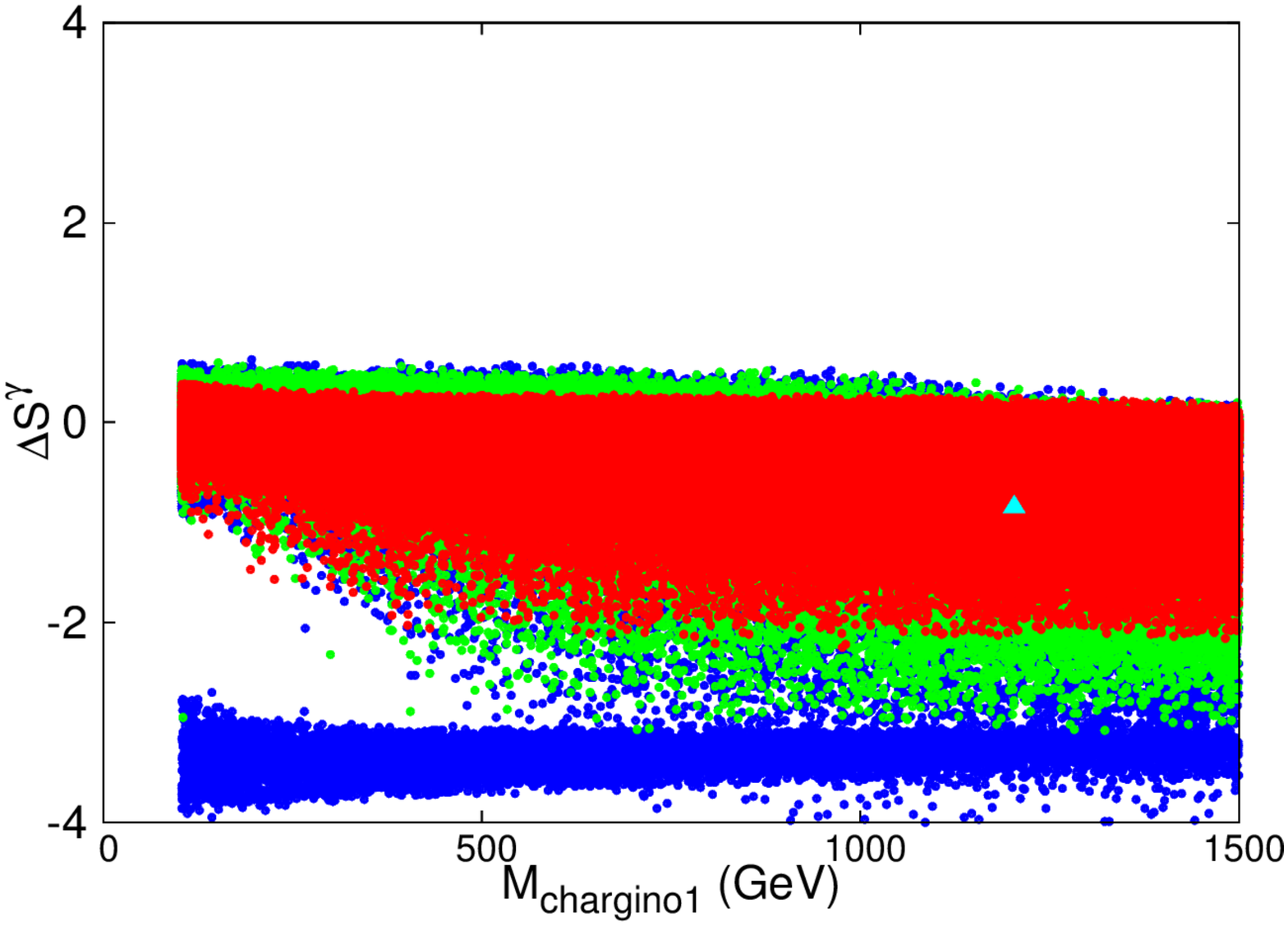}
\includegraphics[height=3.0in,angle=-90]{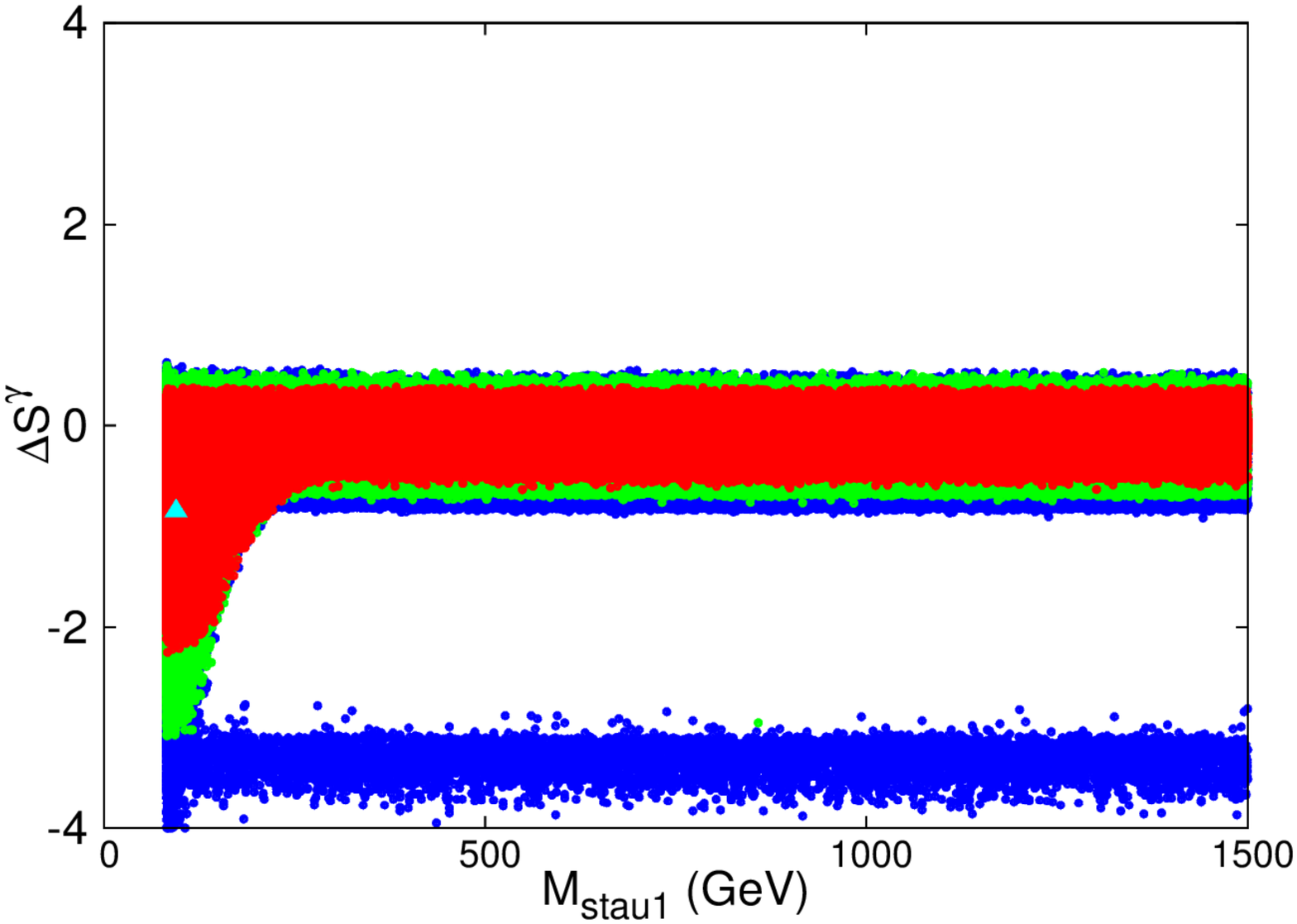}
\includegraphics[height=3.0in,angle=-90]{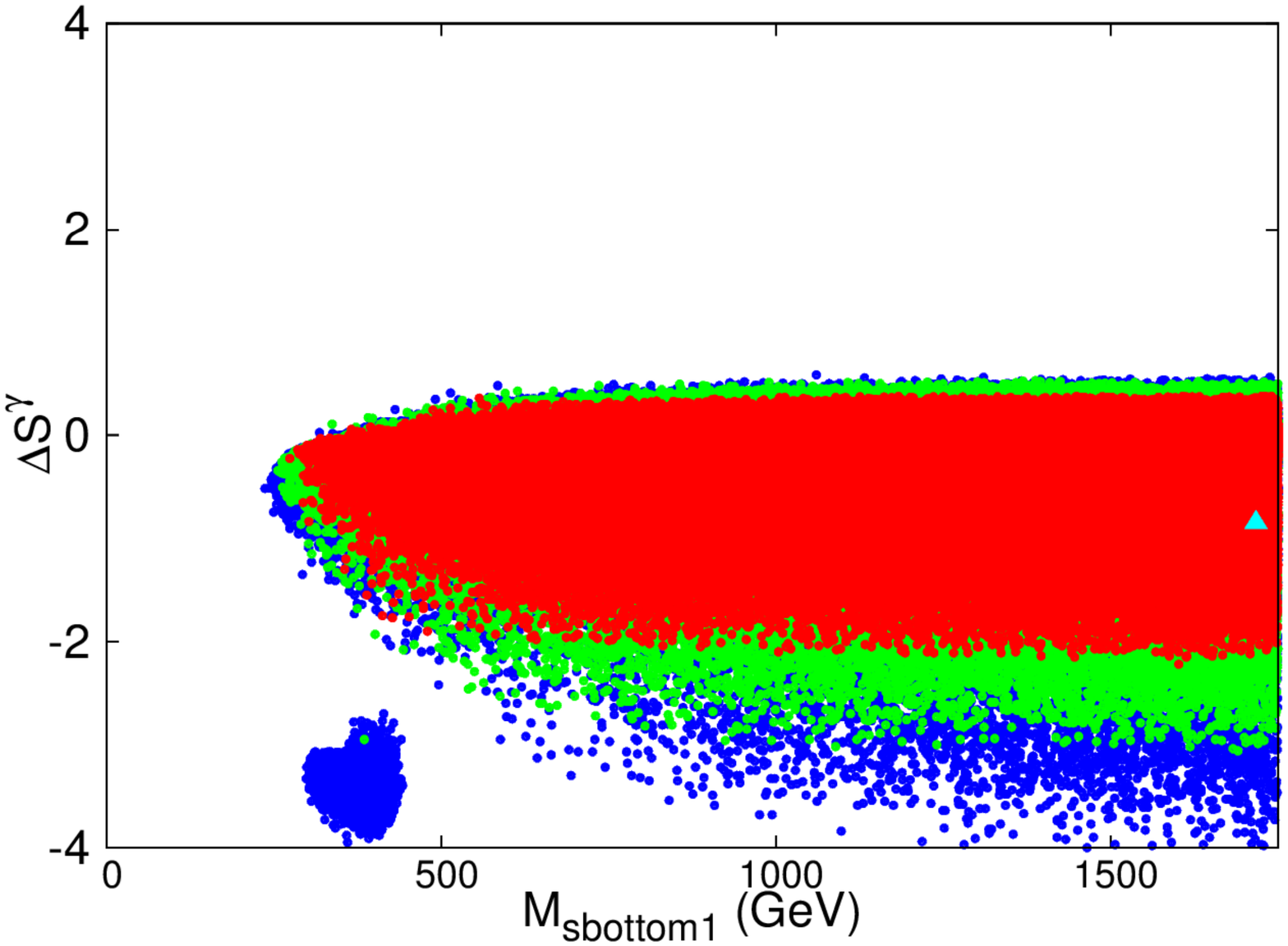}
\includegraphics[height=3.0in,angle=-90]{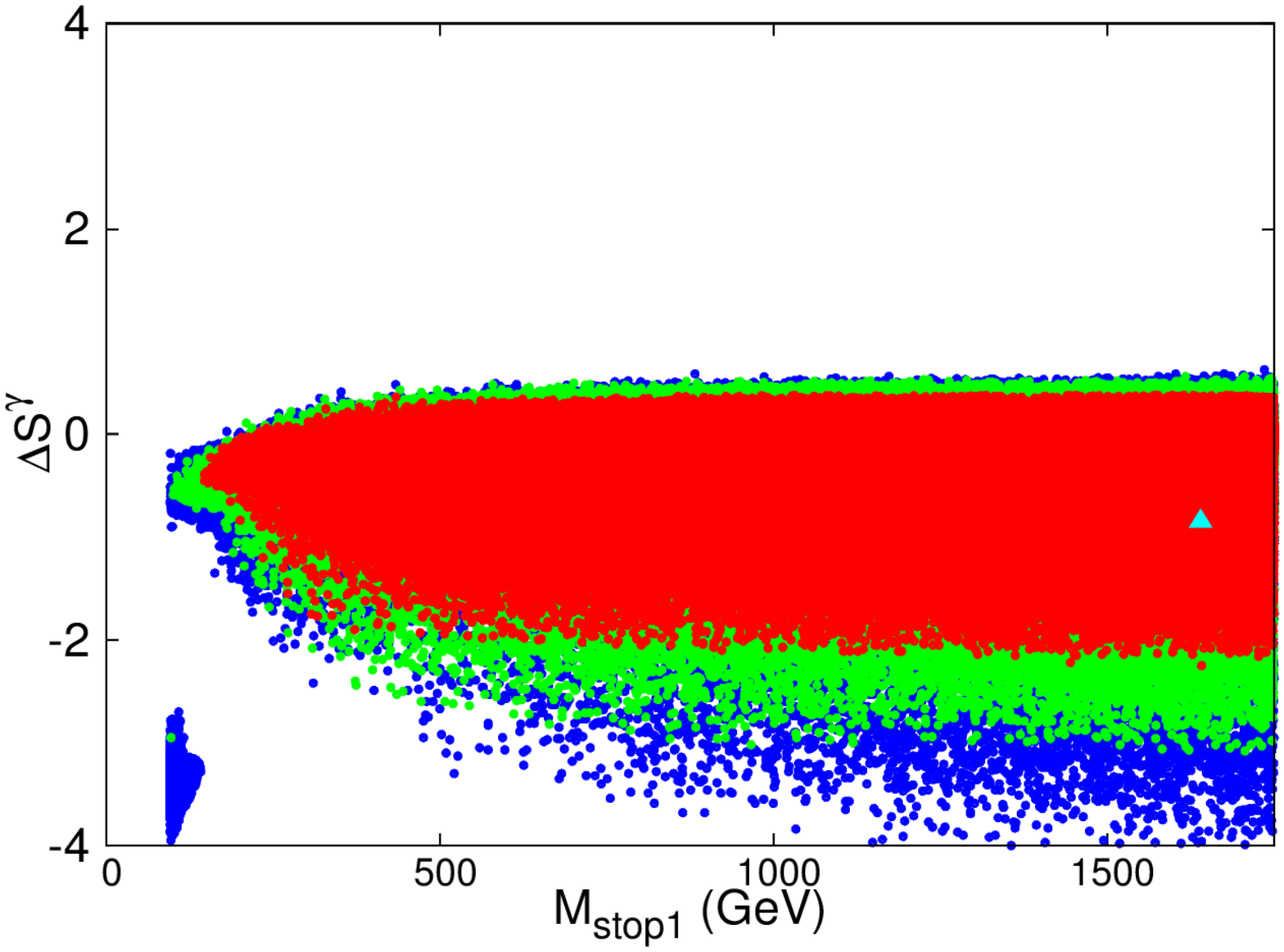}
\caption{\small \label{fig:allsusy} 
{\bf MSSM-3} (All SUSY particles): The confidence-level regions of the fit
by varying $C_u^S$, $\tan\beta$, 
$M_{Q_3}=M_{U_3}=M_{D_3}$, $M_{L_3}=M_{E_3}$, $A_t=A_b=A_{\tau}$, $\mu$
with $M_3=1$TeV, $M_A=300$GeV, $M_2=\pm \mu$, and imposing mass limits 
$|M_{H_1}-125.5\,{\rm GeV}|\leq 6$ GeV,
$M_{\tilde{\chi}^{\pm}_1}>103.5$GeV,
$M_{\tilde{\tau}_1}>81.9$GeV, $M_{\tilde{t}_1}>95.7$ GeV, and
$M_{\tilde{b}_1}>89$ GeV.
The description of the confidence regions is the same as in Fig.~\ref{fig:tanbeta}.
}
\end{figure}

The best-fit values are shown in Table~\ref{tab:all}. 
Note that the lighter stau mass ($94.5$ GeV)
is near to its low mass limit
while all other SUSY particles are heavy, so that the major contribution
to $\Delta S^\gamma$ is from the lighter stau as shown in the middle-right frame
of Fig.~\ref{fig:allsusy}. 
We observe that the stau contribution becomes comparable to 
that of the chargino around
$M_{\tilde\tau_1} = 270$ GeV and.
For the larger values of $M_{\tilde\tau_1}$,
$\Delta S^\gamma$ is saturated to have the values
between $\sim -0.6$ and $\sim 0.4$ at $68\%$ CL where it is
dominated by the chargino loops.

The confidence regions
in the relevant parameter space are shown in Fig.~\ref{fig:allsusy}.
{}From the upper-left frame of Fig.~\ref{fig:allsusy}, we observe
the requirement of $M_{H_1}\sim 125.5$ GeV completely removes
the negative $C_u^S$ region with $|C_u^S-1|\lsim 0.02$ and
$\tan\beta\gsim 3$ at $95\%$ CL.

The majority of allowed parameter space is concentrated at around 
$C_u^S \approx 1$, $-2\lsim\Delta S^\gamma \lsim 0$, 
and $\Delta S^g \approx 0$.
Yet, there is a small island allowed at 99.7\% CL around 
$\Delta S^\gamma \sim -3.5$ and $\Delta S^g \sim -1.5$.
To identify the origin of the island, we note
the following linear relationships between $\Delta S^\gamma$ and $\Delta S^g$:
\begin{eqnarray}
\Delta S^{\gamma}&=&2 N_C Q_b^2 \Delta S^g=\frac{2}{3}\,\Delta S^g\, ~~~
{\rm for~sbottom}\,,\nonumber \\
\Delta S^{\gamma}&=&2 N_C Q_t^2 \Delta S^g=\frac{8}{3}\,\Delta S^g\, ~~~
{\rm for~stop}\,.\nonumber 
\end{eqnarray}
In the chargino and stau cases, $\Delta S^g=0$. These four
correlations 
are represented by the straight lines in the upper-right frame of 
Fig.~\ref{fig:allsusy}.
It is clear that the island is due to the stop loops and it 
disappears completely when 
we require
either $M_{\tilde{t}_1} \gsim  150$ GeV or $M_{\tilde{b}_1} \gsim 450$ GeV, 
as shown in the lower frames..

\begin{figure}[t!]
\centering
\includegraphics[height=3.0in,angle=-90]{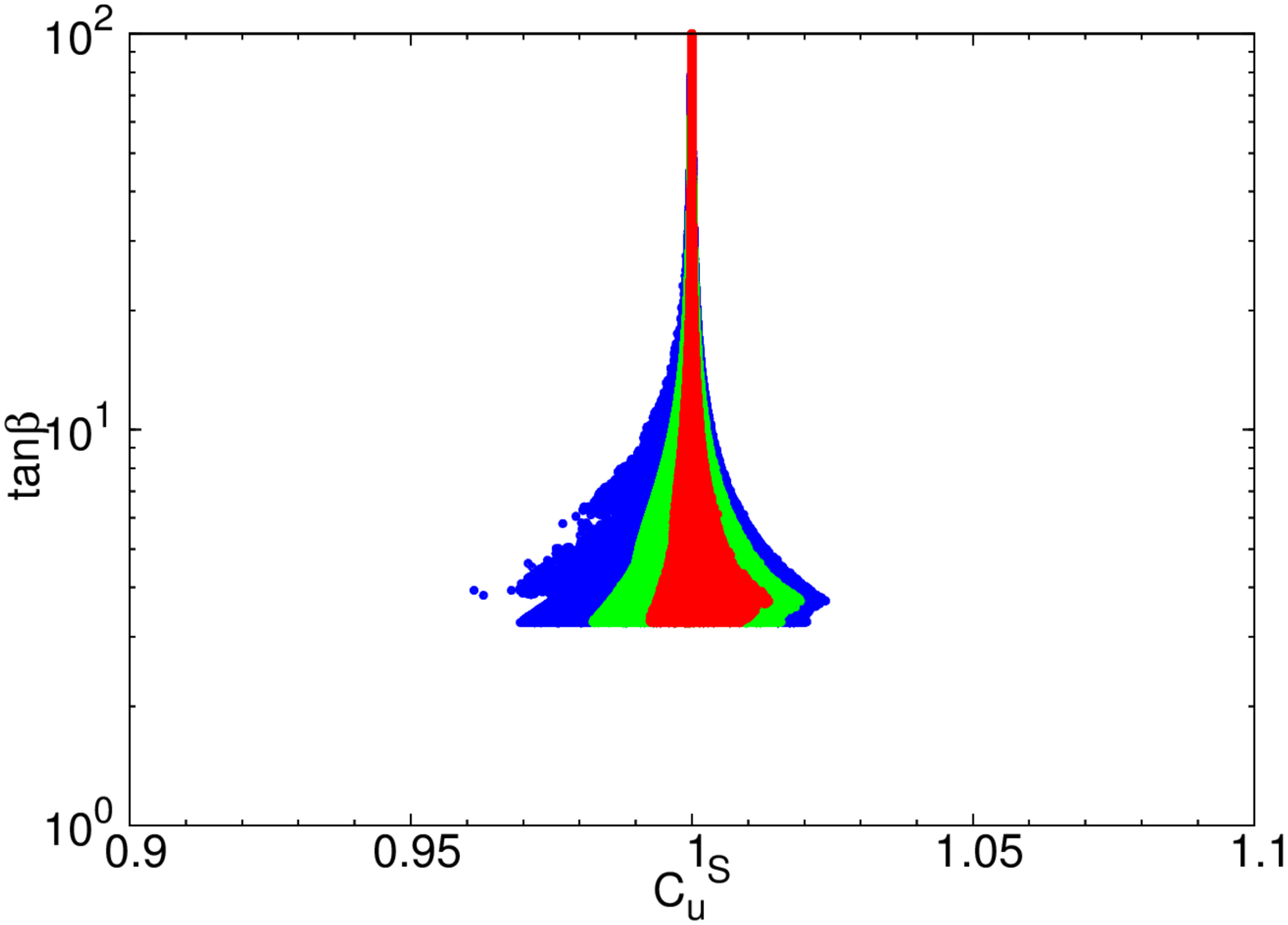}
\includegraphics[height=3.0in,angle=-90]{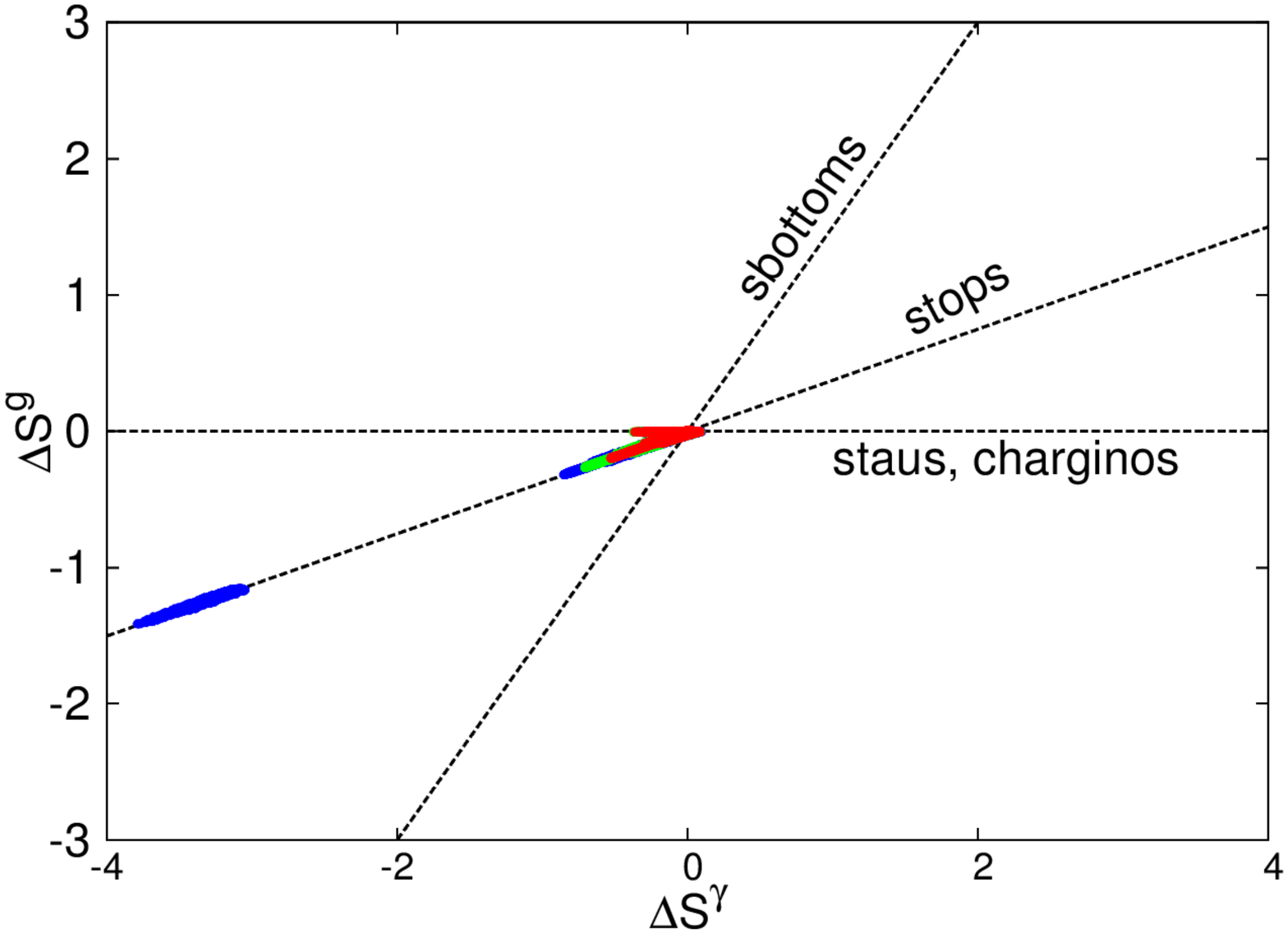}
\includegraphics[height=3.0in,angle=-90]{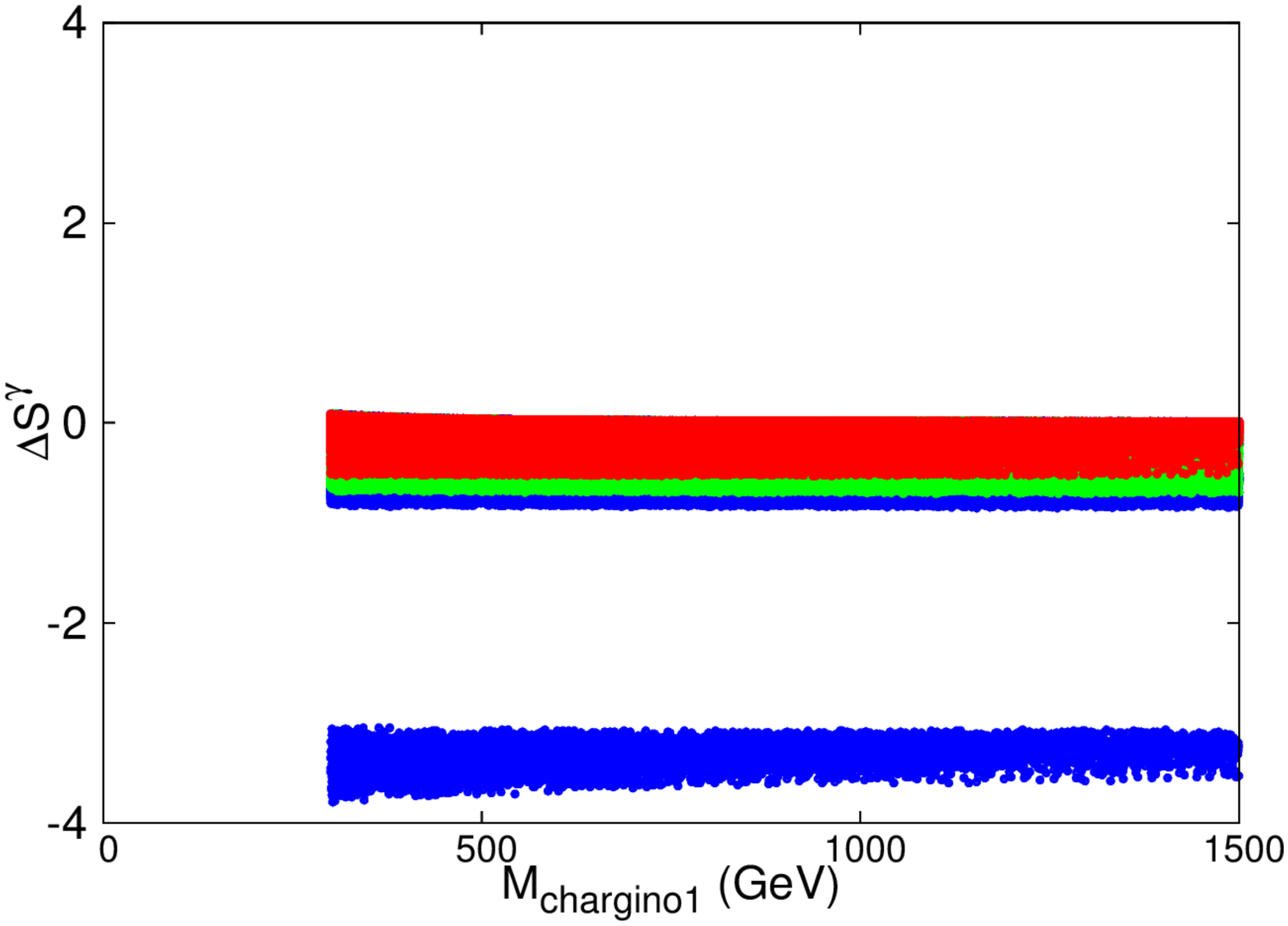}
\includegraphics[height=3.0in,angle=-90]{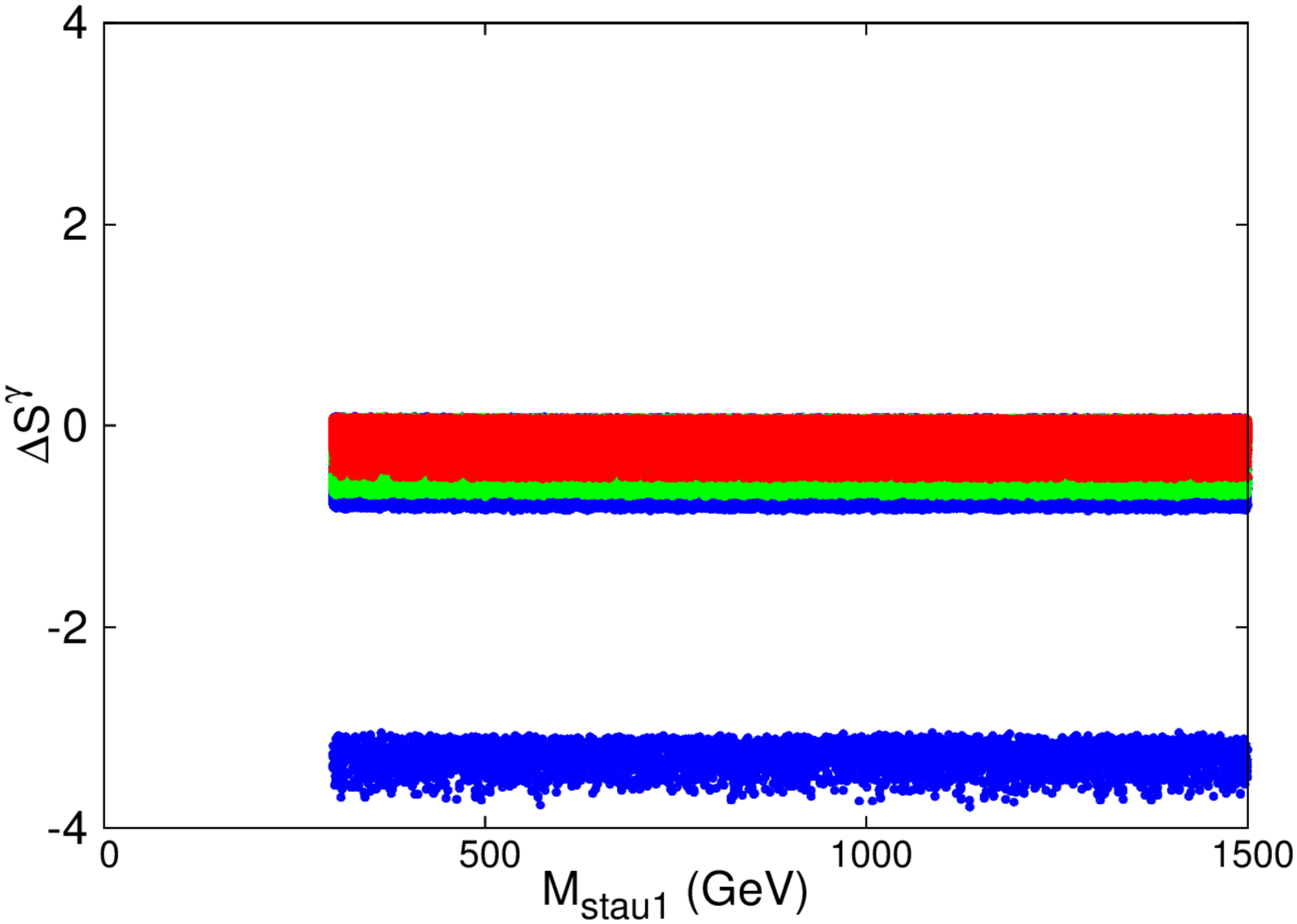}
\includegraphics[height=3.0in,angle=-90]{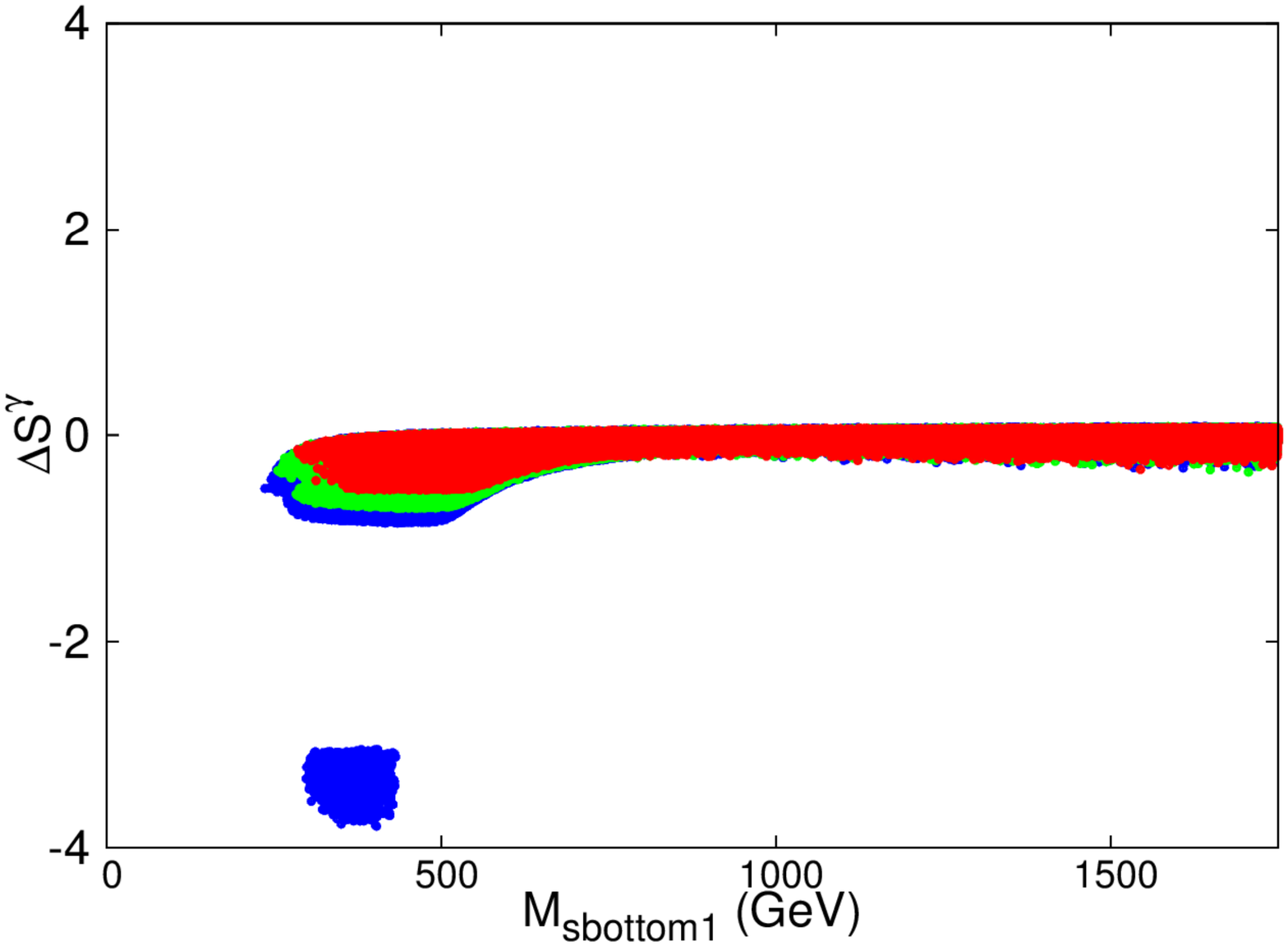}
\includegraphics[height=3.0in,angle=-90]{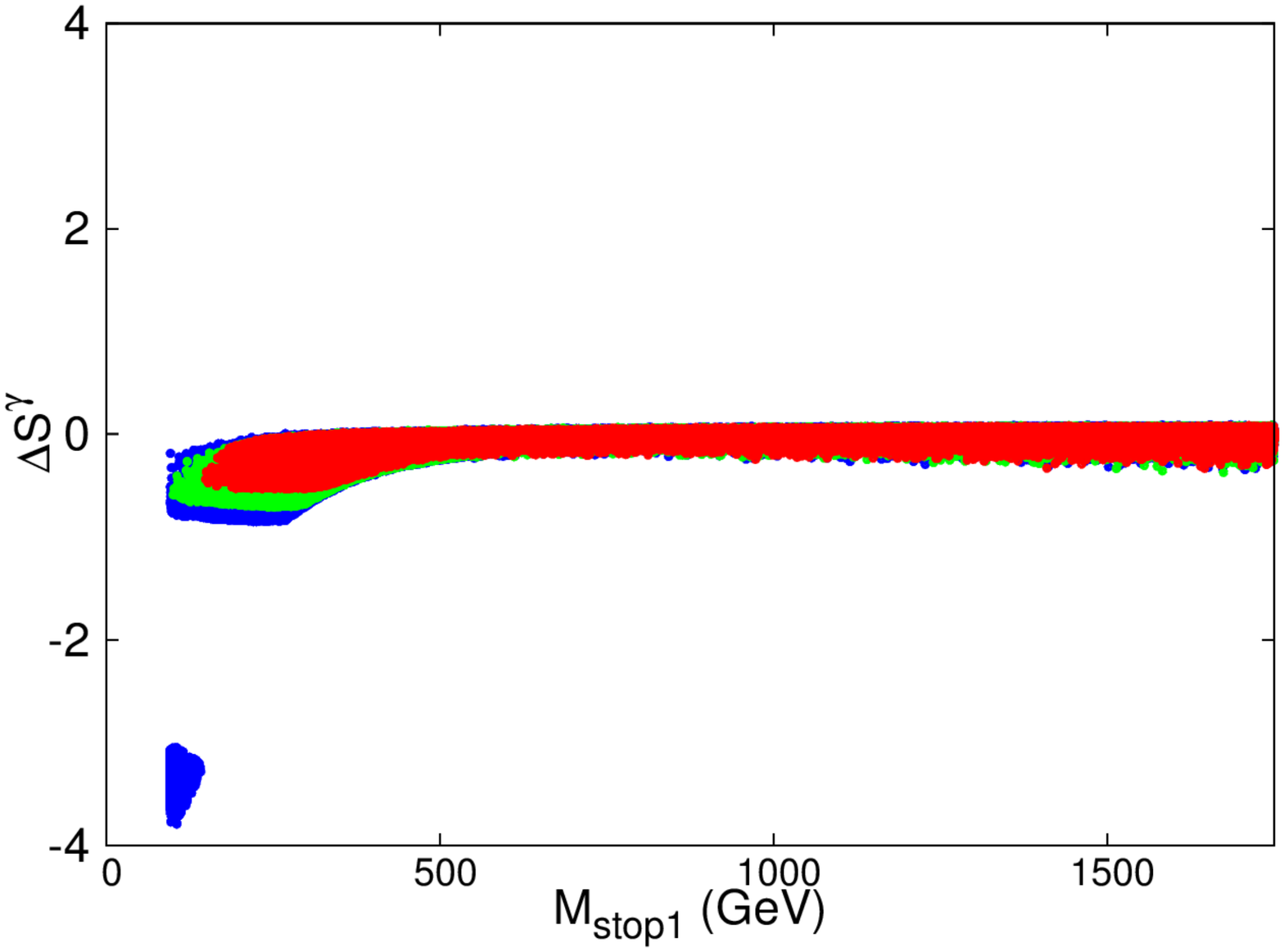}
\caption{\small \label{fig:allsusy3} 
{\bf MSSM-3} (All SUSY particles):
The same as Fig.~\ref{fig:allsusy} but requiring
$M_{\tilde{\chi}^{\pm}_1}>300$ GeV and
$M_{\tilde{\tau}_1}>300$ GeV.}
\end{figure}

In order to examine
how large the squark contributions are or to suppress
the relatively
dominant stau and chargino contributions, we take
$M_{\tilde{\chi}^{\pm}_1}>300$ GeV and $M_{\tilde{\tau}_1}>300$ GeV
and show the results in Fig.~\ref{fig:allsusy3}. We observe that
$|\Delta S^\gamma|\lsim 0.6$ at $68.3\%$ CL independently of the squark masses.
This means that $|\Delta S^\gamma/S^\gamma_{\rm SM}|\lsim 0.1$
with $S^\gamma_{\rm SM}\simeq -6.6$. 
Therefore, unless the $H\gamma\gamma$ coupling is determined 
with a precision better than $10\%$, this may imply
that the Higgs data are not sensitive to the MSSM spectrum at $68.3\%$ CL
when $M_{\tilde{\chi}^{\pm}_1}>300$ GeV and $M_{\tilde{\tau}_1}>300$ GeV
independently of the stop and sbottom masses.
Incidentally, in the middle frames, we observe that the CL regions of
$\Delta S^\gamma$ is almost independent of $M_{\tilde{\chi}^{\pm}_1,\tilde{\tau}_1}$
since it is dominated by the squark loops when
$M_{\tilde{\chi}^{\pm}_1,\tilde{\tau}_1}>300$ GeV.

Furthermore, we observe that
the stau and chargino contributions decrease quickly as their masses 
increase, as
shown in the previous {\bf MSSM-1} and {\bf MSSM-2} fits. 
Also, it worths to note that 
$|\Delta S^\gamma|\lsim 0.2$ when 
$M_{\tilde{\chi}^\pm_1,{\tilde{\tau}_1}} > 500$ GeV, see
Figs.~\ref{fig:char} and \ref{fig:stau} when squarks are very heavy.

Finally, we also find that $|\Delta S^\gamma|\lsim 0.2$
if we take the current $95$\%-CL LHC limits on the stop and sbottom masses
with $M_{\tilde{\chi}_1^0} = 0$ GeV~\cite{pdg}:
$M_{{\tilde t}_1} > 650$ GeV and $M_{{\tilde b}_1} > 600$ GeV,
assuming that charginos and staus are heavy enough and do not 
contribute to $|\Delta S^\gamma|$ more significantly than squarks.

Before concluding, we would like to briefly discuss the SUSY 
impact on future measurements of the Higgs properties 
through the Higgs decay into $Z\gamma$ and the Higgs cubic coupling.
In the {\bf MSSM-1} case, thanks to light charginos,
we have found that the branching ratio of the 125 GeV Higgs boson
to $Z\gamma$ can be enhanced by about $15 \%$ compared to the SM prediction.
On the other hand, in the {\bf MSSM-2} and {\bf MSSM-3}
cases, the SUSY contribution to
the branching ratio is less than $1 \%$.
Meanwhile, in the {\bf MSSM-3} case in which all the masses of 
relevant SUSY particles are specified and an unambiguous estimation 
of the Higgs cubic coupling is possible,
the deviation of the Higgs cubic coupling 
from the SM value $M_{H_1}^2/2v$ ($v\approx 246$ GeV)
is negligible upon
its variation according to the Higgs mass constraint taken
in this work: $|M_{H_1} - 125.5\,{\rm GeV}| < 6$ GeV.

\section{Synopsis and Conclusions}

\begin{figure}[t!]
\centering
\includegraphics[width=2.0in]{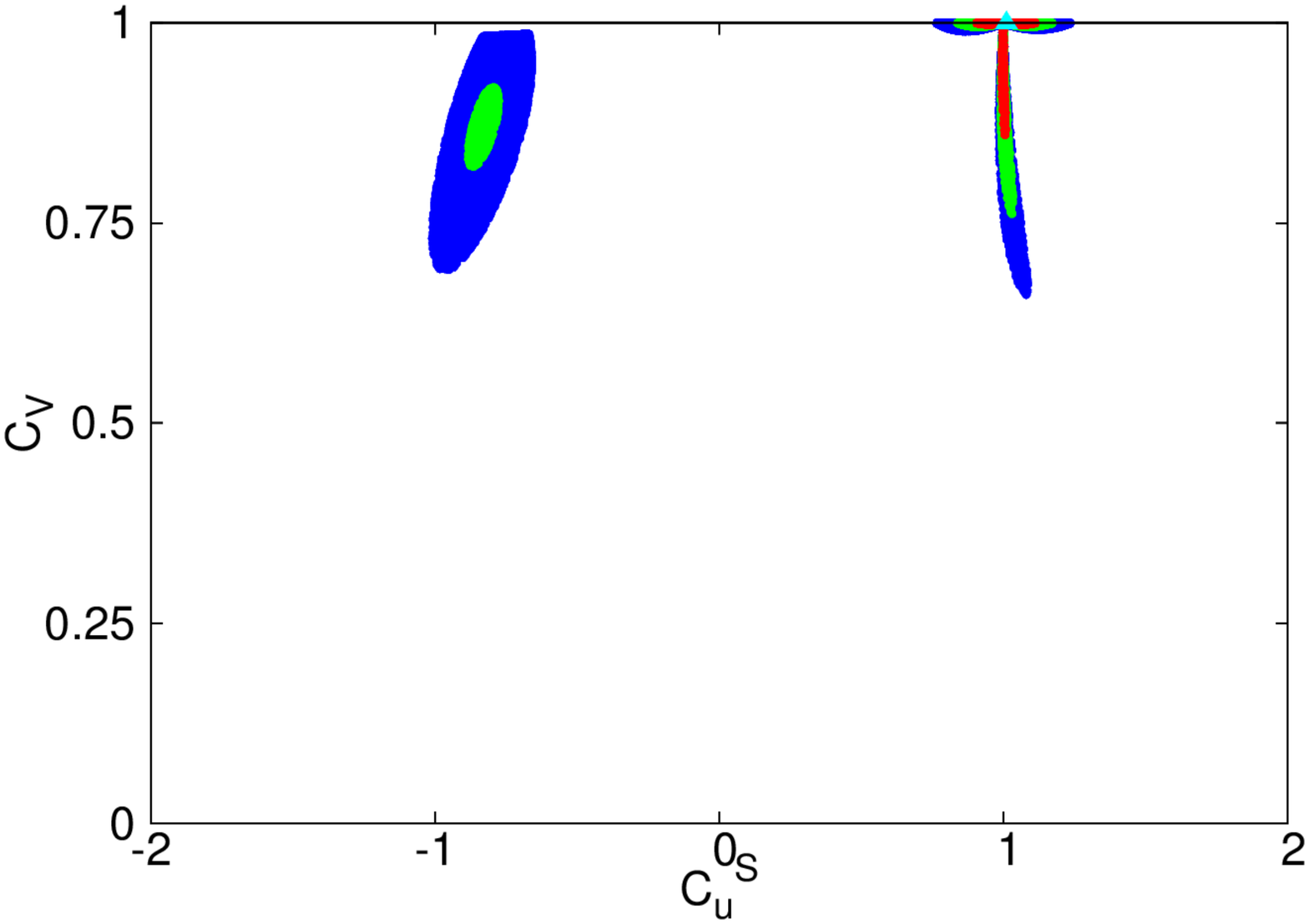}
\includegraphics[width=2.0in]{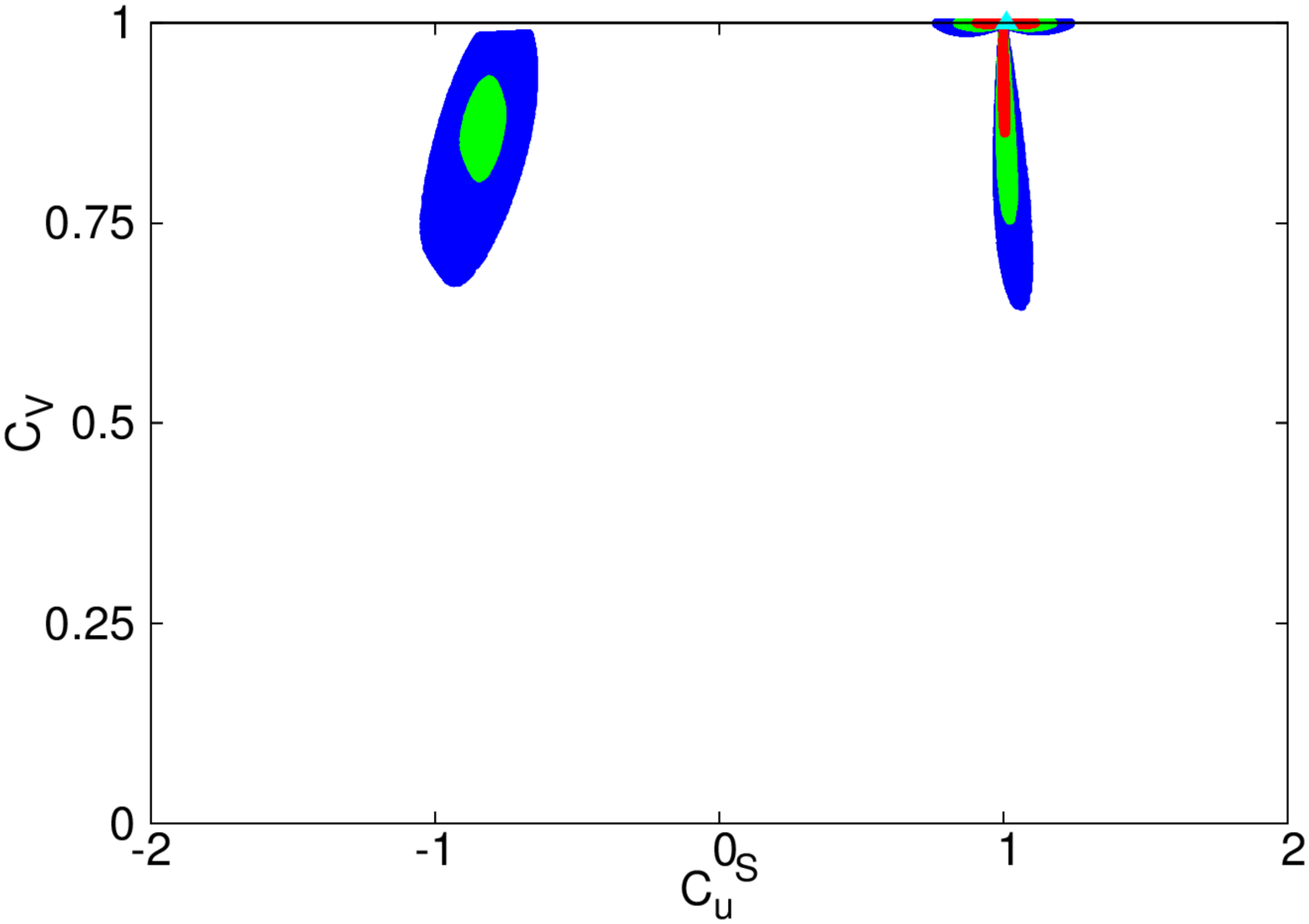}
\includegraphics[width=2.0in]{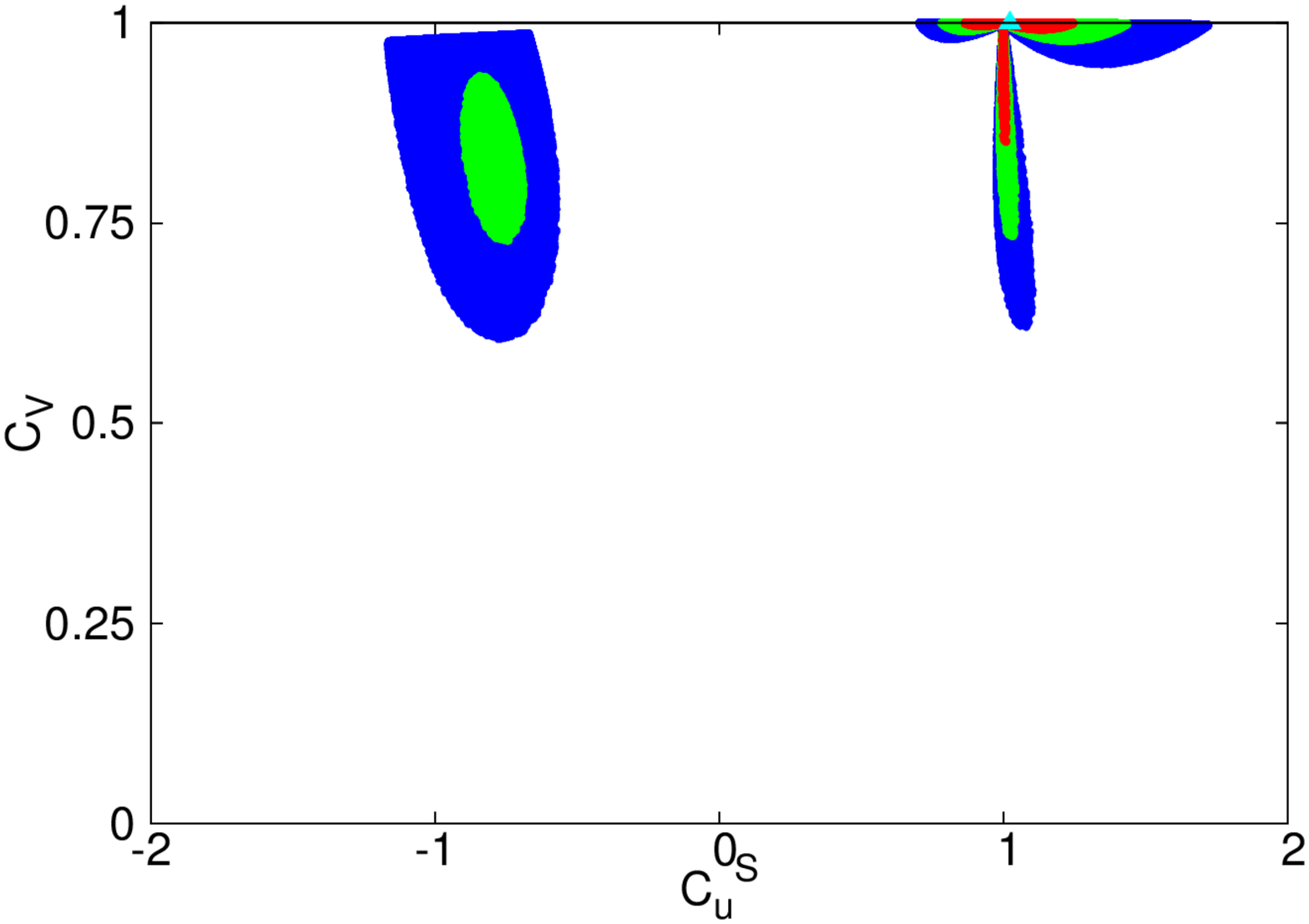}
\includegraphics[width=2.0in]{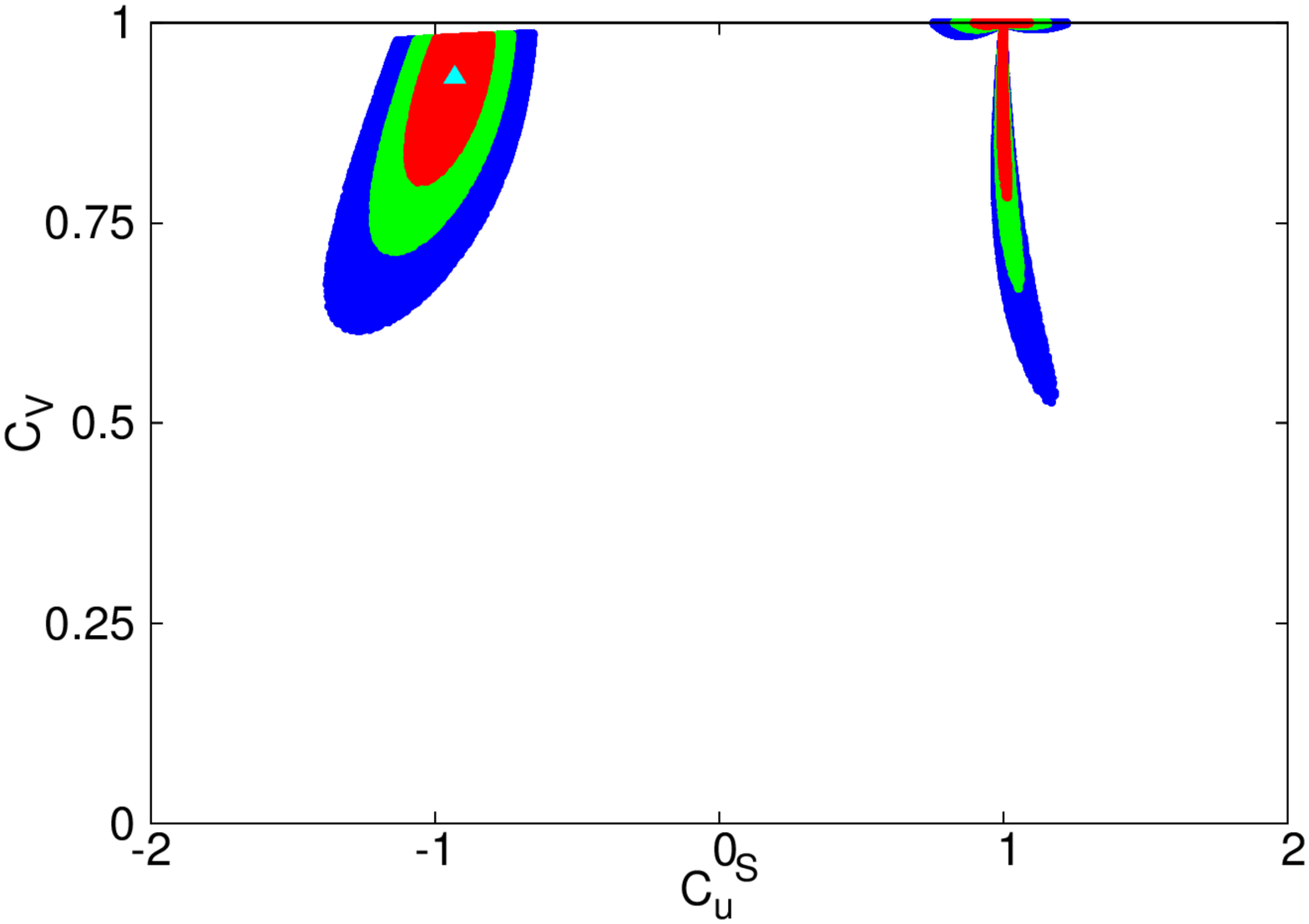}
\includegraphics[width=2.0in]{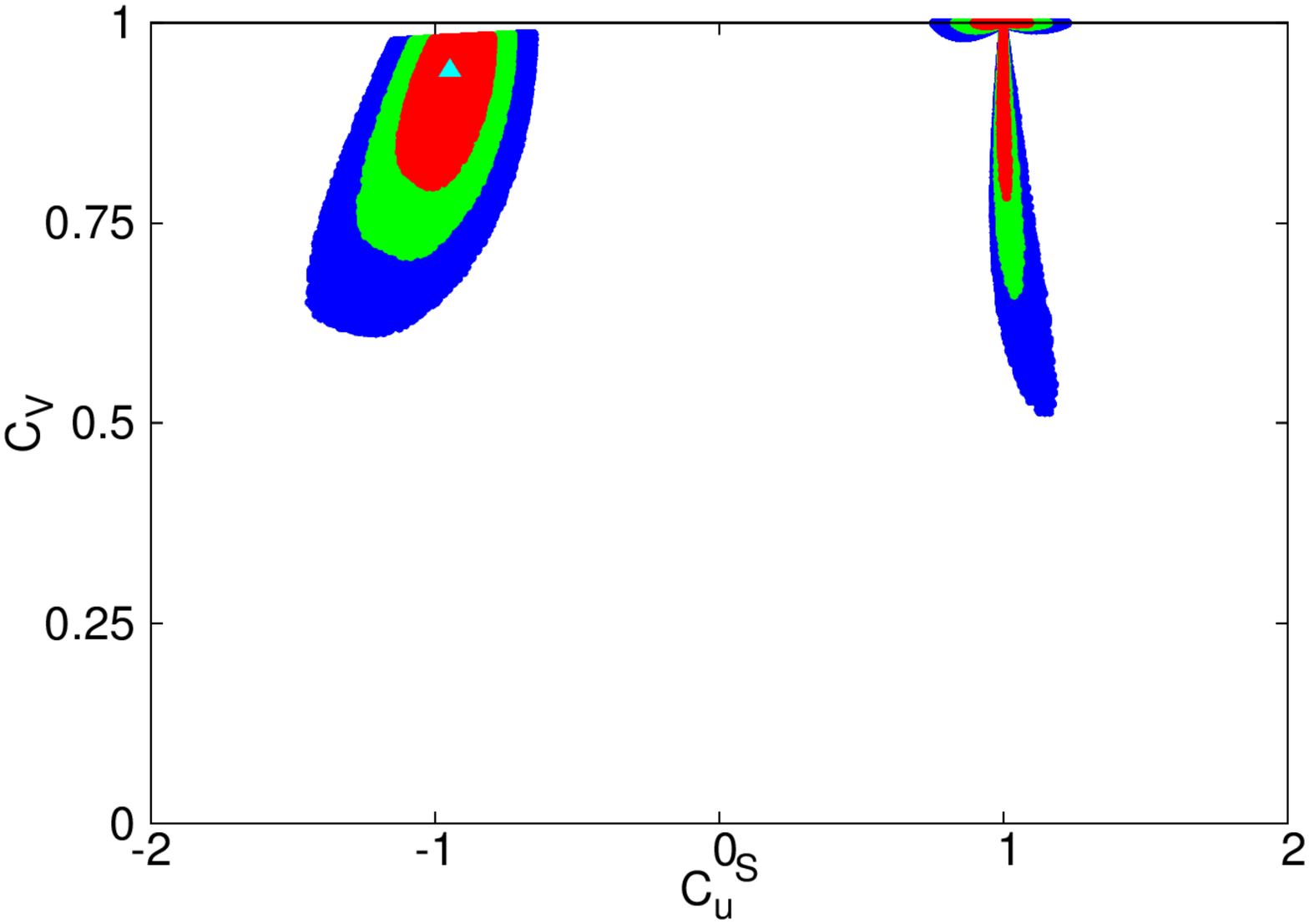}
\includegraphics[width=2.0in]{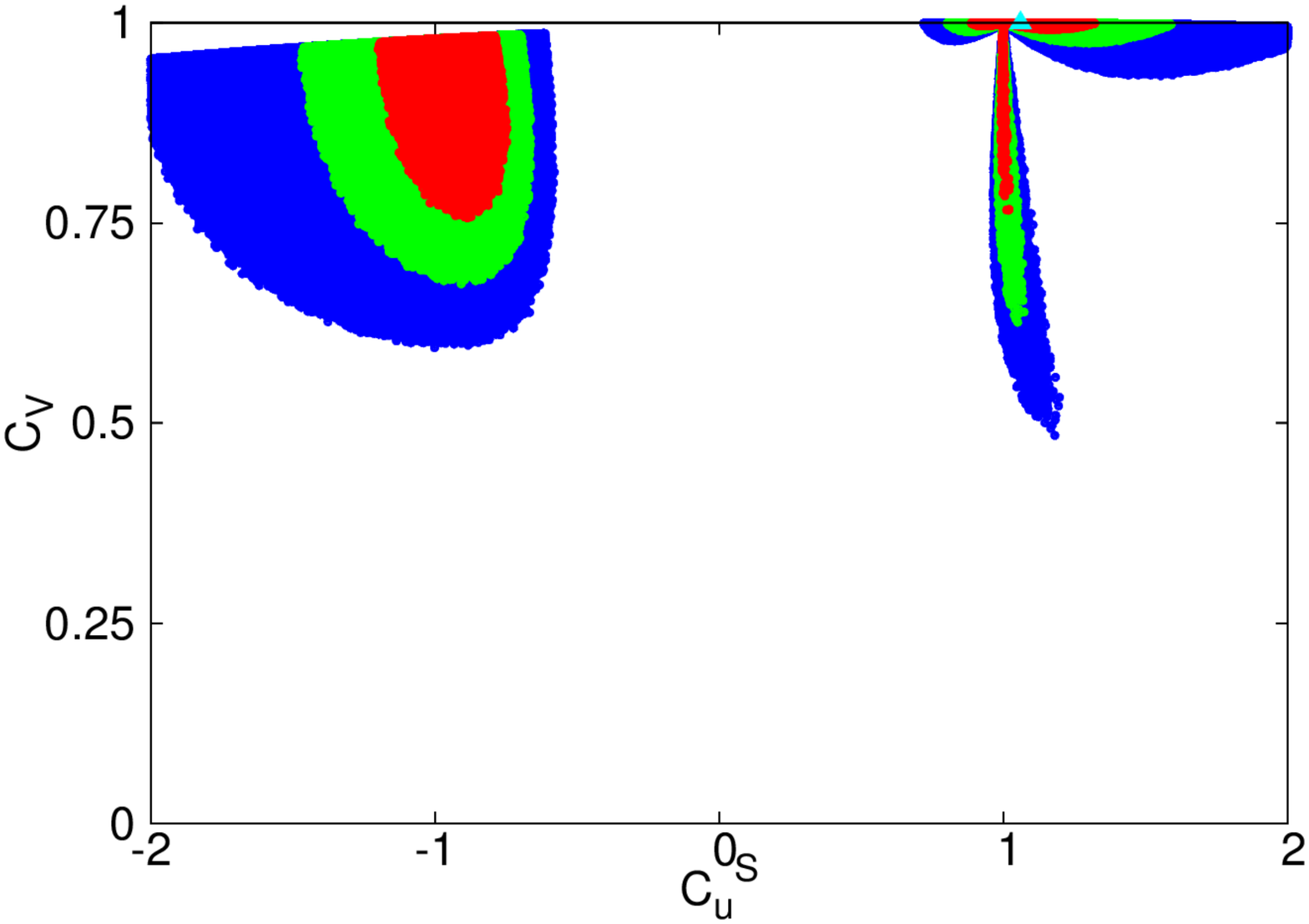}
\includegraphics[width=2.0in]{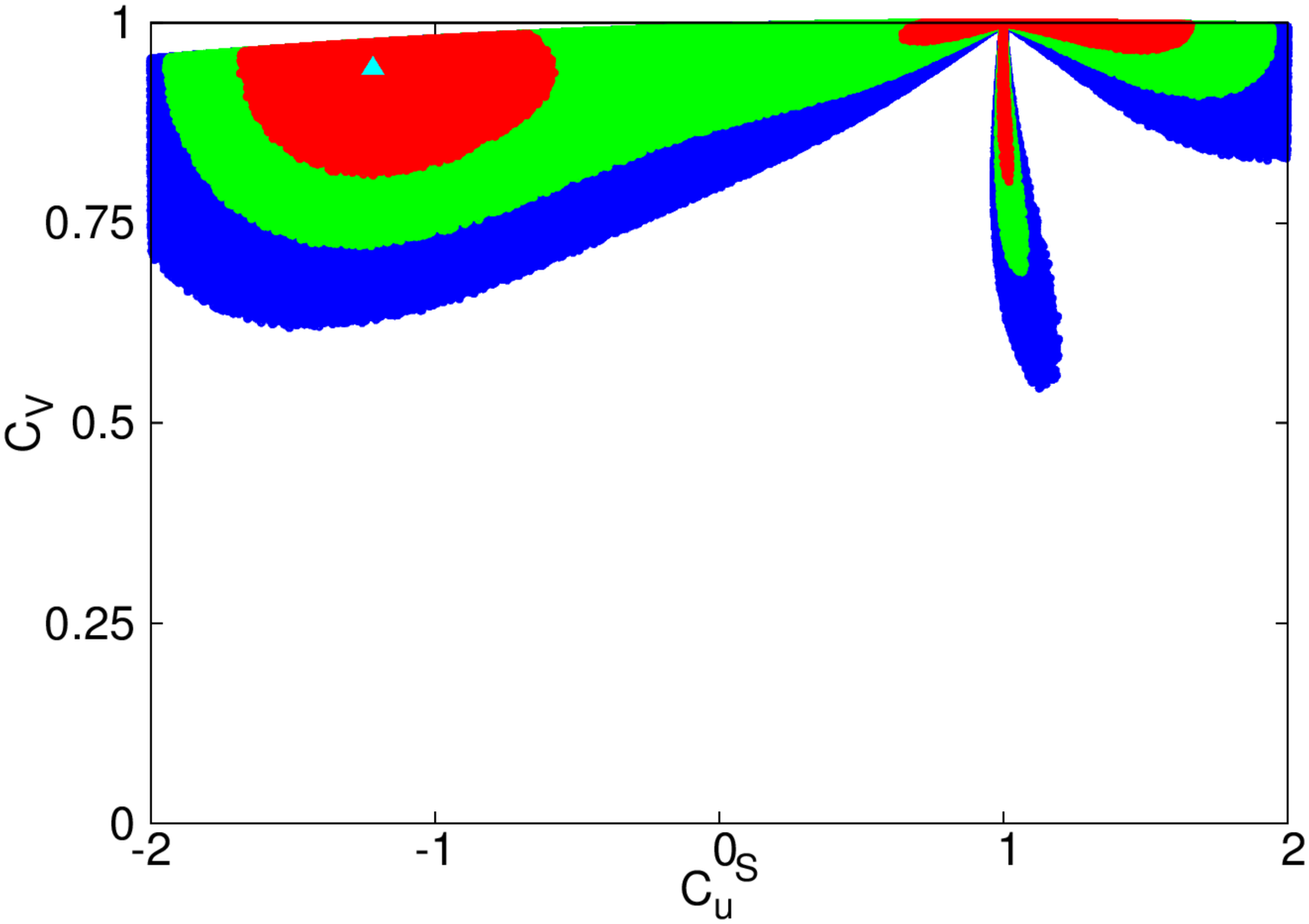}
\includegraphics[width=2.0in]{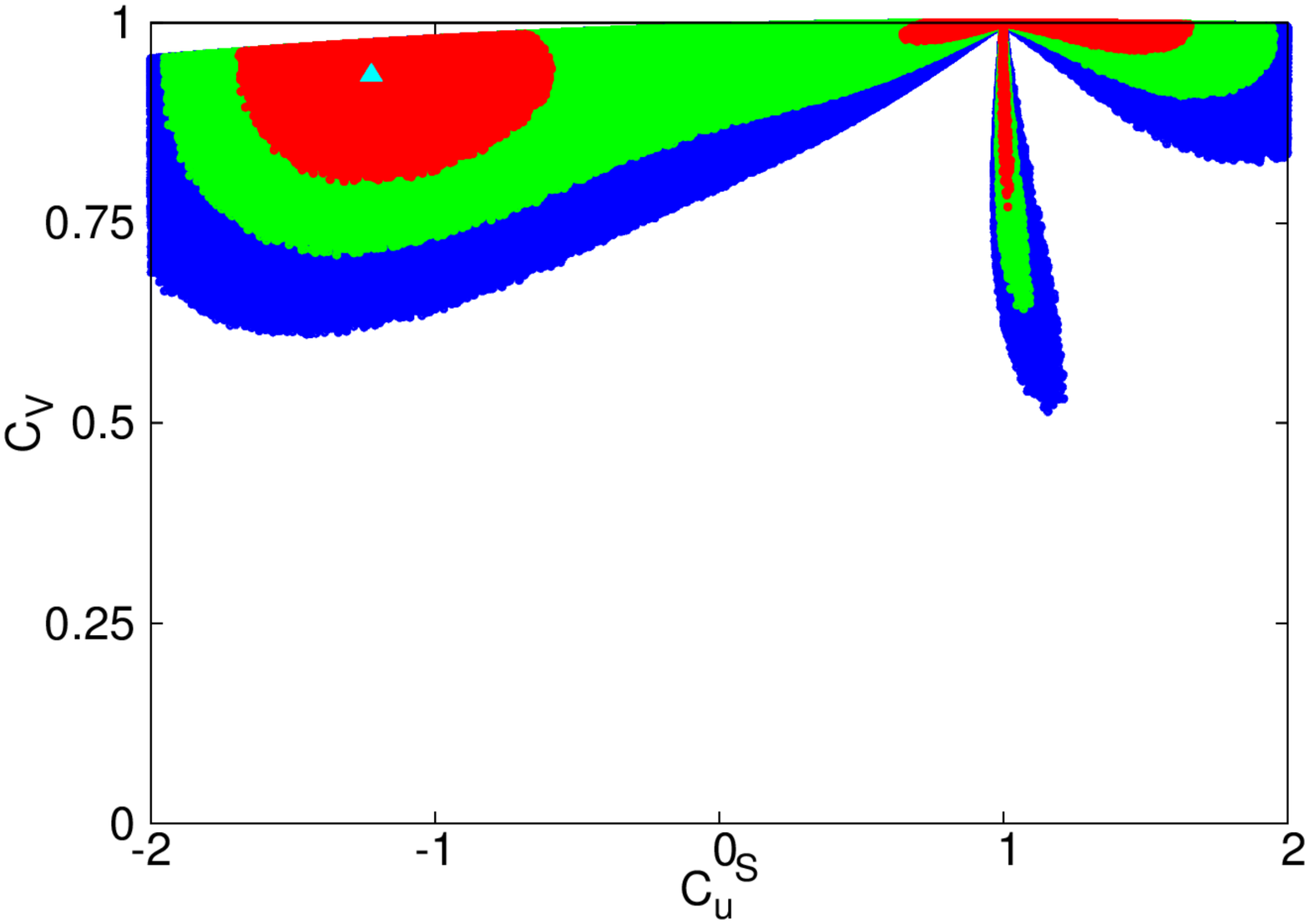}
\includegraphics[width=2.0in]{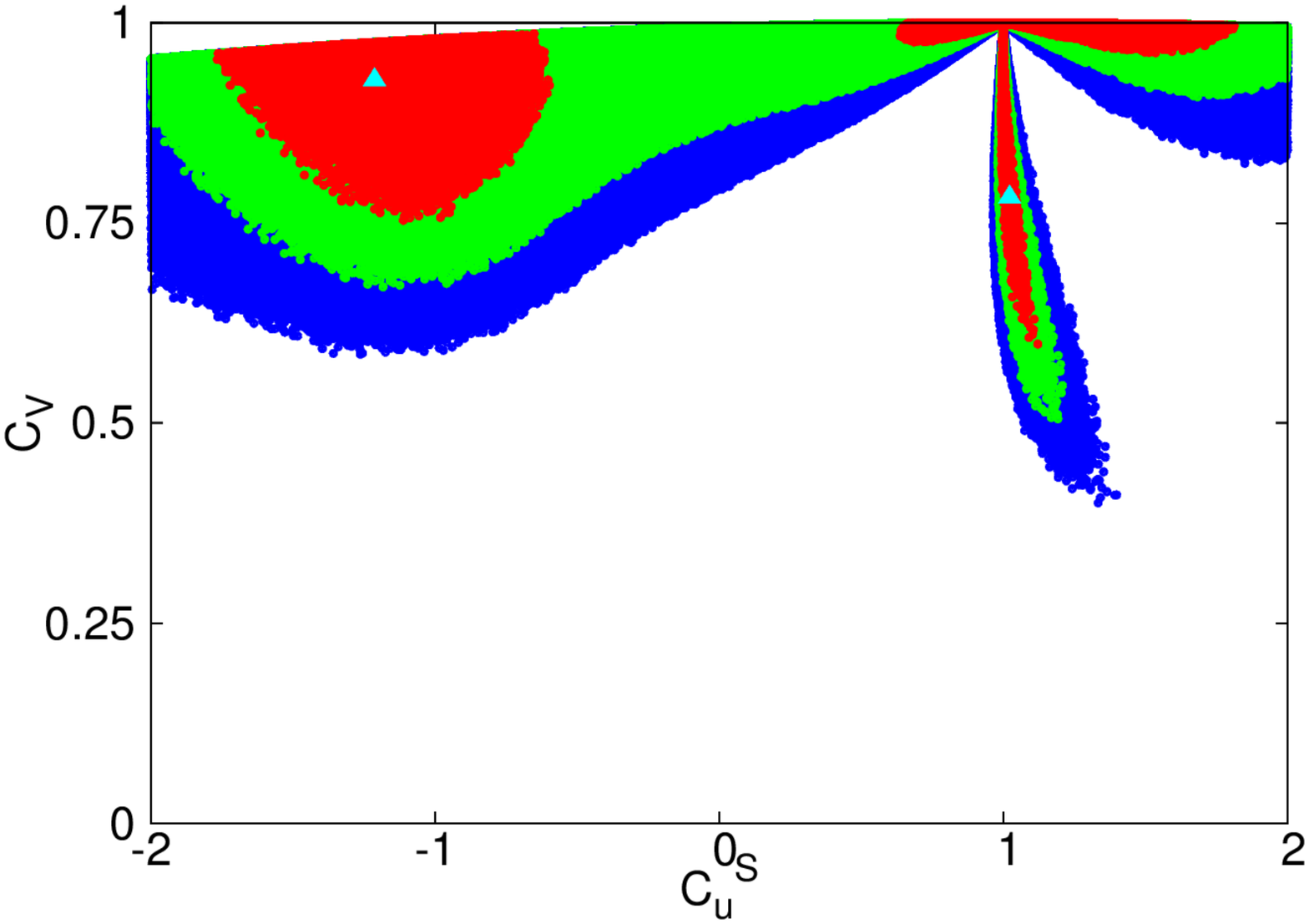}
\caption{\small \label{fig:cv}
The same as in Fig.~\ref{fig:tanbeta} but on
the $(C_u^S, C_v)$ plane.
}
\end{figure}

\begin{figure}[t!]
\centering
\includegraphics[width=2.0in]{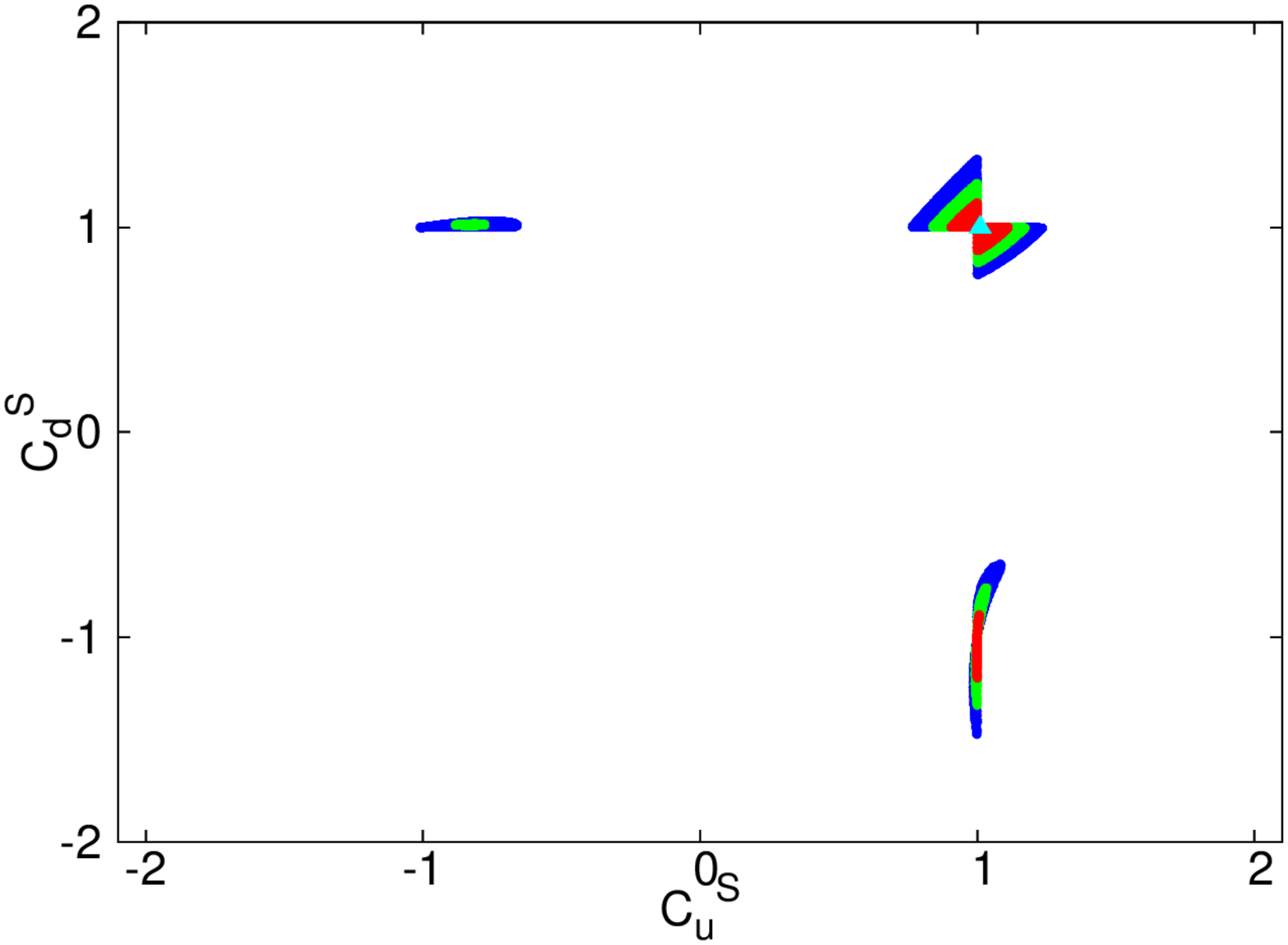}
\includegraphics[width=2.0in]{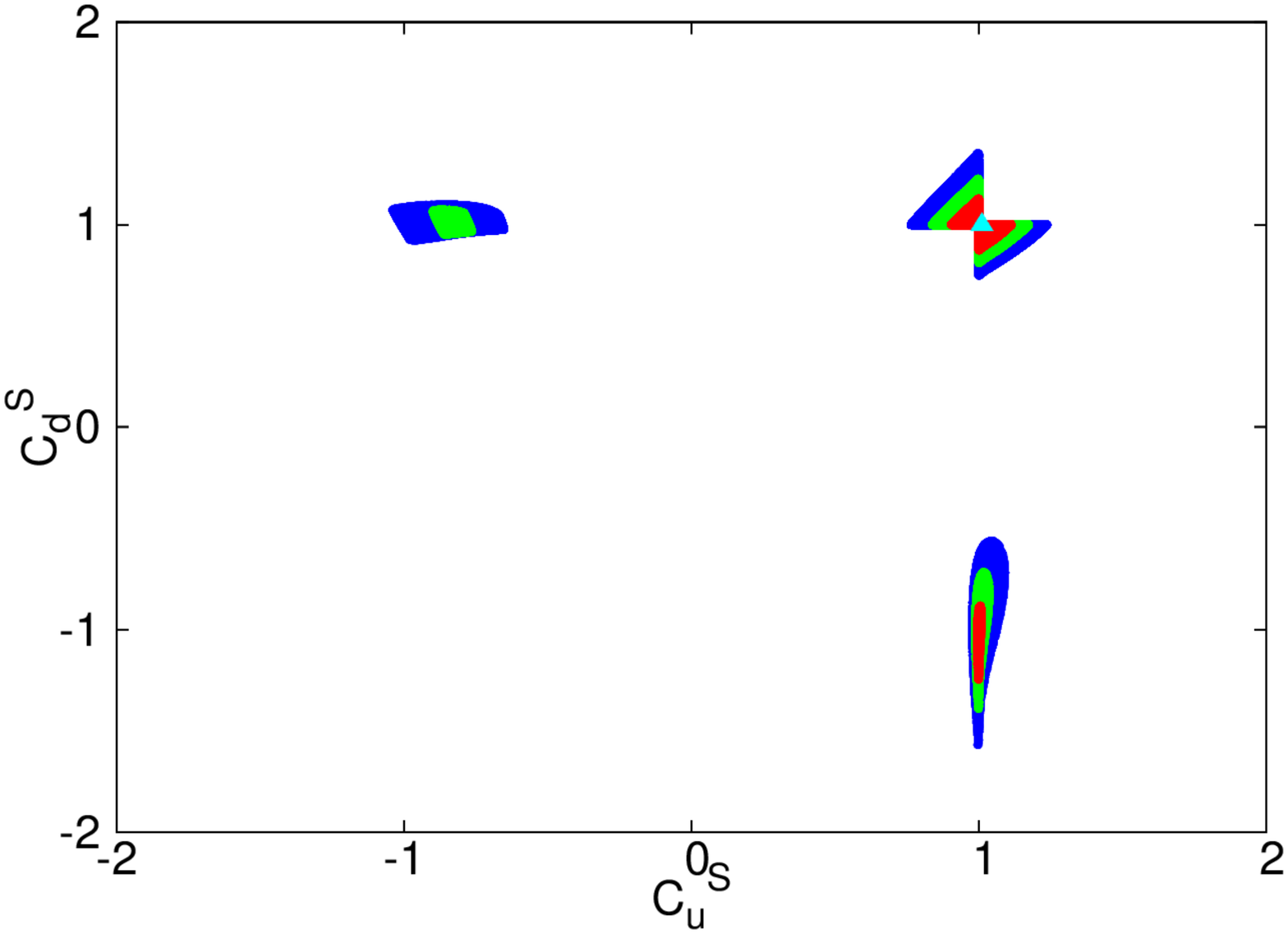}
\includegraphics[width=2.0in]{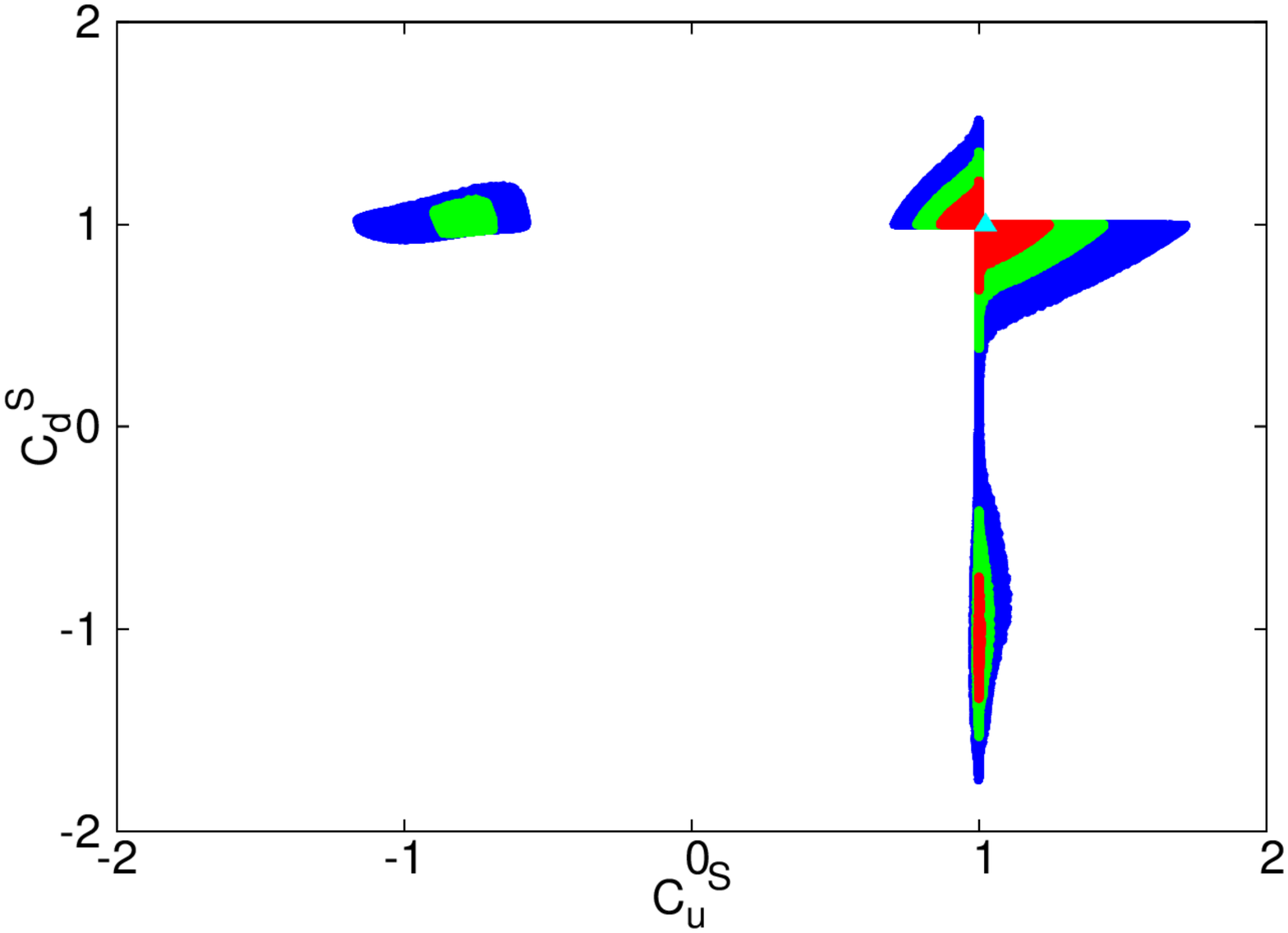}
\includegraphics[width=2.0in]{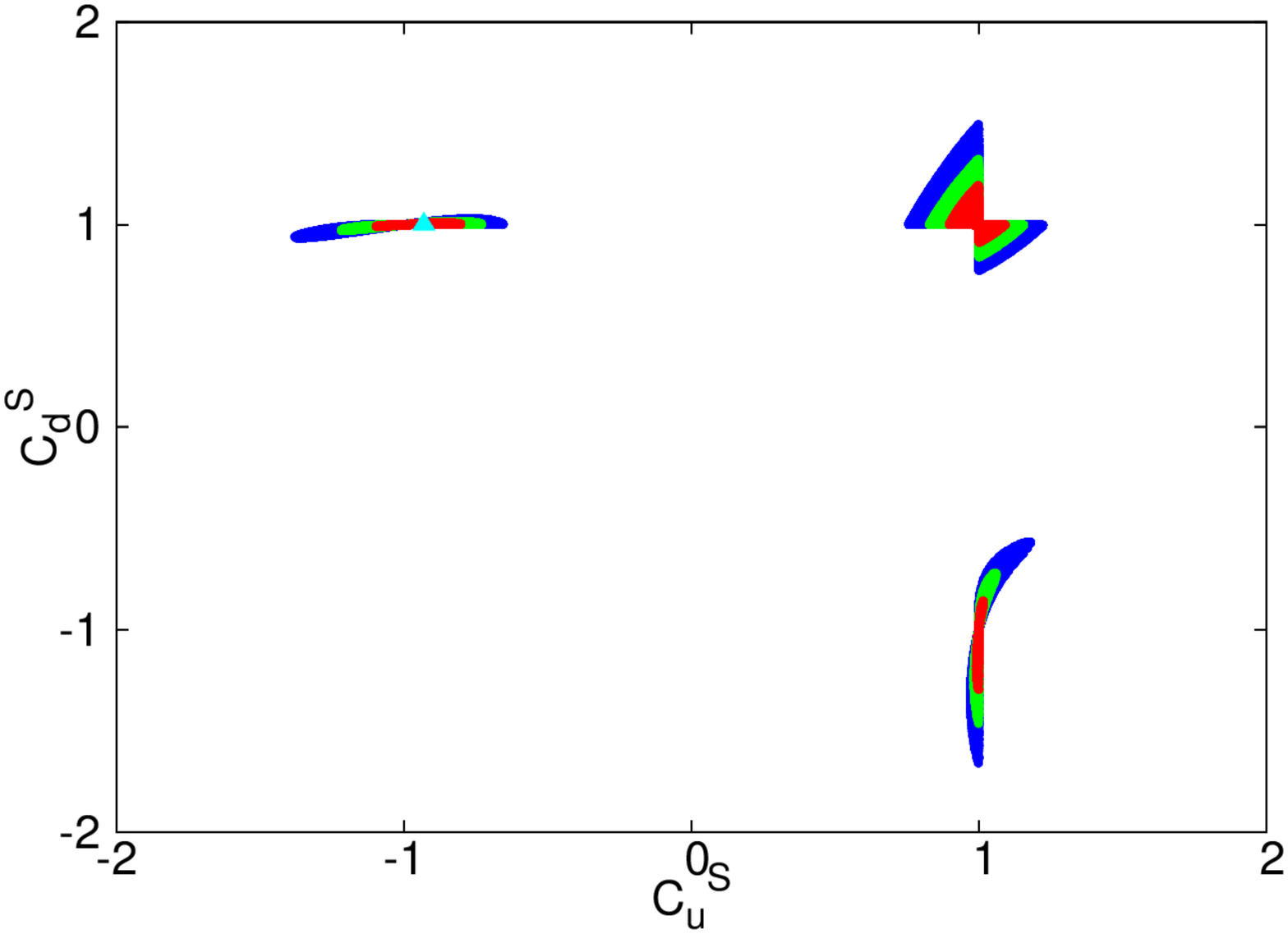}
\includegraphics[width=2.0in]{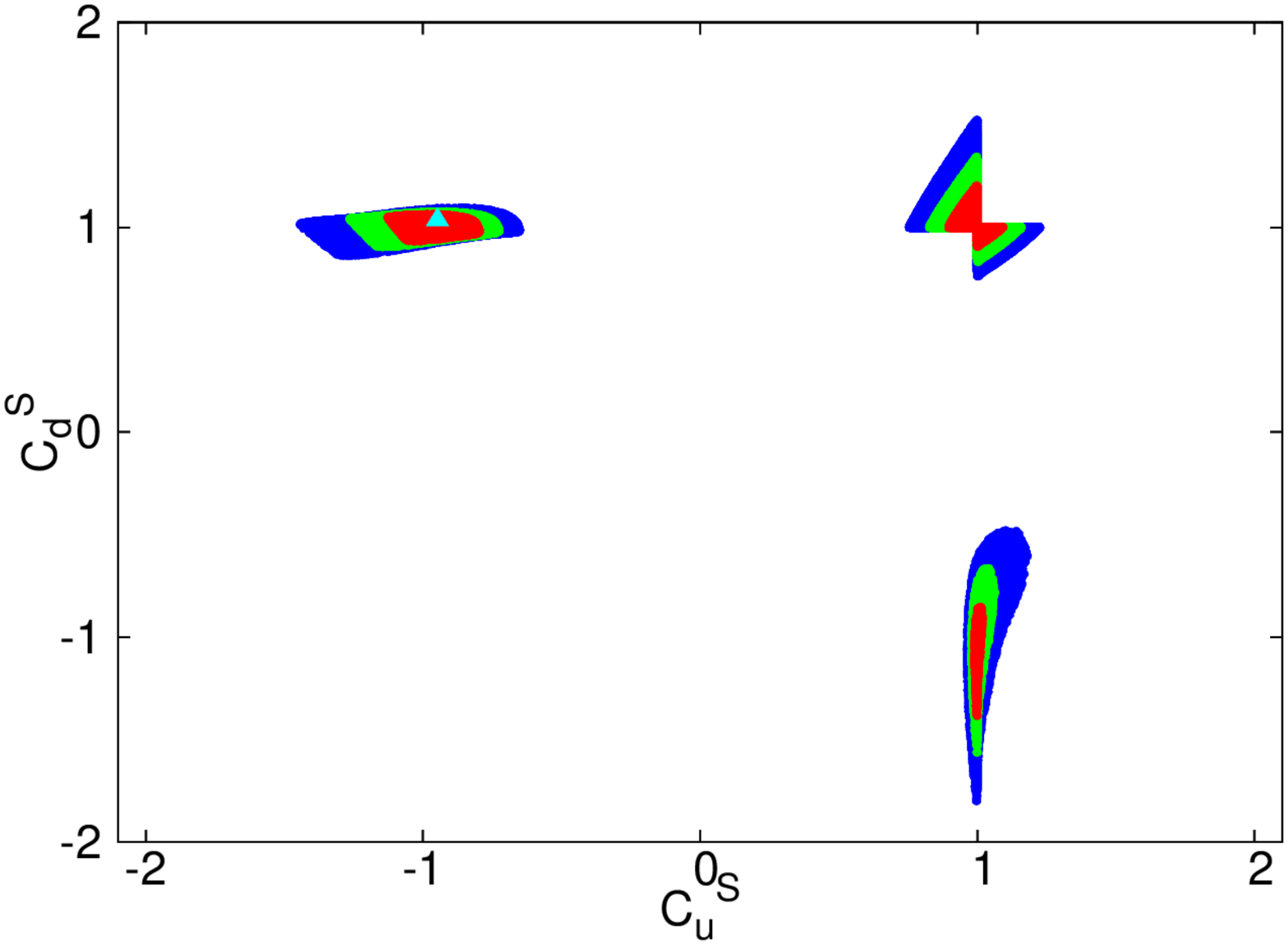}
\includegraphics[width=2.0in]{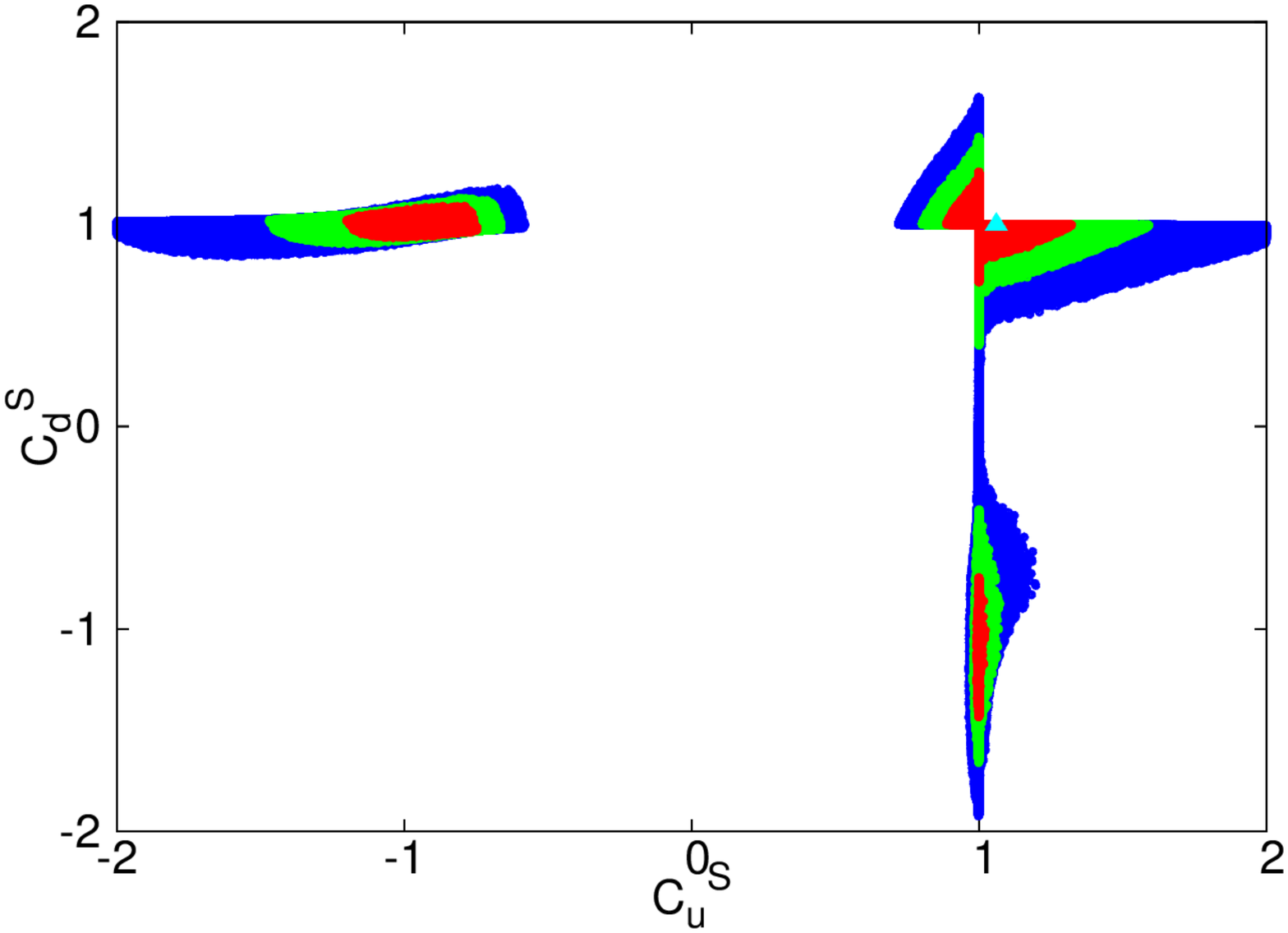}
\includegraphics[width=2.0in]{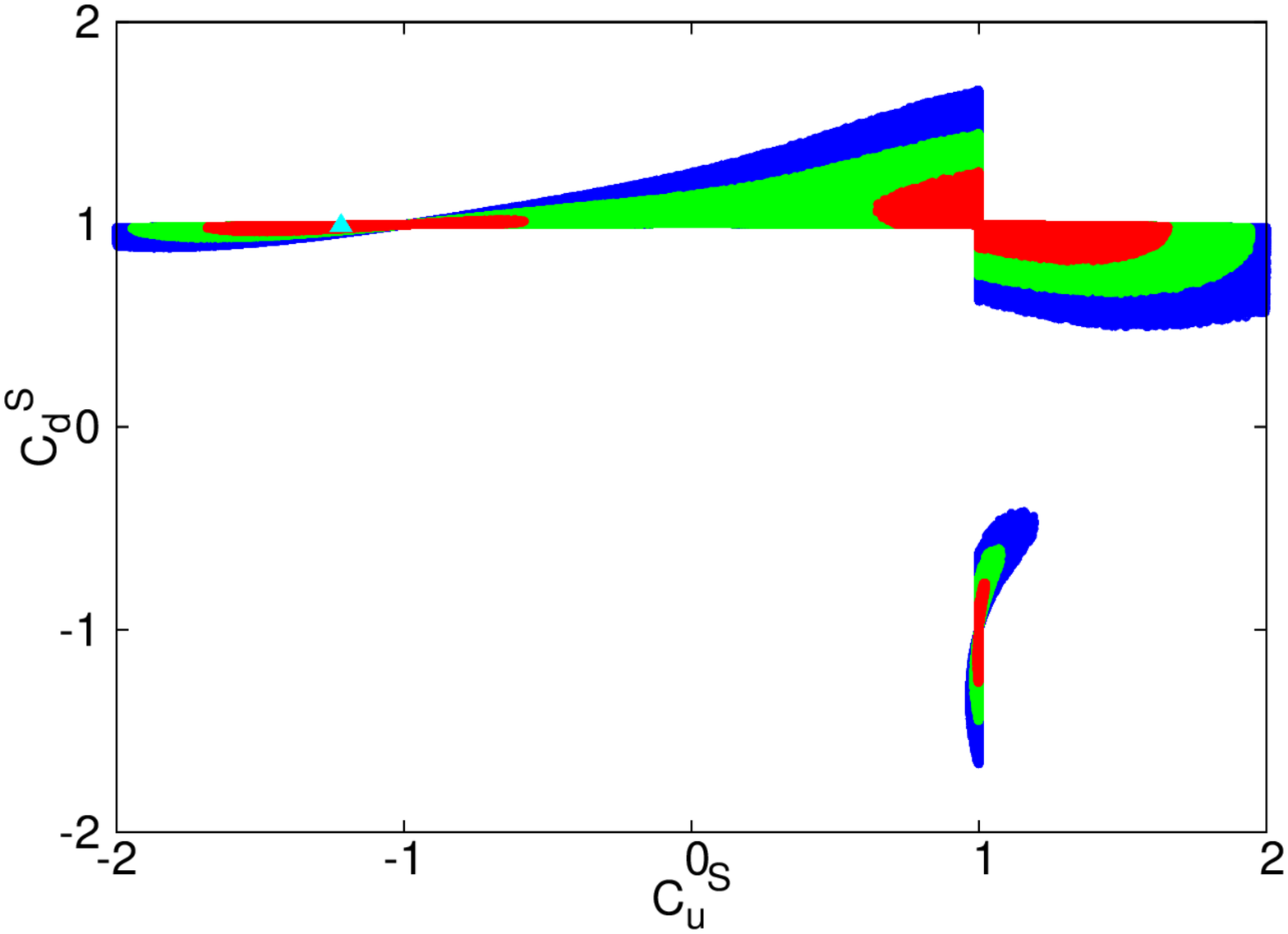}
\includegraphics[width=2.0in]{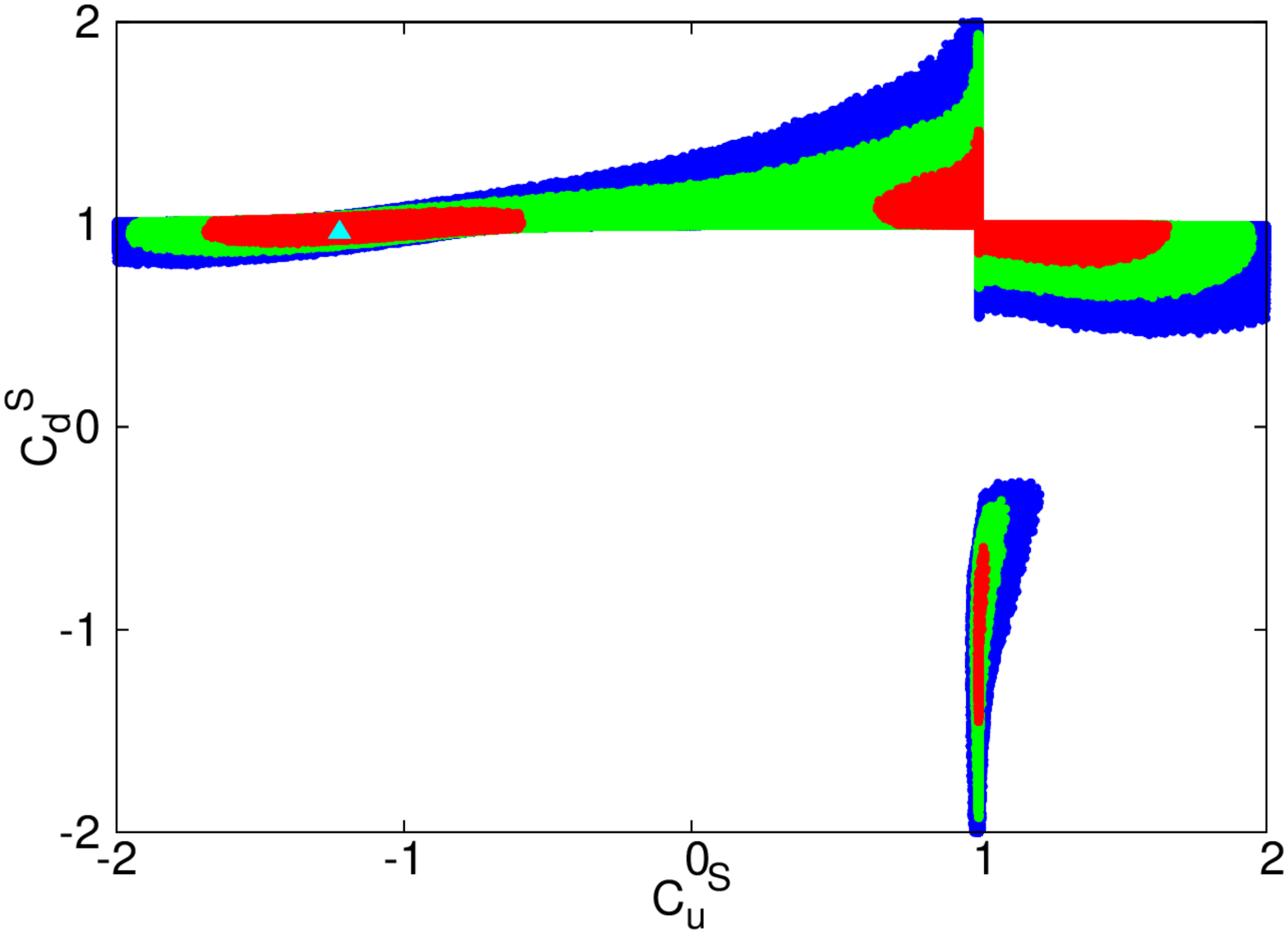}
\includegraphics[width=2.0in]{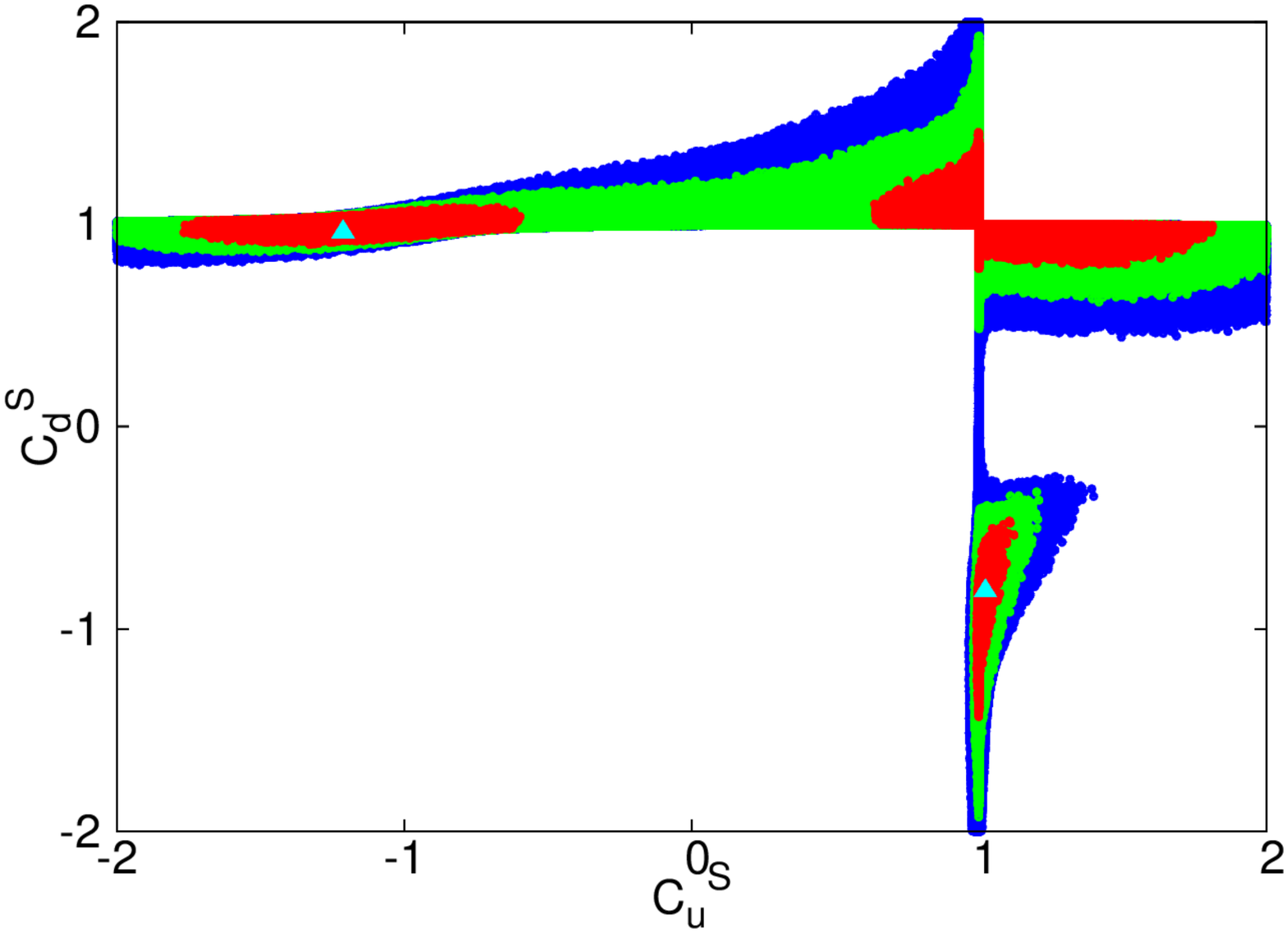}
\caption{\small \label{fig:cd}
The same as in Fig.~\ref{fig:tanbeta} but on
the $(C_u^S, C_d^S)$ plane.
}
\end{figure}

We have analyzed the relevant parameter space in the MSSM with respect to the
most updated data on Higgs boson signal strength. The analysis is 
different from the model-independent one \cite{Cheung:2014noa} mainly 
because $\Delta S^\gamma$ and 
$\Delta S^g$ are related by a simple relation, and up-type, down-type 
and leptonic Yukawa couplings are also related to one another,
such that they are no longer independent.
We have shown in Figs.~\ref{fig:chi2} to \ref{fig:dcpdcg} 
the confidence-level regions in the parameter space for the cases of 
{\bf CPC.II} to {\bf CPC.IV} fits 
by varying a subset or 
all of the following parameters: 
$C_u^S$, $\tan\beta$ (or equivalently $C_v$), $\kappa_d$, $\Delta S^\gamma$,
$\Delta S^g$, and $\Delta \Gamma_{\rm tot}$. 
This set of parameters is inspired by the
parameters of the general MSSM.  
Since the Higgs sector of the MSSM is the same as the 2HDM type II,
the down-type and the leptonic Yukawa couplings are determined once
the up-type Yukawa couplings are fixed.  It implies that $C_u^S$ and
$\tan\beta$ (or equivalently $C_v$) can determine all the tree-level
Yukawa and gauge-Higgs couplings. The effects of SUSY spectrum then
enter into the parameters $\kappa_d$, $\Delta S^\gamma$, and $\Delta
S^g$ through loops of colored and charged particles.

There are improvements in all the {\bf CPC} fits since our analysis of 2HDM
\cite{Cheung:2013rva} a year ago.  The most significant changes in the
Higgs-boson data from 2013 to 2014 were the diphoton signal strengths
measured by both ATLAS and CMS \cite{atlas_zz_2014,cms_aa_2014} while
all other channels were moderately improved.  Overall, all fitted
couplings are improved by about 10\% and the SM Higgs boson enjoys a
large $p$ value close to 1 \cite{Cheung:2014noa}.

The SUSY particles enter the analysis mainly through the loop effects
of the colored and charged particles into the parameters such as
$\Delta S^\gamma$, $\Delta S^g$, and $\kappa_d$ while light
neutralinos with mass less than $M_{H_1}/2$ can enter into $\Delta
\Gamma_{\rm tot}$. We have analyzed the effects of the SUSY spectrum
with the direct search limits quoted in PDG \cite{pdg}.  We offer the
following comments concerning the MSSM spectrum.

\begin{enumerate}
\item 
The effect of $\kappa_d$ on the CL regions is insignificant,
which can be seen easily when we go across from the first column to the 
second column in Figs.~\ref{fig:tanbeta} to \ref{fig:cd}. 
On the other hand, the effect of $\Delta \Gamma_{\rm tot}$ is relatively
large, which can be seen by going across from the second  column to the 
last column in Figs.~\ref{fig:tanbeta} to \ref{fig:cd}. 

\item
Since the mass of the lightest Higgs boson is sensitive to the stop mass,
we especially impose the current Higgs-boson mass limit $M_{H_1} 
\sim 125.5 \pm 6$ GeV (taking on a roughly 3$-\sigma$ level) 
on the parameter space 
in the {\bf MSSM-3} fits with all-SUSY particles.
There are always some underlying assumptions on deriving the mass limits
of stops and sbottoms (also true for other SUSY particles). We have imposed
mild but robust mass limits.

\item 
The {\bf MSSM-1} (chargino) and 
{\bf MSSM-2} (stau) fits
are special cases of {\bf CPC.III.3}
  in which
  $\tan\beta$ (or equivalently $C_v$), $C_u^S$, and $\Delta S^\gamma$
  are varied. Nevertheless, the $\Delta S^\gamma$ is restricted by the
  SUSY parameters $\mu$, $\tan\beta$, and $M_2$ or $M_{L_3,E_3}$ in such a
  way that $\Delta S^\gamma$ is not entirely free to vary.  The
  resulting fits are not as good as the {\bf CPC.III.3} case.

\item In the {\bf MSSM-3} case
in which we consider
  the chargino, stau, stop, and sbottom contributions,
  the preferred $C_u^S$ is very close to 1. The major contribution 
comes from the lightest stau, 
which stands very close to the low mass
limit of 81.9 GeV.

\item The direct search limits on charginos and staus prevent 
the $\Delta S^\gamma$ from becoming too large while those on stops and sbottoms
prevent both $\Delta S^\gamma$  and $\Delta S^g$ from becoming too large.

\item We find that $|\Delta S^\gamma/S^\gamma_{\rm SM}|\lsim 0.1$ when 
$M_{\tilde{\chi}^{\pm}_1}>300$ GeV and $M_{\tilde{\tau}_1}>300$ GeV,
irrespective of the squarks masses.
Note that $S^\gamma_{\rm SM}\simeq -6.6$.

\item Further we observe that
$|\Delta S^\gamma/S^\gamma_{\rm SM}|\lsim 0.03$
when $M_{\tilde{\chi}^\pm_1,{\tilde{\tau}_1}} > 500$ GeV
and $M_{{\tilde t}_1,{\tilde b}_1} \gsim 600$ GeV.

\end{enumerate}

\section*{Acknowledgment}
This work was supported the National Science
Council of Taiwan under Grants No. NSC 102-2112-M-007-015-MY3.
J.S.L. was supported by
the National Research Foundation of Korea (NRF) grant
(No. 2013R1A2A2A01015406).
This study was also
financially supported by Chonnam National University, 2012.
%
%J.S.L thanks National Center for Theoretical Sciences for the
%great hospitality
%extended to him while this work was being performed.

%\section*{Appendices}
%
%\def\theequation{\Alph{section}.\arabic{equation}}
%\begin{appendix}
%%%%%%%%%%%%%%%%%%%%%%%%%%%%%%%%%%%%%%%%%%%%%%%%%%%%%%%%%%%%%%%%%%%%%%%%%%
%%------------------------------------------------------
%\setcounter{equation}{0}
%\section{appendix a }
%\begin{equation}
%appendix eq ...
%\end{equation}
%%
%%---
%
%\end{appendix}

%%%%%%%%%%%%%%%%%%%%%-------------------
%\newpage

\newpage

\clearpage

% ------------- MSSM Spectrum ------------

\end{document}